\documentclass[a4paper,fleqn,usenatbib]{mnras}

\usepackage{newtxtext,newtxmath}

\usepackage[T1]{fontenc}
\usepackage{ae,aecompl}

\usepackage{graphicx}	
\usepackage{amsmath}	
\usepackage{amssymb}	

\title[Weak lensing measurements of the APEX-SZ galaxy cluster sample]{Weak lensing measurements of the APEX-SZ galaxy cluster sample}

\author[M. Klein et al.]{
Matthias Klein,$^{1,2,3}$\thanks{E-mail: mklein@mpe.mpg.de}
Holger Israel,$^{1,3,4}$
Aarti Nagarajan,$^{3}$ 
Frank Bertoldi,$^{3}$ \newauthor
Florian Pacaud$,^3$
Adrian T. Lee,$^{5,6}$
Martin Sommer$,^3$
Kaustuv Basu$^3$
\\
$^{1}$ Faculty of Physics, Ludwig-Maximilians-Universit\"at, Scheinerstr. 1, 81679 Munich, Germany \\
$^{2}$ Max Planck Institute for Extraterrestrial Physics, Giessenbachstrasse, 85748 Garching, Germany\\
$^{3}$ Argelander-Institute for Astronomy (AIfA), University of Bonn, Auf dem H\"ugel 71, 53121 Bonn, Germany\\
$^{4}$ Institute for Computational Cosmology \& Centre for Extragalactic Astronomy, Durham University, South Road, Durham, DH1 3LE, UK\\
$^{5}$ Department of Physics, University of California, Berkeley, California 94720, USA\\
$^{6}$ Physics Division, Lawrence Berkeley National Laboratory, Berkeley, California 94720, USA
}

\date{Accepted XXX. Received YYY; in original form ZZZ}

\pubyear{2019}

\begin{document}
\label{firstpage}
\pagerange{\pageref{firstpage}--\pageref{lastpage}}
\maketitle

\begin{abstract}
We present a weak lensing analysis for galaxy clusters from the APEX-SZ survey. 
For $39$ massive galaxy clusters that were
observed via the Sunyaev-Zel\textquotesingle dovich effect (SZE) with the APEX telescope, we analyse deep optical imaging data from WFI(@2.2mMPG/ESO) and Suprime-Cam(@SUBARU) in three bands. 
The masses obtained in this study, including an X-ray selected subsample of 27 clusters, are optimised for and used in studies
constraining the mass to observable scaling relations at fixed cosmology.
A novel focus of our weak lensing analysis is the multi-colour background selection to suppress effects of cosmic variance on the redshift distribution of source galaxies. We investigate the effects of cluster member contamination through galaxy density, shear profile, and recovered concentrations.
We quantify the impact of variance in source redshift distribution on the mass estimate by studying nine sub-fields of the COSMOS survey for different cluster redshift and manitude limits.
We measure a standard deviation of $\sim 6$\% on the mean angular diameter distance ratio for a cluster at $z\!=\!0.45$ and shallow imaging data of $R\!\approx\!23$ mag. It falls to $\sim 1$\% for deep,  $R=26$ mag, observations. This corresponds to 8.4\% and 1.4\% scatter in $M_{200}$. Our background selection reduces this scatter by $20-40$\%, depending on cluster redshift and imaging depth.
%
%
We derived cluster masses with and without using a mass concentration relation and find consistent results, and concentrations
consistent with the used mass-concentration relation.

\end{abstract}




\section{Introduction}

  Measuring the galaxy cluster mass function, $n(M,z)$, and internal structure of galaxy clusters can help in unveiling the basic nature of dark matter \citep{robertson17}, the initial spectrum of the density perturbations and its evolution from the primordial universe till today \citep{dehaan16,Pacaud18,Costanzi18,Bocquet18}.

The formation of galaxy clusters is known to be sensitively connected to the cosmic expansion rate and the hierarchical structure formation \citep{2002ARA&A..40..539R,2005RvMA...18...76S,2005RvMP...77..207V}. 
Measuring the cluster mass function, therefore, offers a method to obtain constraints on the cosmological parameters additional to other known methods 
such as supernovae type Ia, cosmic microwave background or baryonic acoustic oscillations.
Investigating the evolution of the mass function with redshift can
also constrain the dark energy equation of state parameter $\omega_{\mathrm{DE}}$ \citep{2006astro.ph..9591A}. 

Several methods exist to measure the mass of galaxy clusters based on optical, X-ray and sub-mm observations. Each of them have different advantages and disadvantages with respect to different science goals. In order to use the full constraining power of ongoing or planned cluster surveys for precision cosmology like SPT-3G~\citep{SPT3G}, DES~\citep{DES} or eROSITA~\citep{2010MNRAS.402..191P}, the knowledge of the mass calibration of the cluster obseravbles and biases between these observables are crucial.

In contrast to methods using X-rays or the Sunyaev-Zel\textquotesingle dovich effect (SZE), weak gravitational lensing
probes mass directly without assumptions on the dynamical state of the intracluster medium (ICM).
Especially in cluster mergers the distribution and the dynamical state of the ICM can differ strongly from hydrodynamic equilibrium, 
making it difficult to determine reliable masses with SZE or X-rays. 
For this reason weak gravitational lensing is preferentially used for calibrating mass-to-observable relations, which is the key motivation for the optical follow up observations of the APEX-SZ galaxy clusters \citep{2012Msngr.147....7S,2016MNRAS.460.3432B}.

The APEX-SZ experiment~\citep{2006Dobbs,2011Schwan,2012Dobbs} imaged the SZ decrement of galaxy clusters at 150 GHz using a 280-element Transition-Edge Sensor (TES) bolometer camera on the APEX 12 meter telescope in Atacama Chile~\citep{2006Guesten}. The experiment had $58\farcs$ resolution over a $23\arcmin$ field of view. Apex-SZ observed 42 clusters over a total of 740 hours in the period from 2005 to 2010.

Weak-lensing mass estimation have been actively pursued for numerous galaxy cluster samples (e.g., \citealt{2012arXiv1208.0605A, 2016MNRAS.461.3794O, 2012Postman}) of typical sample size of $\sim$ 20 to 50 clusters.
In this paper, we present the weak gravitational lensing analysis of the APEX-SZ galaxy cluster sample (\citealt{2016MNRAS.460.3432B, 2018Nagarajan}) comprising of 39 galaxy clusters in the redshift range of $z=0.1$ to $z=0.83$. Most of these clusters are X-ray selected. A sub-sample of 27 clusters form a complete galaxy cluster sample in X-ray luminosity and redshift space (see \cite{2018Nagarajan} for details).
One of the main purposes of measuring the cluster masses for this sample is to obtain reasonable constraints on mass-observable scaling relations. In particular, the subject matter of measuring the scaling between the Sunyaev-Zel'dovich effect and the cluster mass is studied in a companion paper by \cite{2018Nagarajan}. Here, we describe the weak-lensing analysis adopted for the cluster mass measurements that form the basis for such work. 

To use weak lensing mass estimates for the mass calibration, the understanding and minimization of systematic effects is important.
Weak-lensing measurements can suffer from systematic effects due to the contamination of the background sources from unlensed foreground or cluster member galaxies.
This can result in an underestimation of the observed Einstein radius by a factor of ~2.5 \citep{2001A&A...379..384C,2005A&A...434..433B}. 
In the current weak lensing literature \citep{2016MNRAS.461.3794O,Melchior17,Dietrich17,Medezinski18}, the selection of galaxies considered to be lensed sources is based on observations in two, three, or five and more bands, reflecting three commonly used selection methods to identify background galaxies. 
While the first two methods rely on a reference catalogue and focus only on the exclusion of cluster member galaxies, the last method uses photometric redshift estimates that allow also to incorporate individual distance estimates for galaxies.

This work presents a three-filter method for background selection, which includes empirically derived photometric redshift estimates for each galaxy based on a comparison to a reference photo-$z$ catalogue. 
This method is a mixed approach between the common three-filter methods (e.g.\cite{Dietrich17,Medezinski18}) and photo-$z$ methods. It reduces the impact of cosmic variance in the observed and reference field on the mass estimates. Furthermore, we give a detailed discussion of the analysis and results using a representative sub-sample of three galaxy clusters in the redshift range from $z=0.15$ to $0.45$, which cover the redshift range of the majority of our cluster sample.

This paper is structured as follows: After a short discussion on cluster selection and data reduction in Section~\ref{clusel}, we give a short introduction to weak lensing theory in Section~\ref{sec:wl}. Section~\ref{sec:bgsel} is focused on the colour properties of galaxies and a lensing optimized selection of background galaxies. In Section\ref{sec:invsys}, we discuss the shear modelling such as, the profile fitting, corrections for contamination and the effect of cosmic variance on our data. Section~\ref{sec:res} shows the lensing results and compares them with measurements from other publications. The last section, Section~\ref{sec:conclu}, presents the conclusions and shows future perspectives.

Throughout this paper we adopt a concordance $\Lambda$CDM cosmology with $\Omega_{m}$=0.3, $\Omega_{\Lambda}$=0.7 and $h\equiv H_{0}/(100 \mathrm{km s}^{-1} \mathrm{Mpc}^{-1})=0.7$.

\section{Cluster selection and data reduction}\label{clusel}

The galaxy clusters for the lensing follow-up observations were selected based on their observations with APEX-SZ SZE detector.
The aim was to cover all SZE detections with $z<1$ using a combination of dedicated observations with the Wide Field Imager (WFI)\citep{1999Baade} and archive data from the same instrument and from Suprime-Cam(@Subaru) with at least three different filter bands. 
This goal was achieved with the only exception of a cluster at $z=0.98$, resulting
in a complete sample up to $z=0.83$. The follow-up observations also includes clusters that were only observed but not detected by APEX-SZ, in order to ensure the completeness of an X-ray selected subsample (\citealt{2018Nagarajan}). 
To illustrate the data analysis we select three clusters as examples. 
The clusters that we choose as examples are selected to reflect the typical redshift range, the difficulties in data reductions, and the optical data quality of the
whole APEX-SZ sample.

The observation strategy with WFI was chosen to make optimal use of archival data in order to minimize the need of additional observations.
For clusters where no archival existed, we observed with WFI in the B-123, V-89 and RC-162 (here after $B$, $V$, $R$) bands. The typical total exposure times
for clusters at $z=0.3$ are $12000$, $4500$ and $15000\,\text{s}$.
In case of existing archival data, we only observed the missing bands or completed
bands which had already some shallower data. This strategy results in a variation of filter bands used for the background selection and shear measurements.
For example for the analysis of RXC 1504, we use $B,V,R$ band WFI observations for the photometry and the Suprime-Cam $V$ band data for shear measurement.

Table~\ref{table:data} shows a summary of the data for the clusters presented in this publication.
\begin{table*}
\caption{Cluster datasets: For the clusters analyzed in this study, 
we list the redshift $z_{\mathrm{NED}}$ as quoted in NED,
the filters (the lensing band is denoted by $\star$), 
exposure time, number $N$ of coadded exposures, and seeing conditions in the lensing band.
In the last column, ``WFI" is the Wide Field Imager at the $2.2\,\text{m}$ MPG/ESO telescope, 
while ``SUP" denotes Suprime-Cam at the Subaru telescope. The subscript "S" is used to distinguish between Suprime-Cam and WFI based data in case of mixed data sets.}
\label{table:data}      
\centering                          
\begin{tabular}{c c c c c c c}        
\hline\hline                 
\hline 
Cluster & $z$ & Filter & Time [s] &N & Seeing [$\arcsec$] & Instrument\\ \hline 
\hline  
A$2744$		&		0.307		&		$B,V,R^\star	$	&		20997		&		40		&	0.87	&	WFI	\\
RXCJ$0019.0-2026$		&		0.277		&		$B,V,R^\star$&		14918		&		30		&	0.82	&	WFI	\\
A$2813$	&		0.292		&		$B,V,R^\star$	&		13497		&		26		&	0.89	&	WFI	\\
A$209$	&		0.206		&		$B,R^\star,Z		$&		2400		&		10		&	0.58	&	SUP	\\
XLSSC $006$		&		0.429		&		$V,R_{\mathrm{C},s}^\star,Z$		&		1800		&		4		&	0.69	&	WFI,SUP	\\
RXCJ$0232.2-4420$		&		0.284		&	$	B,V,R^\star	$	&		13398		&		25		&	0.77	&	WFI	\\
RXCJ$0245.4-5302$		&		0.302		&		$B,V,R^\star	$	&		14697		&		31		&	0.92	&	WFI	\\
A$383$ 		&		0.187		&		$B,V,R_\mathrm{C}^\star		$&		5400		&		18		&	0.69	&	SUP	\\
RXCJ$0437.1+0043$		&		0.284		&	$	B,V,R^\star	$	&		18097		&		33		&	0.82	&	WFI	\\
MS$0451.6-0305$ 		&		0.539		&	$	B,R_\mathrm{C}^\star,Z	$	&		11400		&		26		&	0.83	&	SUP	\\
A$520 $		&		0.199		&		$V,R,I^\star	$	&		3000		&		13		&	0.57	&	SUP	\\
RXCJ$0516.6-5430$		&		0.295		&		$B,V,R^\star	$	&		14548		&		25		&	0.91	&	WFI	\\
RXCJ$0528.9-3927$		&		0.284		&		$B,V,R^\star	$	&		20776		&		48		&	0.85	&	WFI	\\
RXCJ$0532.9-3701$		&		0.275		&		$B,V,R^\star	$	&		13998		&		28		&	0.73	&	WFI	\\
A$3404$	&		0.167		&		$B,V,R^\star	$	&		10078		&		28		&	0.92	&	WFI	\\
Bullet 		&		0.297		&		$B,V,R^\star	$	&		16447		&		35		&	0.77	&	WFI	\\
A$907$	&		0.153		&		$B,V,R_\mathrm{C}^\star$		&		4800		&		16		&	0.53	&	SUP	\\
		&		0.153		&		$B,V,R^\star	$	&		8638		&		24		&	0.80	&	WFI	\\
RXCJ$1023.6+0411$ 		&		0.280	&	$B,V,I^\star$	&		2160		&		9		&	0.62	&	WFI,SUP	\\
MS$1054.4-0321$		&		0.831		&		$V,R^\star,Z	$&		19547		&		33		&	0.73	&	WFI,SUP	\\
MACSJ$1115.8+0129$ 		&		0.348		&	$	B,V,R_s^\star	$&		1200		&		5		&	0.63	&	SUP	\\
		&		0.350		&		$B,V,R^\star	$	&	14497		&		29		&	0.75	&	WFI	\\
A$1300$ 		&		0.308		&		$B,V,R^\star	$	&	16997		&		34		&	0.73	&	WFI	\\
RXC$J1135.6-2019$ 		&		0.305		&	$	B,V,R^\star$		&		14998		&		30		&	0.77	&	WFI	\\
RXCJ$1206.2-0848$		&		0.441		&	$	B,V,I_{\mathrm{C}}^\star$	&		1080		&		3		&	0.73	&	SUP	\\
MACSJ$1311.0-0311$ 		&		0.494		&	$B,V,R_\mathrm{C}^\star$	&		1080		&		6		&	0.62	&	WFI,SUP	\\
A$1689 	$	&		0.183		&		$B,R_\mathrm{C}^\star,I$		&	7984		&		29		&	0.69	&	SUP	\\
RXJ$1347-1145 $		&		0.451		&	$	V,R,Z^\star$		&	2700		&		11		&	0.54	&	SUP	\\
 		&		0.451		&		$B,V,R^\star$		&	16297		&		28		&	0.85	&	WFI	\\
MACSJ$1359.2-1929$		&		0.447	&		$B,V,R^\star	$	&		15032		&		31		&	0.84	&	WFI	\\
A$1835$		&		0.253		&	$B,V,I^\star$		&		1440		&		6		&	0.90	&	WFI,SUP	\\
RXJ$1504$ 		&		0.215		&		$B,V_s^\star,R	$	&		2640		&		11		&	0.78	&	WFI,SUP	\\
A$2163$		&		0.203		&		$V,R_\mathrm{C},I^\star$		&		4500		&		15		&	0.68	&	SUP	\\
A$2204$	&		0.152		&		$V,R_s,I^\star$		&		1680		&		7		&	0.67	&	SUP	\\
RXCJ$2014.8-2430$		&		0.160		&		$B,V,R^\star	$	&		14277		&		27		&	0.80	&	WFI	\\
RXCJ$2151.0-0736$		&		0.284		&		$B,V,R^\star	$	&		14998		&		30		&	0.79	&	WFI	\\
A$2390$		&		0.228		&		$B,R_\mathrm{C}^\star,Z$		&		3700		&		11		&	0.57	&	SUP	\\
MACSJ$2214.9-1359$		&		0.483		&		$V,R_\mathrm{C}^\star,I_{\mathrm{C}}$		&		1200		&		5		&	0.53	&	SUP	\\
MACSJ$2243.3-0935 $		&		0.447		&	$	B,V^\star,Z	$	&		1080		&		3		&	0.75	&	SUP	\\
RXCJ$2248.7-4431 $		&		0.348		&	$	B,V,R^\star	$	&		32995		&		55		&	0.80	&	WFI	\\
A$2537 $		&		0.297		&		$V,R_\mathrm{C}^\star,I$		&		2400		&		5		&	0.69	&	SUP	\\
RXCJ$2337.6+0016 $		&		0.278		&		$B,V,R_\mathrm{C}^\star	$	&		720		&		3		&	0.62	&	SUP	\\
\hline  
\end{tabular}
\end{table*}
The basic data reduction steps (de-biasing, flatfielding, astrometry, absolute and relative photometry, weighted co-addition) 
were conducted with \texttt{THELI}~\citep{2005AN....326..432E,2013ApJS..209...21S} data reduction pipeline following closely the steps described in
\citet{2010A&A...520A..58I}.
In addition to these steps described above, we used the outlier rejection option in \texttt{THELI} with a rejection threshold of 7 or higher for cluster data with a large number
of exposures per filter. This option allows us to remove slow moving asteroids which are difficult to identify in individual images.
For MS0451.6$-$0305, we were able to reuse coadds that were produced and analysed in \citet{2010A&A...514A..60S}.
In the case of RXCJ1347, half of the archival Suprime-Cam data were observed with a field-of-view rotated by $90\degr$.
Since inclusion of these images resulted in a significantly poorer
quality of the astrometric solution, we did not consider this data further.

\subsection{Photometric calibration}

The indirect calibration of the absolute photometric zero point is based on observations of standard fields \citep{2000PASP..112..925S} during photometric nights, where the atmospheric extinction is assumed to be stable over night.
In order to calibrate data sets obtained under unknown or non-photometric conditions and to ensure a highly accurate colour calibration,
we performed a two-step calibration for the WFI data where the last step was also applied to the Suprime-Cam data.

In the first step, we perform a WFI-internal colour calibration, by matching the colours of the stellar locus of each cluster field 
with that of a field observed under photometric conditions. This can be done by simply shifting the position of the main sequence
in colour-colour-space without the need of rotation or stretching.
We then use the stars of the matched fields which are brighter than $R=21.5$ mag as a reference main sequence for the second step.

\begin{figure}
   \centering
    \includegraphics[width=0.99\linewidth]{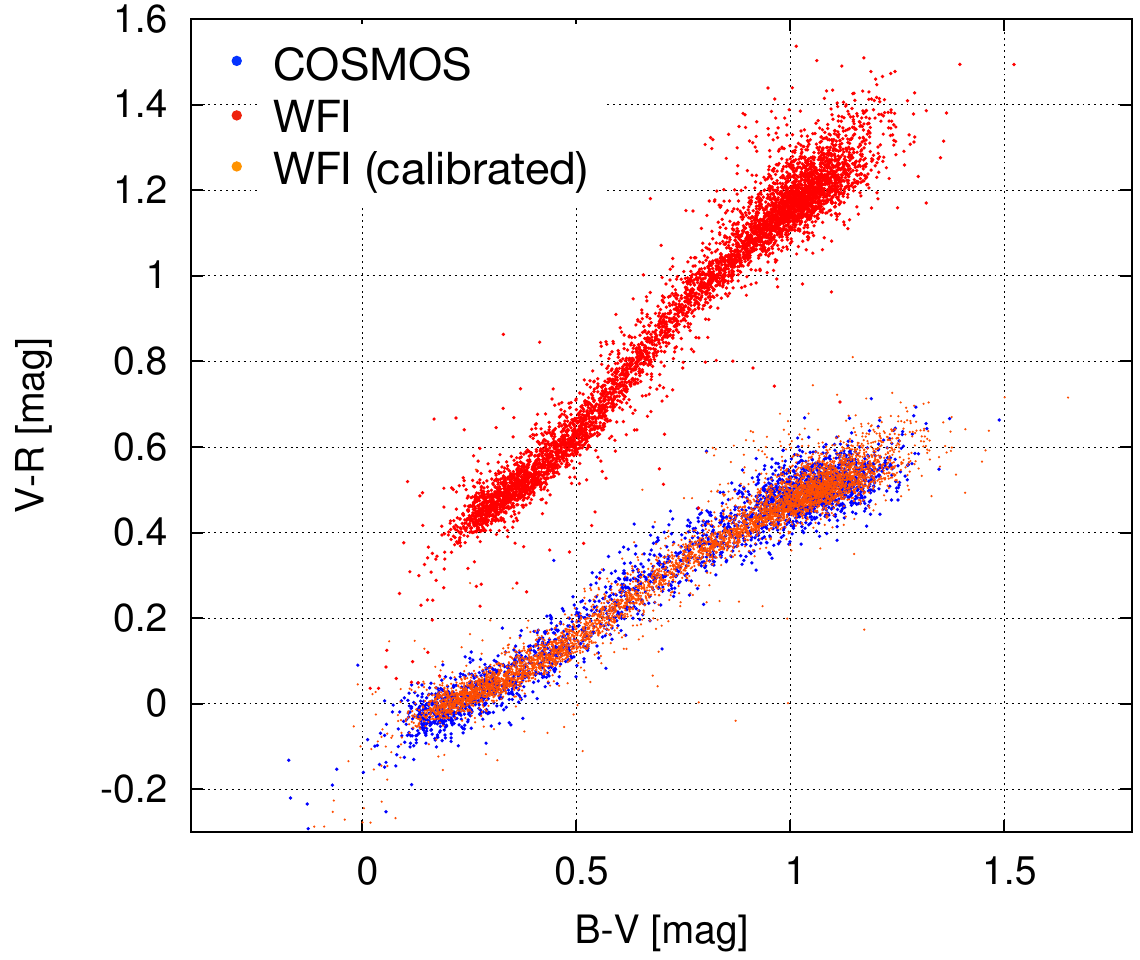}
    \vskip-0.08in
      \caption{Photometric calibration by stellar locus regression (SLR). Blue symbols show stars $R\!<\!22$ mag in the COSMOS field  (using the $B,V,R_{\mathrm{C}}$ bands). Red symbols denote stars $R\!<\!22$ observed with WFI (using the $B,V,R$ bands). Orange symbols denote the same WFI stars after SLR calibration.} \label{FigStars}
 \end{figure}

The second step is a stellar locus regression (SLR) comparable to \citet{2009AJ....138..110H} in order to perform a colour transformation between the WFI or Suprime-Cam bands and the related bands in the COSMOS photo-$z$ catalogue
\citep{2009ApJ...690.1236I}.
We measure the scaling, rotation and shifting terms by matching the prominent features of the stellar locus of the calibrated WFI sequence with that of the COSMOS field .
The resulting excellent agreement can be seen in Figure~\ref{FigStars}.
 
Calibrating the colours against COSMOS makes a background selection
possible which is relatively independent of the instrument used because the colours are converted to the same reference system.
\cite{2009AJ....138..110H} also showed that applying a SLR can also account for galactic extinction if the reference locus is calibrated for that. This is the case for the COSMOS photo-$z$ catalogue. We, therefore, do not apply any additional extinction correction.
The SLR also needs to be applied to Suprime-Cam data. However, as the COSMOS photo-$z$ catalogue is based on Suprime-Cam observations the colour calibration is in most cases straightforward.
Since we calibrate the colours against the same colour system, we use from now on just the shortcuts $B$, $V$ and $R$ band irrespective of whether it is, e.g., the \texttt{V-89} (WFI) or
the  \texttt{W-J-V} (Suprime-Cam) band.

To obtain the absolute zero points, we match our catalogue with the AllWISE catalogue \citep{Cutri14} and perform a stellar locus match in the 
$V\!-\!R$ vs $R\!-\![W1]$ plane, keeping the $V\!-\!R$ colour fixed to the values from the previous calibration.  
AllWISE increases the depth of the Wide-Field Infrared Survey explorer (WISE, \cite{2010AJ....140.1868W}) catalogues.
Here, $[W1]$ denotes the $3.4\,\umu\text{m}$ WISE passband.

By matching the $V\!-\!R$ vs $R\!-\![W1]$ locus to that of our
reference catalogue we ensure that that the overall zeropoint matches the photometric system of the reference catalogue. 
Using $30\arcmin\times 30\arcmin$ subfields of the reference catalogue and applying the same calibration method yields a scatter in the absolute zeropoints of $\sim\!0.05\,\text{mag}$. 
Since we expect correlations between the different subfields, the scatter has to be taken as a lower limit. 
We investigate the impact of the zeropoint scatter on the derived lensing properties in Sec.~\ref{zpscheck}.
Repetititon test show that we are able to recover the relativie zero points to accuracy of 0.013 mag.

\section{Weak gravitational lensing by galaxy clusters}\label{sec:wl}
\subsection{Lensing theory}

This section briefly summarizes the basic theory of weak gravitational lensing by galaxy clusters used in this paper. For a more complete introduction to weak gravitation lensing we refer the interested reader to \cite{2001PhR...340..291B} and \cite{Schneider2006}.

The mapping of a gravitational lens between true position $\vec{\alpha}$ and observed position $\vec{\theta}$ can be described by the Jacobian Matrix
\begin{equation}
 \mathcal{A}(\vec{\theta})=\frac{\partial\vec{\alpha}}{\partial\vec{\theta}}=\left(\delta_{ij}-\frac{\partial^2\psi(\vec{\theta})}{\partial\theta_i \partial\theta_j}\right).
\end{equation}
Here $\psi$ is the lensing potential, which is related to the surface mass density $\kappa$ as
\begin{equation}
 \psi(\vec{\theta})=\frac{1}{\pi}\int \kappa(\vec{\theta^\prime}) \ln|\vec{\theta}-\vec{\theta^\prime}| \mathrm{d}\vec{\theta^\prime}.
\end{equation}

The surface mass density $\kappa$ is generally expressed in units of the critical surface density $\Sigma_{\mathrm{crit}}$, which is defined as
\begin{equation} \label{eq:sigma crit}
\Sigma_{\mathrm{crit}}=\frac{c^2}{4\pi G D_{\mathrm{d}}} \frac{D_{\mathrm{s}}}{D_{\mathrm{ds}}},
\end{equation}

where $c$ is the speed of light, $G$ the gravitational constant, and $D_{\mathrm{d}}$, $D_{\mathrm{s}}$, $D_{\mathrm{ds}}$ are the angular diameter distances between observer and deflector (lens), observer and source, and between deflector and source. 
The second term of this equation describes the strength of the light deflection in dependency of the source distance and is usually referred to as angular diameter ratio
$\beta=D_{\mathrm{ds}}/D_{\mathrm{s}}$.

Using the shear $\gamma_1=(\psi_{,11}-\psi_{,22})/2$, $\gamma_2=\psi_{,12}$, and convergence 
$\kappa=(\psi_{,11}+\psi_{,22})/2$, we can rewrite the Jacobian matrix and get
\begin{equation}
 \mathcal{A}=
 \begin{pmatrix}
      1-\kappa - \gamma_1 & -\gamma_2 \\
     -\gamma_2 & 1-\kappa +\gamma_1
 \end{pmatrix}
 .
\end{equation}
The actual observable in the weak lensing shear measurements is not the shear but the reduced shear, which is given as
\begin{equation} \label{eq:redshear}
 g(\theta)=\frac{\gamma(\theta)}{1-\kappa(\theta)}.
\end{equation}
Using the reduced shear, the lens mapping can be expressed as
\begin{equation}
 \mathcal{A}(\boldsymbol{\theta})=(1-\kappa(\boldsymbol{\theta}))
 \begin{pmatrix}
  1-g_1(\boldsymbol{\theta}) & -g_2(\boldsymbol{\theta}) \\
  -g_2(\boldsymbol{\theta}) & 1+g_1(\boldsymbol{\theta})
 \end{pmatrix}
 .
\end{equation}

In the weak lensing regime with $\kappa$, $|\gamma|\ll 1$, $\text{det}\,\mathcal{A}(\theta)>0$, we can decompose the image distortions into two different effects: 
the image shape distortion and the (de-)magnification.
The reduced shear describes the distortion of the source image caused by the gravitational lens. In the weak lensing case, a circular source image gets distorted to an ellipse with ellipticity $\varepsilon=g$.

The magnification effect which can be described by the inverse Jacobian determinant,
\begin{equation} \label{eq:jacdet}
\mu=\frac{1}{\text{det}\,\mathcal{A}}=\frac{1}{(1-\kappa)^2-|\gamma|^2}=\frac{1}{(1-\kappa)^2(1-|g|^2)}.
\end{equation}
increases the size of the galaxy image by the factor $\mu$ leaving the flux density conserved.

Since the intrinsic size and ellipticity of the individual source are unknown, magnification and shear can only be measured over a sample of sources where the average intrinsic property can be estimated.
In the case of shear measurements, the intrinsic ellipticities of a sufficiently high number of background galaxies are averaging to zero, and thus the observed image ellipticities are then
a direct measure of $g$.

\subsection{The angular diameter distance ratio}

The lensing induced ellipticity $\varepsilon$ scales with the angular diameter distance ratio $\beta$ because $\kappa$ and $\gamma$ do so.  
To estimate $\beta$ for each of the faint
source galaxy, photometric redshifts (photo-$z$) are needed. Most authors that are using three filters do not attempt to derive individual redshift estimates \citep{2012arXiv1208.0597V,2012MNRAS.427.1298H,2010MNRAS.405..257M} and use polygonal shaped regions in color-color space to exclude cluster members. 
Empirically derived photo-$z$'s of individual galaxies using three filters is rather rare in cluster weak lensing studies but was previously used in smaller studies by \cite{2014MNRAS.442.1507G} and \cite{2019Rehmann}.

The standard approach for three-filter observations is to estimate a single mean angular diameter distance ratio $\langle\beta\rangle$. 
This value is obtained by investigating the redshift distribution of a reference photo-$z$ catalogue by applying similar colour, size and flux cuts as in the lensing observations.
This approach can be called \emph{source sheet approximation}, since it assigns one single distance estimate to all galaxies.

Fig.~\ref{betaplot} shows the dependency of $\beta$ on that single source redshift $z_{\mathrm{s}}$ for different cluster redshifts $z_{\mathrm{d}}$. 
For clusters at low redshifts and deep observational data, most of the observed galaxies
will lie on the flat part of the $\beta$ curve. Therefore the scatter in $\beta$ is small even for broad source redshift distributions,
and the noise introduced by the source sheet approximation is negligible. 
At higher redshifts or shallower data most of the source will lie on the steep part of the $\beta$ curve, introducing an additional source of noise to the shear estimate if only an average $\langle\beta\rangle$ is assumed.

The source sheet approximation is prone to biases, which are introduced by the systematic changes of the redshift distribution that correlate with the mass distribution in a cluster field.
First, such a systematic change can be introduced by the magnification effect (Eq.\ref{eq:jacdet}) which is strongly correlated with the mass distribution. 

A second source of such correlated variation
of the redshift distribution can be introduced by targeted observations of clusters. 
Due to vignetting, the noise properties of the image change systematically from the centre to the outskirts. 
This again affects the probability of a source to be detected.
If centred on a cluster, this effect would follow the overall mass profile of the cluster and therefore bias the result.

A third source of bias, related to the lack of precise distance estimates, is the contamination by cluster galaxies. 
Since the distribution of member galaxies correlates with the (projected) mass distribution while the cluster members
systematically change the redshift distribution compared to the reference field, their contamination of the lensing signal has to be minimized or properly accounted for. 
In particular, uncorrected contamination by cluster members systematically affects the observed shear profile.

While the first two effects tend to boost the shear signal towards the centre, 
yielding more concentrated mass distributions than the true one when modelling the shear profile (Sec.~\ref{sec:spm}), 
the last effect suppresses the shear signal towards
the center, resulting in an underestimate of the concentration parameter. 
Low concentration parameters can therefore be a hint of significant contamination by cluster galaxies.

\subsection{PSF correction and shape catalogues}  \label{sec:ksb}
Beside the challenges related to the distance estimates of the lensed sources, accurate shape measurements of these galaxies are of critical importance for unbiased mass estimates.
Important in this context is that the instrumental effects on the observed ellipticities are accounted for and that the measured ellipticity is an unbiased estimate of the true one.

Both instruments used, Suprime-Cam as and WFI are known for their weak lensing capability for more then a decade \citep{2002A&A...395..385C,2002PASJ...54..833M}. Their
smooth variation of the PSF anisotropy can be modelled by a low-order ($3\!\!\leq\!\!d_{\mathrm{ani}}\!\!\leq\!\!5$) polynomial in image coordinates.
Therefore they are well suited for our purpose.

The software and measurement scheme applied to the co-added images in order to get positions, PSF-corrected ellipticities and photometric
data for the source galaxies is mostly identical to those presented in
\citet{2010A&A...520A..58I,2012A&A...546A..79I}.
We, therefore, briefly summarize the methods in order to show the basic steps of the data processing and refer the interested reader to the aforementioned papers for a more detailed discussion.

The photometry was performed using \texttt{SExtractor} in dual-image mode on seeing-equalized images, using the unconvolved lensing band as detection image.
This ensures good quality galaxy colours and proper total flux estimates in the lensing band.
For the lensing measurement we apply  the ``TS'' shear measurement pipeline \citep{2006MNRAS.368.1323H,2007A&A...468..823S,2009arXiv0901.3269H}, 
an implementation of the KSB+ algorithm \citep{1995ApJ...449..460K,2001A&A...366..717E}.
The PSF anisotropy is traced by measuring the brightness distribution of
sources identified as unsaturated stars in the magnitude versus half-light radius $\vartheta$ plots. 

For the clusters observed with WFI, we use the $R$ band for the shape analysis since it is usually the deepest and has the best seeing out of the three bands.
In cases where we have to rely on archival data, we choose the filter which offers the best compromise in terms of seeing, depth and number of images. 
The latter factor becomes important when the number gets small.
Since we apply our PSF correction on the coadded images, we rely on image distortions being smooth over the whole field.
This assumption can be violated at the position of chip gaps if the observation conditions have changed between the exposures.
In some cases such as RXC\,1504, we had to combine the photometry data of WFI with the shear measurements of Suprime-Cam since the WFI $R$ band data were split into two different pointings making it difficult
to correct for the PSF anisotropy.
The residual PSF anisotropy after correction has a dispersion of $0.004\!\leq\!\sigma\!\leq\!0.011$ in the 
lensing image and a nearly vanishing mean value typically two orders of magnitude smaller than the dispersion. 

We keep sources in our lensing catalogue which are fainter than the brightest unsaturated point sources and whose half light radius are larger
than that of stars $\vartheta_{\mathrm{gal}}>\vartheta^{*}_{\mathrm{max}}$. 
Since the estimate of the half light radius gets more and more noisy for fainter sources we also include
faint sources with $\vartheta_{\mathrm{ana}}<\vartheta_{\mathrm{gal}}\leq \vartheta_{\mathrm{max}}^{*}$ for $R>23.5$. 
For sources in the same size range 
but between $R=23.5$ and the magnitude 
$R_\mathrm{max}^*\sim 22$ where the stellar locus in the $\vartheta$ vs.\ $R$ distribution becomes indistinguishable from the cloud of galaxies, 
we exclude all sources which have more then 10 stars in the colour-colour region that is created by the source colours and its errors. 
As a reference, we consider stars in the COSMOS photo-$z$ catalogue.

The choice of the magnitude limit is motivated by the assumption of an absolute magnitude of
$M_{\mathrm{V}}=17\,\text{mag}$ \citep{2004ASPC..318..159H} 
for the faintest main sequence stars, a thin galactic
disc of scale hight of $240\,\text{pc}$ \citep{2008ApJ...673..864J} 
and typical colour of $V-R=1.2$. 
The limit of $R=23.5$ therefore accounts for the majority of stars within the galactic disc
for galaxy clusters at galactic latitudes of $40\degr$ or higher.

\subsection{Shear profile modeling and cluster masses} \label{sec:spm}

In order to obtain cluster masses from our shear measurements we are again following a similar approach as described in \citet{2010A&A...520A..58I} and model the tangential
ellipticity profile $\varepsilon_{\mathrm{t}}(\theta)$ of a cluster with the reduced shear profile
$g(\theta,\Sigma_{\mathrm{crit}};r_{200},c_{200})$ \citep{1996A&A...313..697B,2000ApJ...534...34W} of a NFW \citep{1995MNRAS.275..720N,1996ApJ...462..563N,1997ApJ...490..493N} density profile.
In contrast to \citet{2010A&A...520A..58I} we do not assume a source sheet with common $\Sigma_{\mathrm{crit}}$. Instead we use individual $\Sigma_{\mathrm{crit},i}(\beta_i)$ based on the individually estimated $\beta_i$ for each galaxy $i$.
We derive the best fitting profile parameters $r_{200}$ and $c_{200}$ by minimizing the merit function 
\begin{equation} \label{eq:merit}
\chi^{2}\!=\!\sum_{i=1}^{N}{\frac{\left|g_{i}(\theta_{i},\Sigma_{\mathrm{crit},i};r_{200},c_{\mathrm{NFW}})\!-\!
\tilde{\varepsilon}_{\mathrm{t},i}(\theta_{i})\right|^{2}}
{\tilde{\sigma}_{\!i}^{2}\left(1\!-\!\left|
g_{i}(\theta_{i},\Sigma_{\mathrm{crit},i};r_{200},c_{\mathrm{NFW}})\right|^{2}\right)^{2}}}\quad,
\end{equation}
which is calculated on a regular grid in the $r_{200}$-$c_{\mathrm{200}}$ plane. Here $r_{200}$ corresponds to the radius within which the enclosed mass 
density equals 200 times the critical density $\rho_{\mathrm{c}}(z_{\mathrm{d}})$ at cluster redshift $z_{\mathrm{d}}$.  
The concentration parameter can not be well constrained with our weak lensing data. For we therefore adopt a prior from the mass-concentration relation given in \citep{2013ApJ...766...32B} for the final mass estimate. A detailed discussion on the concentration parameter obtained with and without prior is presented in Sect.~\ref{sec:c200}.
The modified tangential ellipticity $\tilde{\varepsilon}_{\mathrm{t},i}$
is the measured tangential ellipticity multiplied by a global shear calibration factor $f_{0}\!=\!1.08$, identical to that used in \citet{2010A&A...520A..58I}. The error $\tilde{\sigma}_{\!i}$ scales as
$\tilde{\sigma}_{\!i}\!=\!f_{0}\sigma_{\mathrm{\!\varepsilon}}/\!\sqrt{2}$ where $\sigma_{\!\mathrm{\varepsilon}}$ is the dispersion of the measured ellipticities.

The index $i$ runs over all lensing catalogue galaxies with separations within
the fitting range $\theta_{\mathrm{min}}\!\leq\!\theta\!\leq\!\theta_{\mathrm{max}}$
from the assumed cluster center.
The denominator of Eq.(\ref{eq:merit}) accounts for the dependence of the
noise on $g_{i}(\theta_{i})$ itself \citep{2000A&A...353...41S}.
After estimation of $r_{200}$, the cluster mass within that radius can be calculated as follows
\begin{equation} \label{eq:mdelta}
M_{200}\!=\!200 \frac{4\pi}{3} \rho_{\mathrm{c}}(z_{\mathrm{d}})
r_{200}^{3}\quad.
\end{equation}
Calculating $\chi^{2}$ on a grid, some areas of the explored parameter space can result in a model which suggests $g_{i}>0.5$ at $\theta_{i}> \theta_{\mathrm{min}}$. In those cases we change $\theta_{\mathrm{min}}$ to the
value which is fulfilling $g_{i}(\theta_{\mathrm{min}})=0.5$. The impact of this additional constrain is typically negligible and mainly ensures robustness of the results when exploring more exotic areas of the parameter space. It is therefore only needed for the case where the concentration is free to vary. The application of the prior from the mass-concentration relation further ensures a reasonable parameter space.
We centred our profile fits to the position of the BCG and list  $\theta_{\mathrm{min}}$ and $\theta_{\mathrm{max}}$ in Table~\ref{table:res}.

\subsection{Weak lensing convergence maps}

Besides the accurate measurement of the total mass, the mass distribution of the galaxy clusters can also be of interest.
Especially in case of studies of individual clusters, the mass distribution can give insights into the dynamical nature of a cluster.

For the convergence reconstructions we are using the finite-field inversion algorithm \citep{1996A&A...305..383S,2001A&A...374..740S}. 
The mass sheet degeneracy\footnote{
The observed reduced shear is invariant under convergence transformations of the form 
$\kappa\rightarrow\kappa^{\prime} =\lambda\kappa+(1-\lambda)$
for a scalar constant $\lambda$.}
is broken by using the assumption that the mean convergence vanishes along the edge of the images. 
A violation of this assumption would lead to a wrong normalization but would not affect the general shape of the cluster mass distribution.
The lensing convergence or surface mass density $\kappa$ is calculated on a regular grid with a grid size of $\sim5\farcs$. 
On each grid point we use a aperture radius of $\theta_{\mathrm{A}}\!=\!1.5\arcmin$ where the input shear field is smoothed by a truncated Gaussian filter with a FWHM of
 $0.555\,\theta_{\mathrm{A}}$ dropping to zero at $\theta_{\mathrm{A}}$.
By using the assumption of a linear scaling of $\kappa$ with $\beta$ we rescale the observed ellipticites to the average $\beta$. This assumption produces a bias at higher convergence but accounts for changes in image depths and contamination.
For RXC1347 we corrected the contour levels of the WFI map with respect to the average $\beta$  found in the Suprime-Cam data.

\section{Background selection}\label{sec:bgsel}

As stressed in the previous section, distance estimates of the observed galaxies are important to avoid biases and to minimize the number of cluster members contaminating the lensing catalogue.
Unaccounted contamination will cause a dilution of the signal and result in a systematic bias of the mass estimate.
While this might be sub-dominant compared to statistical errors for a single cluster, this systematic effect becomes important for the investigation of large cluster samples.

Current background selections based on two colours typically exclude certain regions in colour-colour space (cc-space). 
One of the most thorough studies of the cc-space for an optimized weak lensing background selection was made by \cite{2010MNRAS.405..257M}.
There, the regions were selected by their mean distance with respect to the cluster center and the number density in colour-colour-space. 
This method allows to find easily the location of the red sequence galaxies for relaxed clusters but not the bluer cluster and field galaxies which are less or not concentrated at the cluster center. 
These galaxies were excluded by introducing a second region, based on overdensity in the cc-plane and the lack of a significant weak lensing signal.
These selection polygons were then justified by evolutionary colour tracks and the application of the colour cuts on a photo-$z$ catalogue. 
However, \citet{2016MNRAS.463.4004Z} developed a photometric method for selecting blue galaxies for their inclusion into lensing catalogues. 

Related to the background selection is also the estimate of $\beta$ for individual galaxies, or,
using the source-sheet approximation, the mean lensing depth $\langle\beta\rangle$.
Colour cut methods usually simply apply
the polygon selection to a reference photo-$z$ catalogue to
derive $\langle\beta\rangle$ and thus do not account for remaining differences between the redshift distributions
of the cluster and reference fields.
A background selection which avoids manual definition of regions and allows for individual distance estimates is therefore highly desirable, especially in case of large cluster samples.

\subsection{CC-Diagram of COSMOS}

\begin{figure*}
   \centering
\includegraphics[width=1\linewidth]{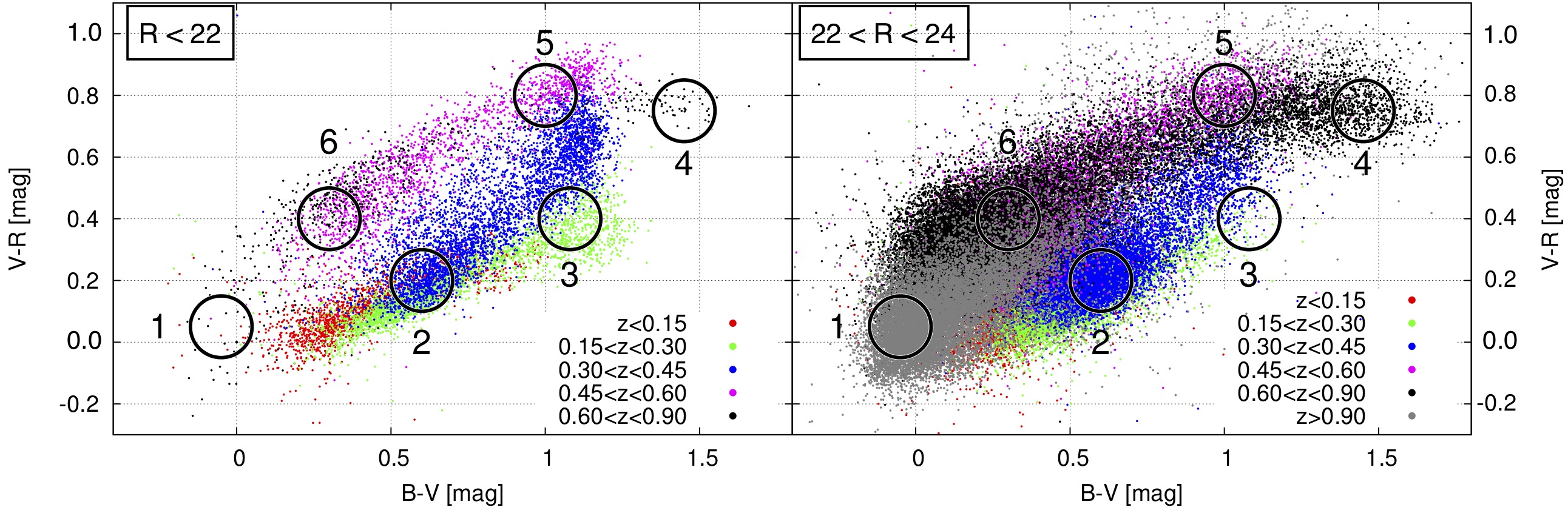}
\vskip-0.05in
      \caption{Distribution of COSMOS photo-$z$ galaxies in colour-colour space. \emph{Left panel:} Galaxies brighter than $R=22$.
\emph{Right panel:} Galaxies $22<R<24$. 
Several redshift slices are colour-coded. 
Circles mark different regions for which redshift distributions are shown in Fig.~\ref{Figure 2}.}
         \label{Figure 1}
 \end{figure*}
\begin{figure*}
  \includegraphics[width=\textwidth]{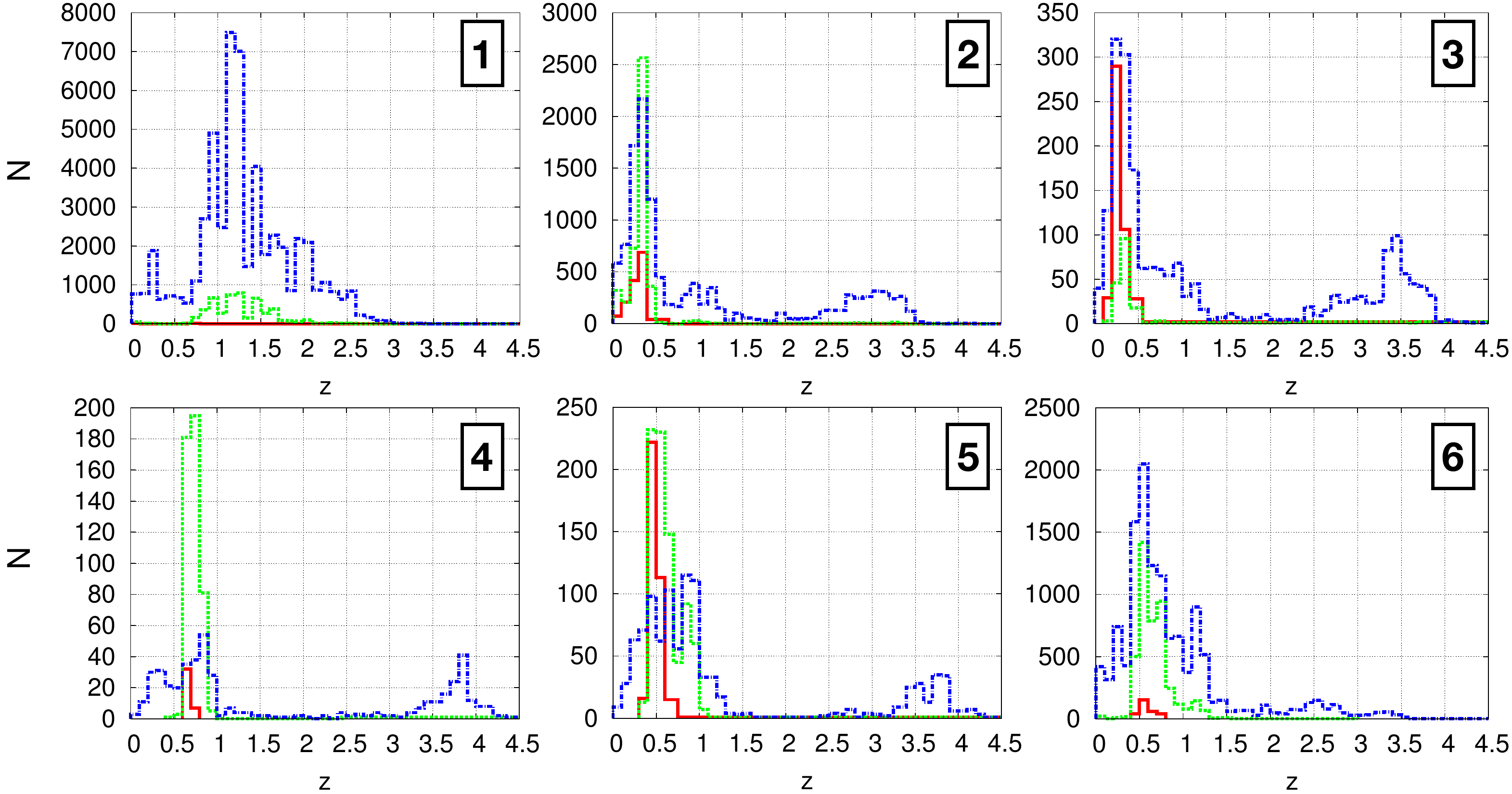}
  \vskip-0.05in
     \caption{Redshift distribution of galaxies falling in the colour-colour space regions marked in Fig.~\ref{Figure 1}. 
Red histograms show galaxies brighter than $R=22$, green histograms galaxies with $22<R<24$, and blue histograms galaxies fainter than $R=24$.}         \label{Figure 2}
\end{figure*}

To help the reader to visualize the properties of galaxies in colour-colour-magnitude space, we shortly discuss these on the example of the $B,V,R$ filter set with COSMOS.

As reference photo-$z$ catalogue we chose the COSMOS photo-\textit{z} catalogue \citep{2009ApJ...690.1236I} because it covers a huge wavelength range with $30$ bands, including bands close or identical to our observational data. In terms of depth and image quality, the used version of the COMSOS catalogue is comparable or deeper than most of our data.
The panels of Fig.~\ref{Figure 1} show the colour-colour diagrams of galaxies in the COSMOS catalogue for magnitudes
$R<22\,\text{mag}$ and $22<R<24\,\text{mag}$.
The latter roughly reflects the typical depths of our WFI observations.
Different plotting colours indicate different redshift bins. 

In these panels one can recognise regions which are populated with galaxies with a tight redshift distribution,
and other regions with a mixture of different redshifts. 
Investigating the distribution of galaxies of a single redshift slice, we see that they are arranged in elongated stripes in the cc-diagram. 
These stripes reflect the diversity of galaxy types,
starting at the blue corner (low $B-V$ and $V-R$ values) with strongly star forming galaxies and ending at the red side with elliptical galaxies with no significant star formation.

The fact that we observe that kind of redshift distribution in the cc-diagram gives us the opportunity to identify red as well as blue cluster members at a certain redshift.
In order to highlight how the redshift distribution
varies with position in colour-colour and magnitude space, 
we show in Fig.~\ref{Figure 2} the redshift distribution of galaxies in six different regions and in three different magnitude bins.
For the two brighter magnitude bins, regions with very tight redshift distributions exists, such as Regions 3 and 4. 
Others, like Region 1 and 5, show a broader
distribution in all magnitude bins. 
One can also see a magnitude dependency in several regions.
Region 1 has basically no sources brighter than $R=22\,\text{mag}$ but is the densest region in higher magnitude bins.
For all except for Region 1, going from the brighter to the fainter subsample, 
one can see a broadening of the distribution as well as the rise of a second broad peak at $z>2.5$.
The broadening of the distribution comes partially from the increasing
photometric errors of the catalogue, but the rise of the high-$z$ peak is also caused by real sources at that redshift.
These sources are indistinguishable from lower redshift sources for the filter set we used here.
For sources fainter than $R=24\,\text{mag}$ the broadening of the redshift distributions results in a mix of a vast redshift range.
However, this is not harmful for our purposes: 
the typical image depth before lensing cuts usually does not exceed $R\sim 24.5\,\text{mag}$,
while size and shear measurement quality cuts further reduce the number of faint sources. 


To exclude sources with strong bimodal redshift distributions where one peak is at or below the cluster redshift while the second is at high redshifts we calculate the probability of a galaxy to be background
galaxy $p_{\mathrm{g}}$ for each galaxy $g$.
Requiring a minimum $p_{\mathrm{g}}$ helps to avoid such cases in the lensing catalogue. 
Additionally it allows to include galaxies with high probability of being background galaxies, even if the average distance is only slightly above the cluster redshift.

In addition to the limitations due to the available filter set, some limitations arise from the used photo-$z$ catalogue.
Before introducing a background selection, these limits of the used photo-$z$ catalogue have to be investigated. 
One of the major points of concern is the cosmic variance. 
The volume probed in the two square degree field of COSMOS is smaller for lower redshift ranges, which makes the influence of cosmic variance more prominent.
For example, there is no massive galaxy group or cluster below redshift $z<0.2$ in this field, resulting in the absence of red galaxies of this redshift in the COSMOS catalogue.
The lack of these galaxies may influence the background selection but should not effect the mean redshift distribution of the background sample. 
Cosmic variance at higher redshifts 
and its effect on the mean lensing depth will be discussed in Sect.~\ref{cosmvar}.

\begin{figure}
   \centering
   \includegraphics[width=\linewidth]{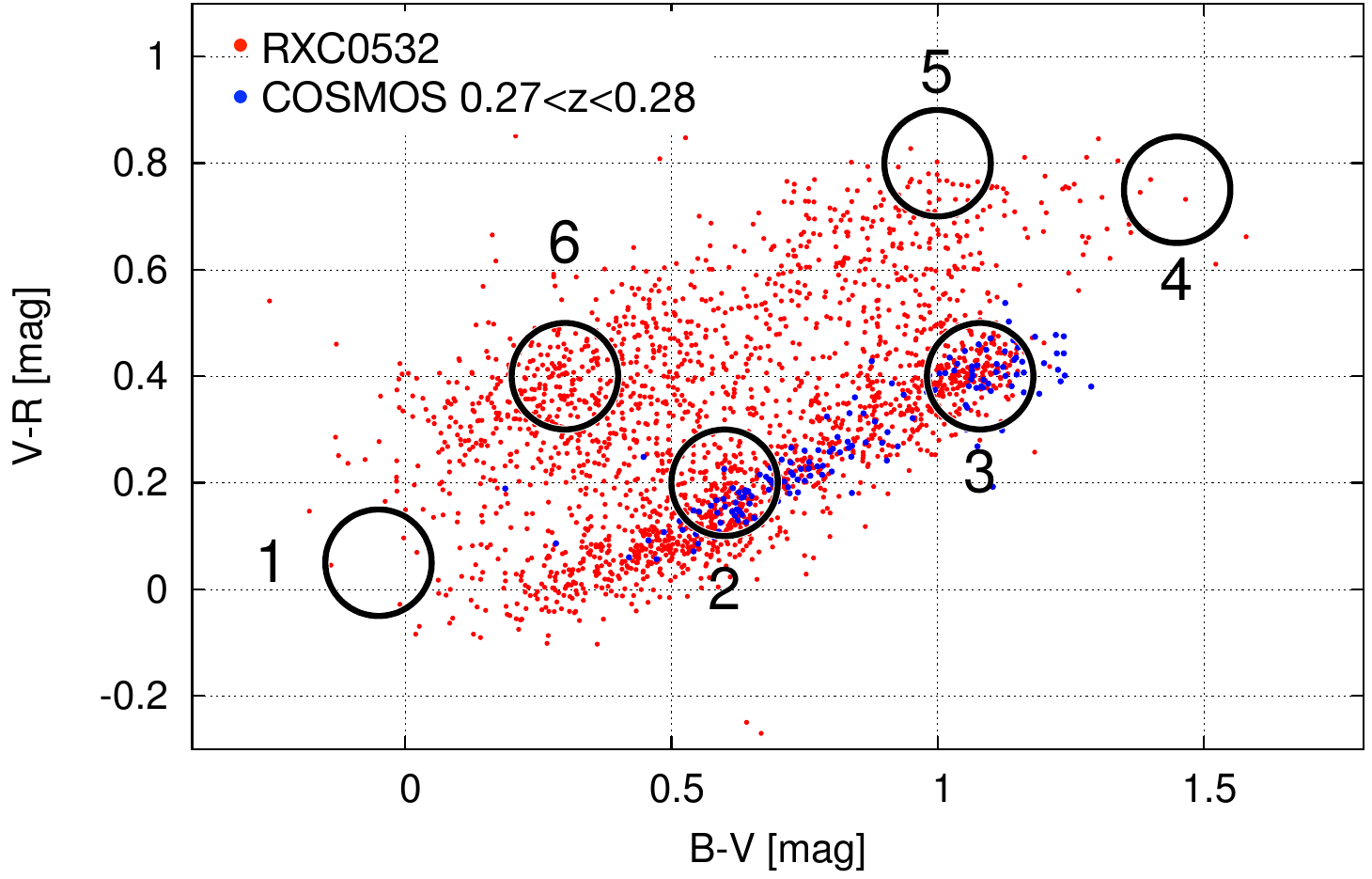}
   \vskip-0.08in
      \caption{Colour-colour diagram of the RXCJ\,0532 cluster field, 
showing galaxies brighter $R=22\,\text{mag}$ (red symbols).
Blue symbols show galaxies in COSMOS with $0.27\!<\!z\!<\!0.28$ and brighter $R=22\,\text{mag}$. 
The circles mark the same regions as in Fig.~\ref{Figure 1}.}
         \label{Fig0532field}
 \end{figure}

Figure~\ref{Fig0532field} shows the calibrated cc-diagram of RXCJ\,0532 for galaxies brighter than $R=22$. 
Overplotted in blue are galaxies from COSMOS with redshifts of $0.27<z<0.28$. 
Comparing with the left panel of Fig.~\ref{Figure 1}, one can see the overdensity of galaxies for redshifts lower than $z=0.3$, revealing the presence of a massive cluster. 
The overdensity at Region 3 marks the location of the red cluster galaxies for a cluster at redshift $z=0.275$.

\subsection{Background selection based on COSMOS} \label{backgroundsel}
To convert the photo-$z$ information in colour-colour-magnitude space (ccm space) into an background selection, 
we choose the angular diameter distance ratio $\beta \equiv D_{\mathrm{ds}} / D_{\mathrm{s}}$ 
and the purity estimator $p$, which gives the probability of being a background galaxy, as selection
criteria, and calculate these values for each galaxy individually. 

Since the shear signal scales with $\beta$, the $\beta$ criterion enables us to exclude regions of ccm space which carry no or only low signal, 
typical for regions dominated by foreground and cluster galaxies. 
Because the presence of a massive cluster alters significantly the redshift distribution at low redshifts compared to the distribution in the COSMOS field, 
the purity estimator can put only a lower limit to the contamination.
Since we directly calculate $\beta$, we do not need to derive individual redshifts for the galaxies in the lensing catalogue. 
Nevertheless, the method described below can be used in a similar way to calculate individual redshifts as well.

To calculate $\beta_{\mathrm{g}}$ for each galaxy $g$ in our lensing catalogue, we apply the following scheme to each source. 
First we define an elliptical cylinder in ccm space centred on the position of the galaxy in ccm space. 
The elliptical base is defined by $\sigma_{\mathrm{c1}}$ 
and $\sigma_{\mathrm{c2}}$ which are the measurement errors of the two colours measured for this galaxy.
Its height is defined by 
the magnitude uncertainty in the lensing band.

We now use this cylinder to select galaxies in ccm space from the photo-$z$ catalogue. 
In order to ensure sufficient statistics we require at least $50$ selected galaxies. 
If not enough galaxies are found, we increase the size of the cylinder by $10\%$ in each direction unless the minimum number 
of sources or a maximum size (2.5 mag in either colour axis) is reached. 
In the latter case, we treat the measurement as outlier and set $\beta_{\mathrm{g}}=0$.  
Finally, we calculate $\beta_{\mathrm{g}}$ as the weighted mean of the angular diameter distance ratios (cf.~Fig.~\ref{betaplot}) $\beta(z_{\mathrm{d}},z_{k})$ of the selected galaxies in the photo-$z$ catalogue, 
with $z_{k}$ the photo-$z$ of the $k$th source inside the cylinder, 
and $z_{\mathrm{d}}$ the cluster (deflector) redshift.
\begin{equation} \label{eq:betasel}
\beta_{\mathrm{g}}\!=\!\frac{\sum_{k=1}^{N}w(\Delta c_{1,k},\Delta c_{2,k})\beta(z_{\mathrm{d}},z_{k})}
{\sum_{k=1}^{N}w(\Delta c_{1,k},\Delta c_{2,k})}
\end{equation}

As weight function $w(\Delta c_{1,k},\Delta c_{2,k})$ we use a two-dimensional Gaussian function 
centred on the colours $c_{1,g}$ and $c_{2,g}$ of the lensing catalogue galaxy $g$,
with standard deviations of $\sigma_{\mathrm{c1}}$ and $\sigma_{\mathrm{c2}}$. 
The two parameters
$\Delta c_{1,k}$ and $\Delta c_{2,k}$ are the differences in colour of the photo-$z$ galaxy $k$ with respect to 
$g$,
\begin{equation} \label{eq:deltac}
\Delta c_{1,k}\!=\!c_{1,k}-c_{1,g}, ~\Delta c_{2,k}\!=\!c_{2,k}-c_{2,g}.
\end{equation}
We restrict our weighting to the colour-colour plane 
since the magnitude-dependence within the cylinder is much weaker than the colour dependence. 
A full 3D weighting was tested, but with no significant improvement and higher sensitivity on the zeropoint calibration. 
We therefore choose the simpler treatment for our distance estimate.

In the same way as $\beta_{\mathrm{g}}$, we can calculate
$p_{\mathrm{g}}$ by simply replacing $\beta(z_{k})$ with the
Heaviside step function $\Theta(z_{k}-z_{\mathrm{d}}):$
\begin{equation} \label{eq:psel}
p_{\mathrm{g}}\!=\!\frac{\sum_{k=1}^{N}w(\Delta c_{1,k},\Delta c_{2,k})\Theta({z_{k}-z_\mathrm{d}})}
{\sum_{k=1}^{N}w(\Delta c_{1,k},\Delta c_{2,k})}.
\end{equation}

\begin{figure}
   \centering
   \includegraphics[width=\linewidth]{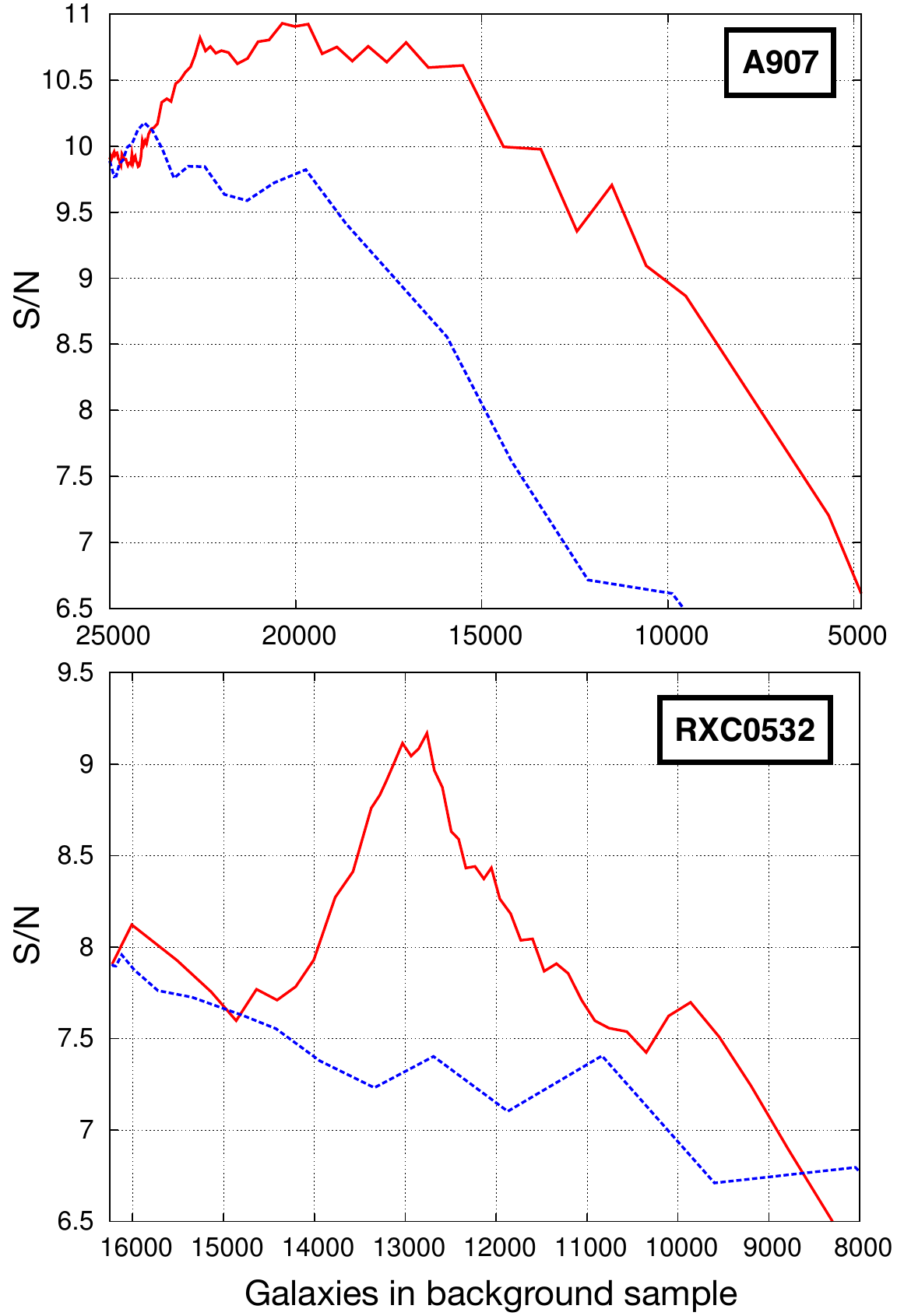}
   \vskip-0.05in
      \caption{S/N at the cluster position taken from the S-statistics maps vs remaining background galaxies using $\beta_{g}$ for selection (red) and magnitude cut (blue) for A907 (top) and RXC0532 (bottom).
The abscissa start at the complete catalogue of candidate background galaxies derived in Section~\ref{sec:ksb}, i.e.~$\beta_{\mathrm{cut}}\!=\!0$. Going to the right, $\beta_{\mathrm{cut}}$ increases, resulting in a progressively purer but smaller catalogue.}
         \label{Figure 5}
 \end{figure}

A first selection can be made by investigating the dependency of the signal-to-noise of the lensing detection 
as measured in S-statistics maps \citep{2007A&A...462..875S} on a minimum cut in $\beta$. 
We can find a $\beta_{\mathrm{cut}}$ which maximizes the S/N at the position of the cluster peak, this we call ``peak S/N''. 
It marks the point from which 
on the decrease of the number of background galaxies becomes dominant against the effect of a cleaner catalogue on the signal to noise.
Figure~\ref{Figure 5} shows the peak S/N over the number of galaxies left in the lensing catalogue for A907 ($z=0.1527$) and RXC0532 ($z=0.2745$). 
Instead of using the peak S/N one can also use the integrated
S/N over a certain aperture or S/N threshold as an estimator. For the majority of the clusters presented in this paper the so found $\beta_{\mathrm{cut}}$ is 
insensitive against which of the estimators is used, but for extreme 
merger with several shear peaks one can find different values. Those clusters then have to be investigated individually.
Despite the fact that the redshift of A907 is lower than the closest cluster in the COSMOS field and therefore red galaxies at that redshift may be lacking, 
the background selection still yield a S/N increase by $0.8$ compared to a magnitude cut or no cut at all.
In the case of RXC0532 (Figure \ref{Figure 5}, bottom), one can see a clear peak corresponding to a cut at $\beta_{\mathrm{cut}}=0.27$. 
For comparison we also plotted the S/N curve for a simple magnitude cut selection. It excludes galaxies brighter a certain
magnitude value. Since both method depends only on one parameter we can compare both methods based on the number of galaxies left in the lensing catalogue.
Note that the so-found $\beta_{\mathrm{cut,max}}$ is just the lowest reasonable cut that can be made, since lower values yield lower S/N. It provides the reference value for the final estimate of $\beta_{\mathrm{cut}}$ that will be discussed later.

 Generally, given the imaging bands we chose, the redshift distribution changes smoothly in the cc-plane. 
In principle, there is a risk that $\beta$ could be biased high in highly bimodal redshift distributions, where one peak may lie at or below the cluster redshift. 
Such galaxies might not be excluded if we use $\beta$ as the only selection criterion.
However, most contaminant foreground galaxies have a higher probability to be located in regions of the cc-plane where the median redshift of the populations is low in the first place.
Therefore, a cut in $\beta$ is sufficient for most of our cluster redshifts and colour combinations.
But to avoid stray low-$z$ interlopers in high-$z$ regions of the cc-plane, additionally, we explicitly exclude regions with high contamination (low $p$) using the purity estimator $p$ and include sources with $p=1$.

\begin{figure}
   \centering
   \includegraphics[width=\linewidth]{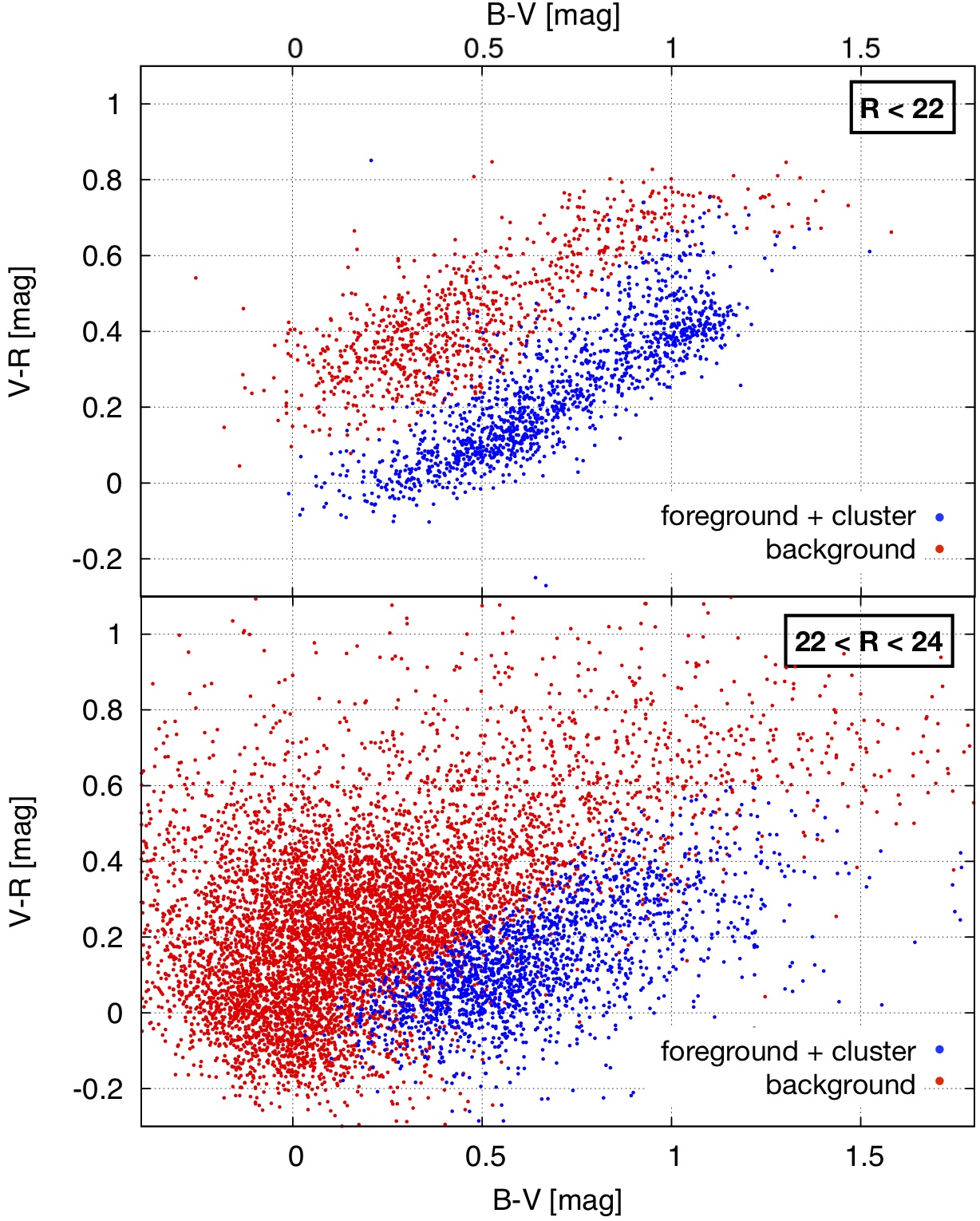}
   \vskip-0.05in
      \caption{\emph{Top panel:} Galaxies in the RXCJ0532 field brighter than $R\!=\!22$ mag. Blue points mark excluded foreground and cluster galaxies; red points mark background galaxies. \emph{Bottom panel:} The same, but for the magnitude range $22<R<24$ mag.}
         \label{FigSel1}
 \end{figure}

Figure~\ref{FigSel1} illustrates the background selection in colour-colour space for two magnitude ranges for the galaxies in the RXC0532 field. Marked in blue are galaxies which are rejected from the lensing catalogue. 
A detailed discussion about remaining contamination can be found in section~\ref{Dilution}. The values of $\beta_\mathrm{cut}$, $p_\mathrm{cut}$, $\langle \beta \rangle$ and $\langle p \rangle$ can be found 
in Table \ref{table:res}.

\begin{figure*}
 \centering
\includegraphics[width=0.99\textwidth]{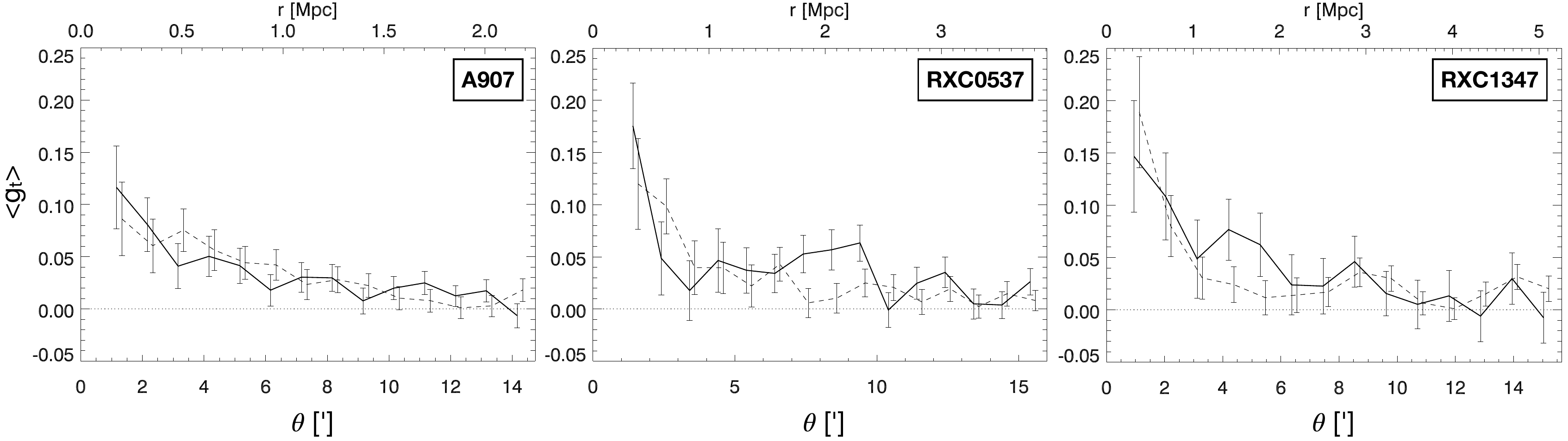}
\vskip-0.05in
 \caption{Binned tangential shear for A907 (left), RXC0532 (middle) and RXC1347 (right). The continuous line represents the low-redshift subsample, the dashed line the high-redshift background subsample.}\label{shearcomp}
\end{figure*}

\subsection{Accounting for second order effects} 
Some systematic effects on the weak lensing mass estimate can be accounted for if the mass or the shear profile is already known.
We, therefore, decided for an iterative approach where the previous mass estimate is used to account for the impact of a broad redshift distribution on our estimate of $\beta_{g}$, and to account for difference in redshift distributions of the cluster field and the reference field.
  
 \subsubsection{Correction for differences in redshift distributions}
 As of now, the estimate of $\beta_{g}$ assumes that the redshift distribution of the reference catalogue at a given position in ccm space reflects the true redshift probability distribution for that galaxy. 
Beside variations due to cosmic variance, this assumption becomes increasingly incorrect the more the ccm regions of cluster galaxies and background galaxies start to mix up. As discussed in the previous section, we assume those regions get excluded by our S/N optimization step, as long as the redshift distribution does not get bimodal. Nevertheless, we try to account for this effect in a second iteration of the mass measurement, using the mass from the first measurement.

We estimate the impact of the cluster galaxies on the redshift distribution in a given colour-colour region by splitting the observed field into several radial bins, depending on our first measurement of $r_{200}$. For each radial bin, we create a smoothed galaxy density map in colour-colour space. We normalize these maps by dividing them by the map obtained for the outermost bin,which is set to $r>1.5r_{200}$ or at least outermost $10\%$ of the sources in case of nearby massive clusters. Under the assumption that the outermost bin is close to the cosmic average, the resulting maps show the difference of the redshift distribution within this bin to the cosmic average.

We only aim to account for the strongest effect of the cluster on the redshift distribution, which is caused by the cluster galaxies. More subtle effects such as that caused by the weak lensing magnification effect are ignored.

For any galaxy $g$ in a radial bin $b$ that lies within a colour-colour region with a significant positive peak,
we correct the estimated $\beta_{g}$, by dividing the initial $\beta_{g}$ by $v_{b}(c_{1},c_{2})$, the
corresponding value of the normalized maps.
This correction assumes each positive peak in those maps to be caused by cluster member galaxies,  
which do not carry a shear signal. 
To ensure that we do not select possible clusters at higher redshift than our targeted clusters, we visually investigate the selected regions and compare them with the expected location of cluster members.
Performing the correction within several radial bins ensures that radial trends such as the probability of being a cluster member as well as changes in the mixture of galaxy types are captured.

 \subsubsection{Accounting for broad redshift distributions}
 The expected reduced shear caused by a cluster can be calculated via Eq.~\ref{eq:redshear} if $\beta$ is precisely known.
 In absence of a precise estimate, one has to account for a broad distribution in $\beta$. 
 Using $\beta_{g}$ derived as a weighted mean over reference sources assumes linear dependence of the reduced shear with $\beta$. This only valid for small $\kappa$ and $\gamma$ and becomes increasingly incorrect towards the cluster center.
 Although this effect is assumed to be small, we aim to correct for it during the second iteration of the mass estimate.
 Under the assumption that our first mass estimate of a cluster is approximately correct, we can calculate the expected reduced shear for a given source with cluster-centric distance $\theta$ and angular diameter distance $\beta$. This allows to recalculate the $\beta_{g,2}$ such that it satisfies the equation
 \begin{equation} \label{eq:gstep2}
 g(\theta_{g},\beta_{g,2})=\!\frac{\sum_{k=1}^{N}w(\Delta c_{1,k},\Delta c_{2,k})\,g(\theta_{g},\beta(z_{d},z_{k}))}
{v_{b}(c_{1},c_{2})\sum_{k=1}^{N}w(\Delta c_{1,k},\Delta c_{2,k})}.
\end{equation}
In contrast to Eq.~(\ref{eq:betasel}), we are now estimating our second step $\beta_{g,2}$ as the value that gives the same reduced shear as the weighted average of the expected reduced shear, calculated from the reference sources at the position of galaxy $g$. Further we include the correction term $v_{b}(c_{1},c_{2})$ from the previous subsection to account for redshift distribution differences caused by the cluster itself.

\section{Investigating sources of systematics and scatter}\label{sec:invsys}
\subsection{Remaining contamination and contamination correction}\label{Dilution}
The background selection by imposing a cut $\beta_{\mathrm{cut}}$ that optimizes the signal does not preclude contamination by cluster and foreground sources.
We therefore perform two types tests to identify remaining contamination by cluster galaxies, first on individual clusters and finally on a stack using all clusters.

\subsubsection{Shear profiles from low-z and high-z background galaxy samples}
To check for remaining unaccounted contamination, we split the background sample into a low-$\beta$
and a high-$\beta$ subsample of equal size and investigate the tangential shear of these subsamples. Since we assume that the low-$\beta$ (low-$z$) part will be affected more by contamination than the high-$\beta$ subsample, we expect it to have lower tangential shear, especially towards the cluster centre where the density of cluster galaxies is higher. To account for the different $\beta$ of the two background samples for a given cluster, we use the assumption that the approximation $g_{+} \approx \gamma_{+}$ holds and therefore the reduced shear scales simply with $\beta$.
We test this for all clusters in our sample. Figure~\ref{shearcomp} shows the binned tangential shear of these subsamples for the three example clusters.
In most cases, both samples agree well with each other, in particular there is no hint of a increasing dilution towards cluster centre. 

\subsubsection{Galaxy density profiles}
\begin{figure*}
 \centering
  \includegraphics[width=1\linewidth]{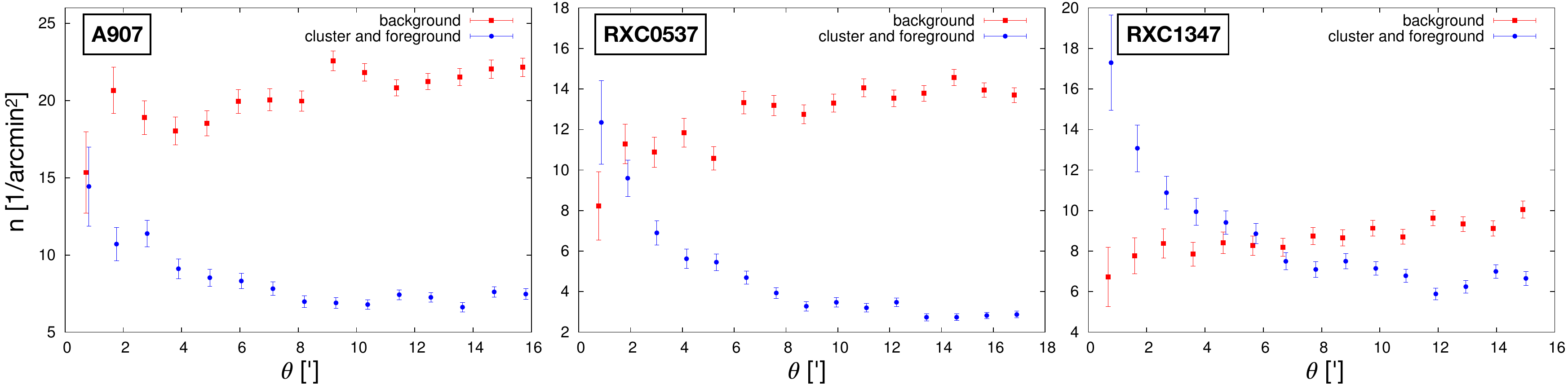}
  \vskip-0.08in
   \caption{Binned galaxy number density for A907 (left), RXC0532 (middle) and RXC1347 (right). Blue symbols show sources excluded from the lensing catalogue; red symbols show galaxies included in the lensing catalogue.}
\label{ndense}
\end{figure*}
A frequently used tracer for contamination of the background galaxy sample is the number density of sources in dependency of the radial distance to the cluster centre (e.g.\cite{2012arXiv1208.0605A,2015MNRAS.449..685H,Dietrich17}.
The usual assumption is that a clean sample would have a constant number density and contamination by cluster galaxies would show up by an increase of
the number density towards the cluster center.
However, several physically and observational effects complicate the interpretation of these plots.
The magnification effect can create a similar positive or a counteracting negative number density profile,
depending on the redshift, brightness and size distribution of the background galaxy sample. In addition, the obscuration by cluster and foreground galaxies, holes in the dataset created by masked areas, changes in the depth of the field, and small scale clustering can impact the radial density profile, if not accounted for.
Finally, one has to account for the next-neighbour filtering in the lensing catalogue which excludes close source pairs from the catalogue to avoid wrong shape and flux measurements. Since the total number density of sources and therefore the probability of close pairs increase towards the cluster center, this filter has also a radial dependency.
This effect is difficult to model, therefore some authors like \citet[][hereafter A14]{2012arXiv1208.0605A} use a  less filtered source catalogue for their contamination estimate rather than the final lensing catalogue. 
But since close pairs not only effect individual shear measurements but also the photometric quality, the lensing catalogue might be slightly cleaner than the other catalogue used to derive the galaxy density profile, since it avoids sources with blending issues. 
The number density plots of the lensing catalogues seen in Fig.~\ref{ndense} are accounting for most of the measurement effects except the next-neighbour filtering and small scale clustering. 
All plots show a depletion of background galaxy density profiles towards the cluster center, which can be interpreted as being caused by either the next-neighbour filtering, the magnification effect, or a combination of these effects. The lack of an increase towards the cluster center at least supports the result of the previous test that there is no obvious cluster contamination left.

\subsubsection{Stacked profiles}
\begin{figure*}
 \centering
  \includegraphics[width=\linewidth]{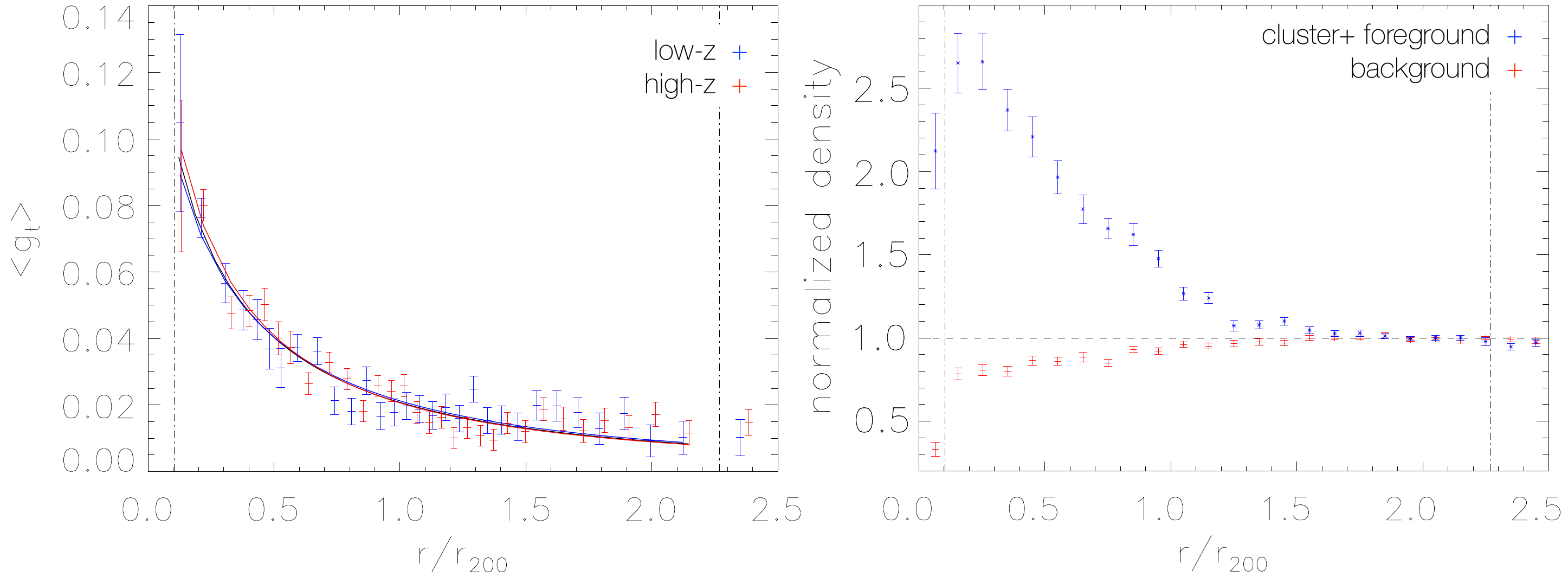}
\vskip-0.05in
 \caption{Left: Similar to Fig.~\ref{shearcomp} but stacking the shear profile of all clusters. The shear of low-z (blue) and hig-z (red) background galaxy sample is scaled to the average $\beta$ of the full sample. The corresponding fits to the different samples are shown as continuous lines with the corresponding colour. The black line shows the fit to the full sample.
Right: Similar to Fig.~\ref{ndense} but using stacking all cluster catalogues. Blue symbols show sources excluded from the lensing catalogue; red symbols show galaxies included in the lensing catalogue. The number density of the stacked cluster sample is normalized to one at $2\,r_{200}$. The vertical dash-doted line indicated the typical fitting range for the shear measurement.}
\label{stackedprof}
\end{figure*}

Individual profiles yield an insight to the cluster-by-cluster performance of the background selection, but do not offer sufficient statistics to search for small amounts of contamination.
For this reason both test were repeated using a stack of all clusters.
We stack the background catalogues, scaling the radial distances by our estimates of $r_{200}$.
For the shear profile test we split the background galaxy catalogue of each cluster by its median
$\beta$ to ensure that each cluster contributes equally to each subsample.
To account in a more precise way for the difference in the average $\beta$ of the subsamples, we first fit a NFW-based shear profile to the total background sample. 
We then scale the measured ellipticity of each source by the ratio $g(\beta_g,r)/g(\langle\beta\rangle,r)$, the expected shear for the galaxy at distance $r$ using $\beta_g$ over the expected shear at the same position using the average $\beta$ of the full background sample. The resulting shear profiles are shown in Fig.~\ref{stackedprof} (left panel).
Shear profile fits yield consistent results for the low-$\beta$ and high-$\beta$ stacked background samples, 
suggesting either a very low unaccounted contamination or that the contamination in both subsets are essentially of the same order.

In case of the stacked galaxy density profile, we normalize the density profile of each cluster by dividing the profile by the median density beyond $1.5\,r_{200}$. The right panel of Fig.~\ref{stackedprof} shows the median of the normalized densities of the individual clusters for each radial bin.
Similar to the individual profiles shown in Fig.~\ref{ndense}, the stacked density profiles of background galaxies show a clear decrease towards the cluster centre. The density of excluded sources increases by a factor of $2.6$ towards the cluster centre.


Despite the encouraging results in Figs.~\ref{shearcomp}, \ref{ndense} and \ref{stackedprof}, we expect some remaining minimum contamination of up to $\sim\!14\%$ judging by the 
mean of the purity estimator as can be seen in Table~\ref{table:res}. This contamination of $\sim\!14\%$ is mainly due to foreground galaxies and should not be related to the presence of a cluster. The remaining contamination is therefore included in our reference catalogue and is of the same amount as for the cluster fields. When calculating $\beta$, we therefore automatically include the correct amount of contamination by foreground sources yielding an unbiased estimate of $\beta$, and with that of the cluster mass. 

\subsection{Mean lensing depth and cosmic variance}\label{cosmvar}
The reduced shear $g_{i}(\theta_{i})$ exerted by a lens on the image of a 
background source depends on the ratio of angular diameter distances $\beta$, and the resulting
cluster masses via Eq.(\ref{eq:mdelta}) scale roughly linear with it.

A point of concern in weak lensing is the influence of cosmic variance on the determination of the mean lensing depth.
Where unaccounted cosmic variance in a cluster field results in an additional source of noise, the effect of cosmic variance in the reference photo-$z$ catalogue would systematically affect the mass estimate.
The cosmic variance on the mean redshift of a COSMOS-sized field can be estimated to be approximately $3\%$ \citep{2006APh....26...91V}. To backup this approximation
and to explore how this impacts our measurements, we utilise
the COSMOS and the CFHTLS deep fields \citep{2006A&A...457..841I}.

Our background selection estimates $\beta_g$ for each galaxy. Therefore, it is straightforward to calculate $\langle\beta\rangle$ for the whole cluster field.
Cosmic variance can act in two ways on the background sample: First, it can change the number density of sources in a certain region in ccm space. Second, it can alter 
the redshift distribution of a given ccm region with respect to the photo-$z$ catalogue used for our selection.
The first effect will not affect the estimate of $\langle\beta\rangle$ or $\beta_i$ but the second one does.

In order to estimate the level of scatter in $\langle\beta\rangle$ that is induced by the cosmic variance and how far our background selection can reduce the scatter, we explored the behaviour of this quantity in $9$ individual subfields of the COSMOS field.
The size of $30\arcmin\times30\arcmin$ of the subfields matches the typical field sizes of our observations. We apply our background selection on these fields by assuming the typical cuts in $\beta$ for three different cluster redshifts.
Afterwards, we measure the mean lensing depth  $\langle\beta_{\mathrm{meas}}\rangle_i$ based on our method and the mean lensing depth $\langle\beta_{\mathrm{true}}\rangle_i$ 
obtained by using directly the COSMOS redshifts of each galaxy for each field $i$.

\begin{figure*}
   \centering
    \includegraphics[width=\linewidth]{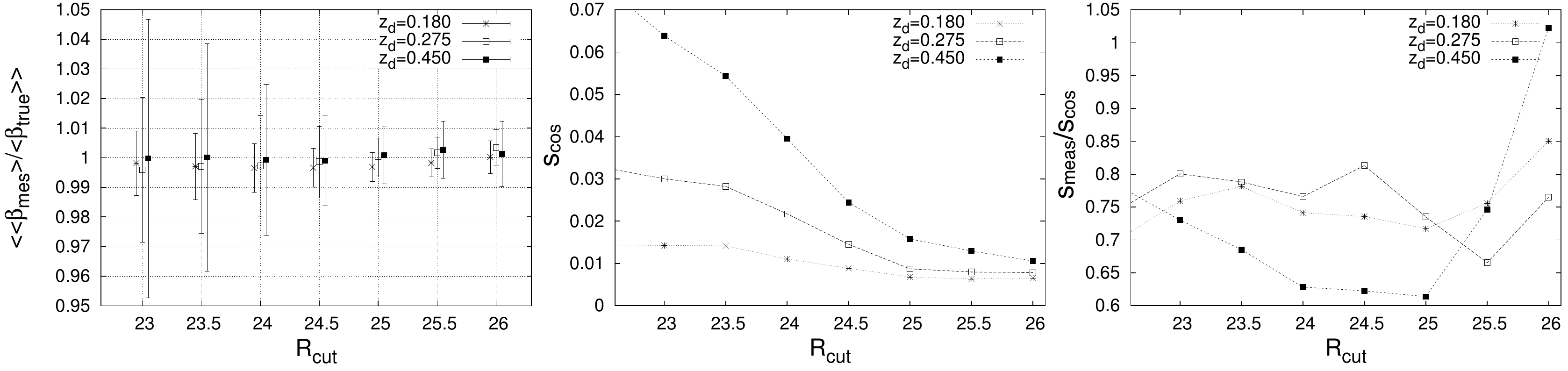}
\vskip-0.04in
      \caption{Tests on COSMOS subfields. \textit{Left:} Mean ratio of mean lensing depth measured by our method over true mean lensing depth. The errorbar indicates the field to field scatter ($s_{\text{meas}}$). \textit{Middle:} Scatter in true mean lensing depth between fields divided by the mean over all subfields ($s_{\text{cos}}$). \textit{Right:} Ratio of measured scatter $s_{\text{meas}}$ over cosmic variance induced scatter $s_{\text{cos}}$. All measurements are for three cluster redshifts and as function of limiting magnitude $R_{\mathrm{cut}}$
         }\label{cosvar1}
 \end{figure*}

Left panel of Fig.~\ref{cosvar1} shows mean value of $\langle\beta_{\mathrm{meas}}\rangle_i$/$\langle\beta_{\mathrm{true}}\rangle_i$ of the nine subfields for different limiting magnitudes and redshifts. The errorbars indicate the standard deviation between the fields. On average our method recovers the true mean lensing depth within $0.5\%$, but significant scatter is found between subfields.

To investigate if this scatter is introduced by cosmic variance or by noise in our method, we measure scatter of the mean true lensing depths between the subfields. The middle panel of Fig.~\ref{cosvar1} shows the standard deviation between fields over the average of the mean lensing depths over all subfields $s_{cos} = \sigma_{\langle\beta_{true}\rangle_i}/\langle\langle\beta_{true}\rangle_i\rangle$. The scatter introduces through cosmic variance on the mean lensing depth is a string function on imaging depth and cluster redshift, but falls below 4\% even for $z=0.45$ for catalogs reaching depths greater R=24. At this point it is important to note that this test assumes approximately similar imaging depths for all subfields in COSMOS and that all fields are uncorrelated. Given the size of the COSMOS field, we expect some correlation between subfields potentially causing an underestimation of the scatter.

To measure by how much our method to derive individual $\beta_g$ reduces the impact of the variance between fields we compare the scatter between recovered and measured mean lesning depth $s_{meas} = \sigma_{\langle\beta_{meas}\rangle_i}/\langle\langle\beta_{true}\rangle_i\rangle$ with the scatter found between subfields ($s_{cos}$). In right panel of Figure \ref{cosvar1} we show the ratio $s_{meas}/s_{cos}$, again for different cluster redshifts and limiting depths. We find that the scatter between measured and true mean lensing depth is $20-40\%$ smaller than that caused by cosmic variance between fields.

For individual clusters the scatter, compared to the typical measurement scatter of $\sim25$, is small enough to be ignored if the imaging depth is high and the cluster redshift low enough. But assuming that the $20-40\%$ reduction in scatter is maintained also for the full COSMOS field, the overall systematic uncertainty on mean lensing depth that comes from the limited size of the COSMOS field is reduced too.

 In order to examine the significance of the trends we see in Fig.~\ref{cosvar1}, we mimic repeated observations of the same subfields. To this end, we randomly varied the colours of the individual sources in the photo-$z$ catalogues within their photometric errors. We ignore an additional scatter that may come with an imperfect zeropoint calibration.

We created $20$ realizations of the same subfield for three of our nine subfields used in our previous test.
The scatter of $\langle\beta_{\mathrm{true}}\rangle_{i}-\langle\beta_{\mathrm{meas}}\rangle_{i,j}$ over all realizations $j$ of a subfield $i$ is of about one order of magnitude smaller than the scatter found for 
$s_{\mathrm{meas}}$, indicating that the remaining $60-80\%$ scatter seen in the right panel
of Fig.\ref{cosvar1} is driven by variance of the redshift distributions and not by the scatter in the photometry of the observation.

\subsubsection{Cosmic variance on the scales of the reference field}
To further investigate the impact of cosmic variance on the mean lensing depth on scales of the size of our reference catalogue, we make use of the four CFHTLS deep fields. They consist of four well separated one square degree fields with five-band ($u$,$g$,$r$,$i$,$z$) photometry allowing to derive individual source redshifts \citep{coupon09}. 
We calculate $\beta$ for all sources, assuming a cluster redshift of $z=0.3$. We measure the mean $\beta$ for all fields separately and for the merged catalogue for four $r$-band limiting magnitudes between $r=23$ and $r=24.5$ mag applying a cut in photo-$z$ of $z>0.5$ to mimic the background selection.
The faintest magnitude limit was chosen to ensure that all fields are complete at that magnitude. We find a scatter of $2.5$-$2.7$\% among the mean $\beta$ of the subfields with respect to the mean of the merged catalogue.
As the COSMOS catalogue is about twice the size of a single CFHTLS deep field, the expected impact of cosmic variance is therefore lower. Further, as shown in Fig.~\ref{cosvar1}, our method is able to compensate for $25-30\%$ of the scatter introduced by cosmic variance. Naively imposing a $\sqrt{2}$ scaling in area and a $25\%$ reduction thanks to our method results in an estimated $1.4\%$ uncertainty caused by cosmic variance within the limited size of the reference catalogue.

\subsection{Mean lensing depth and photometric calibration}\label{zpscheck}
In this section we investigate the impact of the limited precision of the photometric calibration on the recovered mean lensing depth, using a subset of the COSMOS catalogue as input to our pipeline to estimate angular diameter distance ratios $\beta_g$.

\subsubsection{Dependency on photometric zero point}
\begin{figure*}
\centering
\includegraphics[width=0.99\linewidth]{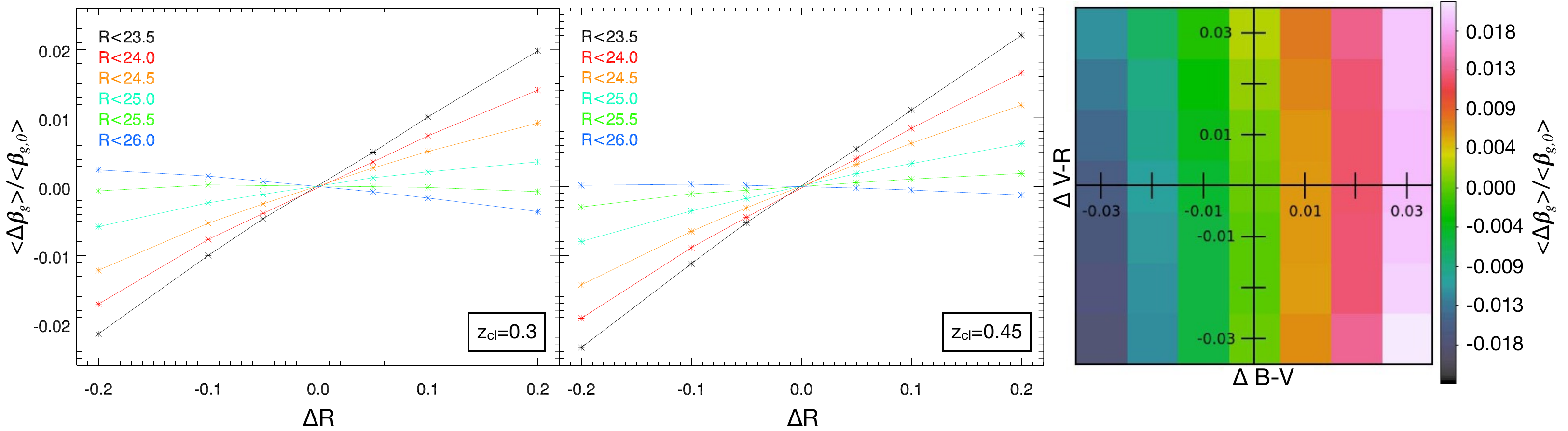}
\vskip-0.06in
\caption{Dependency of estimated $\beta_{\mathrm{g}}$ on photometric calibration (zero point calibration and color calibration). \textit{Left \& middle:} Mean difference $<\Delta \beta_g>$ between $\beta_{\mathrm{g}}$ derived with and without altering the R-band magnitude by $\Delta$R over mean $\beta_{\mathrm{g}}$ of the unaltered galaxy sample, for different depths and cluster redshifts.\textit{Right:} Dependency on shifts in color for a $z=0.3$ cluster. }\label{fig:betaonZPcalib}
 \end{figure*}

To investigate the dependency of $\langle\beta_{\mathrm{meas}}\rangle$ on the estimated ZPs, we use a subset of $40000$ randomly selected galaxies from the COSMOS catalogue and vary the overall ZP by $\pm0.05$, $\pm0.1$ and $\pm0.2$ magnitudes. Those catalogues are then used as input to our pipeline to estimate $\beta_g$ assuming two different lens redshifts of $z\!=\!0.3$ and $z\!=\!0.45$. The output catalogue is then matched with the catalogue obtained without altering the ZPs. We calculate the difference $\delta\beta_g$ for each source and finally measure the ratio $\langle \delta\beta_g \rangle / \langle\beta\rangle$ for different $R$ band limiting magnitudes. The result of this test is shown in Fig.~\ref{fig:betaonZPcalib}.

We find that the sensitivity of our estimate of the mean lensing depth is mainly a function of imaging depth. The differences between the $z\!=\!0.3$ and $z\!=\!0.45$ cases is very small and reaches a maximum of $0.5$\%.
The difference between estimated mean lensing and true mean lensing depth depends almost linearly on zeropoint for offsets of up to $\pm0.2$ and decreases with increasing image depth. 
Beside these dependencies, the overall impact of zeropoint shifts on the mean lensing depth stays small for the typical depth of our weak lensing observations. We note that we do not expect a systematic offset in zeropoints in our data, since we used a stellar locus based on our reference catalogue.

\subsubsection{Dependency on colour offset}

From Section \ref{clusel}, we expect a scatter of $\sim\,0.013\,\text{mag}$ in the calibration of the colours. To investigate how the mean lensing depth depends on colour offsets, we repeat the previous test, but varying the colours instead of the overall ZPs. We vary the colours by $\pm0.01$, $\pm0.02$ and $\pm0.03$ mag and compare the estimated  $\langle\beta_{\mathrm{meas}}\rangle$ to that obtained without shift of relative zeropoints. 
The right panel of Fig.~\ref{fig:betaonZPcalib} shows the outcome of this test, assuming a cluster at $z\!=\!0.3$. 
The colour-coded cells display $\langle \beta_{\mathrm{g},\Delta} - \beta_{\mathrm{g},0}\rangle / \langle \beta_{\mathrm{g},0}\rangle$,
the ratio of the mean difference between $\beta_{\mathrm{g}}$ evaluated with colour shift and without colour shift, over the average $\beta_{\mathrm{g}}$ without colour shift.
The impact of relative zeropoint shifts is stronger for $B-V$ shifts compared to $V-R$ shifts. 
For colour shifts of $\sim\,0.013\,\text{mag}$ the average lensing depth changes by $1.3$\%.

By construction we do not expect the overall or the relative zeropoints to be systematically off from the reference catalogue. Therefore colour and zeropoint shifts only contribute to the overall mass--to--weak lensing mass scatter. Presuming no correlation between the scatter in zeropoints and colours, the overall scatter in $\beta$ can be estimated to be $\sigma\approx\sqrt{1.3^2 + 0.6^2}=1.4$~\%. Where $0.6$\% comes from the 0.05 mag scatter of the absolute zeropoint for $R=24$ depth. This is small compared to the expected scatter between true mass and weak lensing based mass and is of the same order as the scatter introduced by cosmic variance.

\subsection{Shear calibration bias}\label{shearcalibbias}
The used KSB+ shape measurement pipeline is known to recover biased measurements for sources with low signal-to-noise ratio \citep{2006MNRAS.368.1323H,2001A&A...366..717E}. \citet{2010A&A...520A..58I,2012A&A...546A..79I} used a signal-to-noise threshold of $4.5$ above which a constant shear calibration factor of $1.08$ (or $8$\%) is assumed. The high number of clusters in this sample allows us to investigate the signal-to-noise dependent shear calibration bias in greater detail. For each cluster, we measure the radius $r_{200}$ for different signal-to-noise thresholds and normalize them by dividing each $r_{200}$ by the median over all thresholds.
For each threshold we derive the median over all clusters, the standard deviation, and the error on the mean, by dividing the standard deviation by the number of clusters.

\begin{figure}
   \centering
   \includegraphics[width=\linewidth]{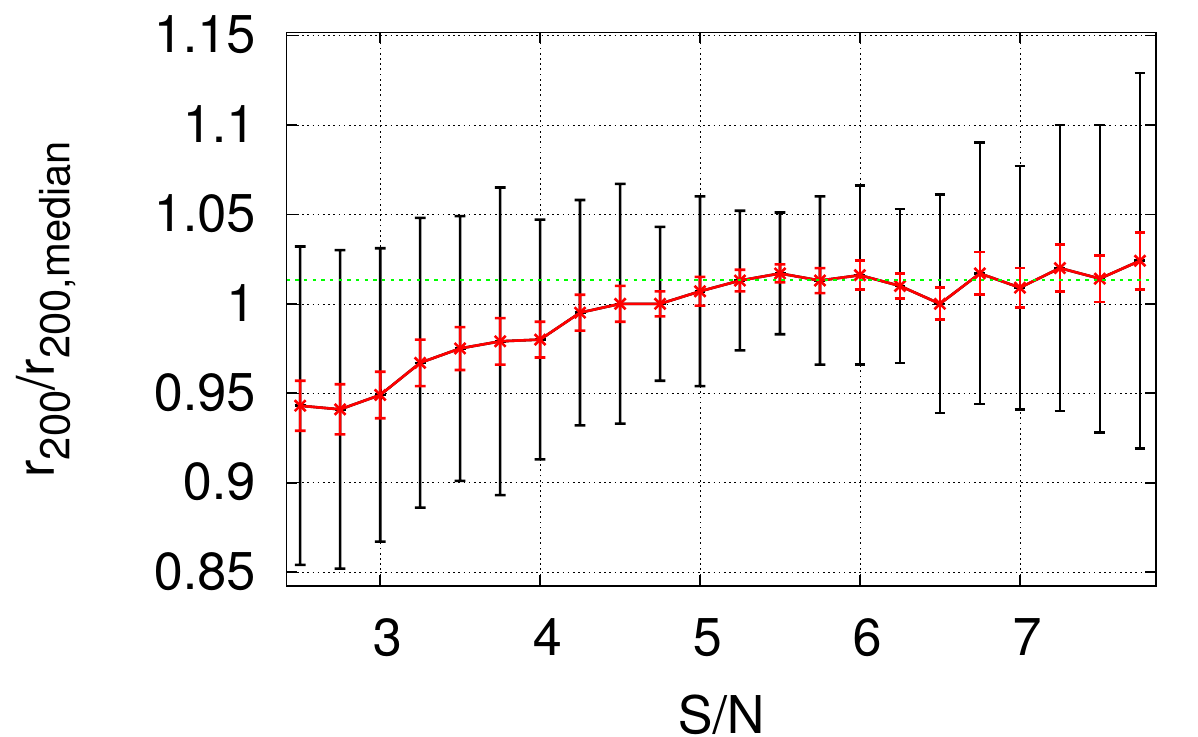}
  \vskip-0.03in
      \caption{Mean virial radius as a function of the signal-to-noise threshold. Black error bars indicate the standard deviation of the estimated virial radii while red error bars show the error on the mean value. The mean value above signal-to-noise greater than five is shown as green dotted line.}  \label{scbias}
 \end{figure}

Figure \ref{scbias} shows how the normalised scale radius $r_{200}$ depends on a signal-to-noise threshold.
The black error bars show the standard deviation of the distribution of values around the mean, while red error bars show the error on the mean.
The mean scale radius stays constant beyond a threshold of $5.25$-$5.5$. Using galaxies with lower signal-to-noise values can underestimate the scale radius by up to $5$\%.
For our analysis we therefore choose a threshold of $5.5$ for our conservative model and $4.5$ for a signal-to-noise optimised model. Following \citet{2010A&A...520A..58I,2012A&A...546A..79I} we assume a systematic uncertainty on the shear calibration bias of 0.05, corrsponding to $4.6$\%.

To convert this uncertainty to cluster mass, we investigate the response of the mass estimate on variations on the assumed shear calibration bias. 
For that, we created synthetic, NFW-based, shear profiles spanning the full redshift and mass range of our sample using the \citet{2013ApJ...766...32B} $c$--$M$ relation.
We find that the response of the cluster mass on the uncertainty of the multiplicative shear calibration factor to be described by a linear relation with slope $1.4$.
We further find a small approximately linear dependency of the response on cluster mass and redshift, where we find $1.35$ to be the lowest slope (reached at low mass and low redshift) and $1.45$ as the highest slope (high mass and high redshift).
Given the redshift and mass distribution of the cluster sample, the value of $1.4$ is a sufficiently good approximation for the vast majority of the sample. 
An uncertainty of $4.6$\% in shear calibration therefore translates into $6.5\%$ in cluster mass. 
We note that the same response also holds, to first order, for uncertainties in mean lensing depth $\langle\beta\rangle$. 
The uncertainty of $1.4\,$\% associated to cosmic variance in the cosmos field, corresponds to $\sim2\%$ uncertainty in mass.
 
\subsection{Selection induced bias}
\begin{figure}
   \centering
   \includegraphics[width=\linewidth]{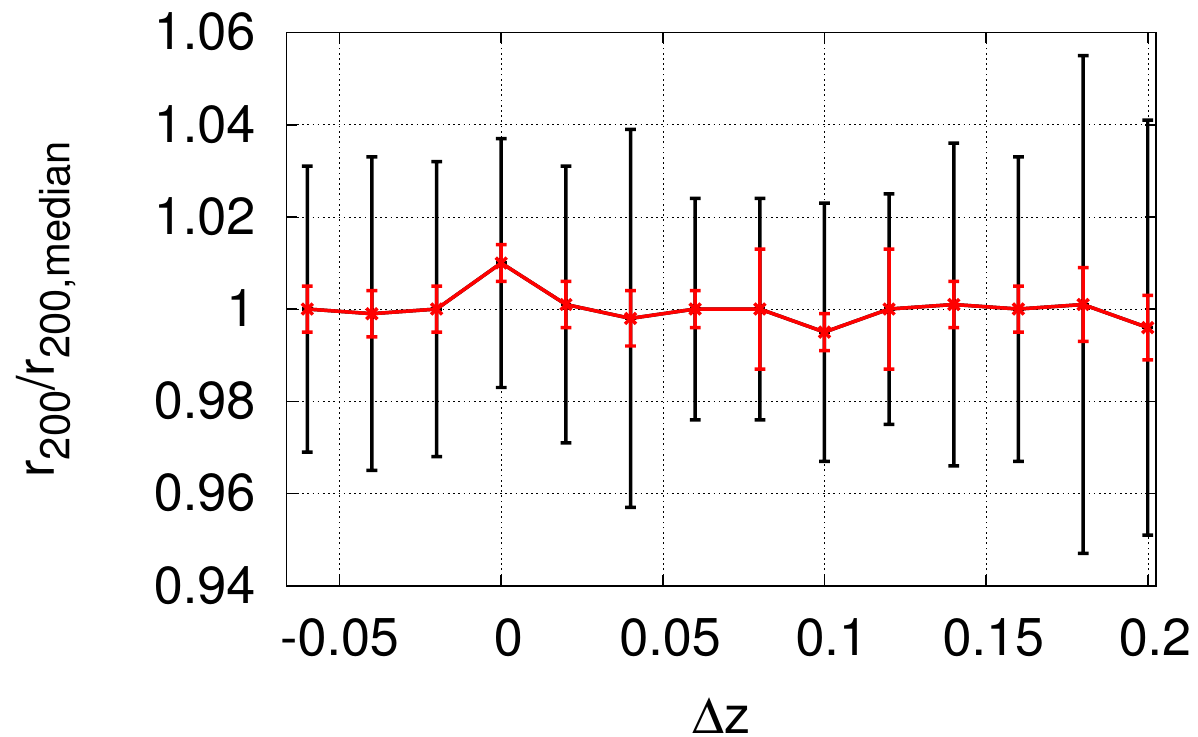}
   \vskip-0.05in
      \caption{Mean virial radius as a function of $\beta_{\mathrm{cut}}$, parametrised in $\Delta z = z_{\mathrm{cut}} - z_{\mathrm{cut,max}}$. Black error bars indicate the standard deviation of the estimated virial radii while red error bars show the error on the mean value.}
         \label{betabias}
 \end{figure}

The background selection relies on the optimization of the shear signal of the cluster. Since the shear signal is a noisy property, there is some chance to use a peak boosted by noise to
select the value for $\beta_{\mathrm{cut}}$. This results on average in mass estimates biased high compared to the true lensing mass. We use the same method as described in the previous subsection to investigate the selection induced bias. 
We vary $\beta_{\mathrm{cut}}$, parametrized by the offset $\Delta z$ from the redshift $z_{\mathrm{cut,max}}$ (corresponding to $\beta_{\mathrm{cut,max}})$ which maximizes the lesning signal-to-noise (cf. Fig.~\ref{Figure 5}). 
As visible in Fig. \ref{betabias}, the scale radius is biased high at offset $\Delta z = 0$, meaning exactly at $\beta_{\mathrm{cut,max}}$. 
To both sides of this value, the scale radius stays constant.
The fact that even a $\beta_{\mathrm{cut}}$ slightly below $\beta_{\mathrm{cut,max}}$ gives similar masses than $\beta_{\mathrm{cut}}>\beta_{\mathrm{cut,max}}$ indicates
that the background catalogue is already free of unaccounted cluster members and that the increase in the signal-to-noise ratio of the lensing detection arises mainly by cleaning the catalogue from foreground sources.
The constant value of the scale radius at higher $\beta_{\mathrm{cut}}$ also supports the good quality of our lensing analysis, since average redshift as well as average signal-to-noise of galaxies in the background sample changes significantly with increasing $\beta_{\mathrm{cut}}$.
We choose  $\delta z = 0.04$ for our conservative model, to avoid the bias caused by the selection of $\beta_{\mathrm{cut,max}}$. For the signal-to-noise optimized model, we choose $\delta z = 0.0$, this counteracts the underestimation caused by the chosen signal to noise threshold. 
While the conservative model is intended to be used for future scaling relation studies, the signal to noise optimized selection is used for the construction of $\kappa$-maps.

\subsection{Profile fit based systematics}
Simulations have shown that weak lensing masses, derived by fitting a single radial NFW basded shear profiles, show significant intrinsic scatter of $\approx20$\% and a mass dependent bias due halo triaxiality, correlated and uncorrelated large scale structure \citep{2011ApJ...740...25B,2012MNRAS.421.1073B,2018MNRAS.tmp.1333L}. 
Given the number of clusters in this sample, this scatter places a systematic limit of $20\%/\sqrt{39}\sim\!3.5$\% on the accuracy of the mass scale for the overall sample.
The NFW profile is a good approximation of the dark matter density profile within the virial radius, but gets increasingly discrepant from the truth while going beyond the virial radius.
Therefore, A14 restrict the fitting range to $0.75$ to $3$ Mpc.
However, the same study showed that extending the fitting range out to $5$ Mpc results in a insignificant decrease of the mass by $1.3^{+1.2}_{-1.0}$\%. Our fitting range extends on average between $0.2$ and $4.1$ Mpc, which is larger than that used in A14 but also fitting closer to the cluster core. Re-measuring masses within $0.75$ to $3$ Mpc, we find a median mass ratio $M_{\mathrm{Default}}/M_{0.75-3.0}=0.989\pm0.033$ or $-1.1\pm3.3\%$. Given the small difference  between both fitting ranges investigated in A14 we do not expect that the fitting ranges chosen by us yield different results with respect to A14.
We therefore assign the same systematic uncertainty of $3$\% in mass to the shear profile fitting.

\subsection{Summary on sources of systematics and scatter on the weak lensing analysis}
Taking the work of A14 as a template, we summarize the different contributions to the overall systematic uncertainty, the total uncertainty is then assumed to be the square root of the sum of squared uncertainties. 
The uncertainty on shear calibration of $0.046$ \citep{2010A&A...520A..58I} contributes a mass uncertainty of 6.5\%.
The scatter between true mass and WL-inferred mass is assumed to be $\sim\!20$\% \citep{2011ApJ...740...25B,2012MNRAS.421.1073B} for a single cluster.
Given our sample size this results in a $3.5$\% uncertainty on the sample-wide mass scale. 
The choice of fitting a NFW profile within $4.1$ Mpc results in an uncertainty of $3$\%.
Finally, from the scatter between different CFHTLS fields we expect the uncertainty of our reference catalogue due to cosmic variance to be $1.4$\% in $\beta$, corresponding to 2\% scatter in mass. This results in an overall systematic uncertainty of $8$\%, where the main contribution comes from the shear calibration.

\section{Results}\label{sec:res}
The main focus of this paper is to provide the mass measurements for the APEX-SZ cluster sample. We therefore first describe and discuss the results for the entire sample in Section \ref{sec:resultsall}. 
Later, in Section \ref{indiclusters}, we discuss in detail the obtained results for the three example clusters from the sample.
We stress that the measurement errors given throughout this work include only statistical errors from the fit to the shear profiles. 
\subsection{Results using the full cluster sample}\label{sec:resultsall}
In this subsection we concentrate on few properties related to the full or a large subset of the cluster sample. The inferred weak lensing masses can be found in Table \ref{table:res}.

\begin{table*}
\caption{Lensing results. Columns (2) to (5) show detection-optimizing distance ratio cuts 
$\beta_{\mathrm{cut}}$, the lower purity cuts $p_{\mathrm{cut}}$ (\ref{backgroundsel}), the mean distance ratios
$\langle\beta\rangle$ and mean purity $\langle p_g\rangle$,of the background galaxy sample. In Columns (6) to (8), we
give the radii $r_{200}$, concentration parameters $c_{200}$, and cluster masses 
$M_{200}$ obtained from the NFW fits to the shear profile (\ref{sec:spm}).
Assuming these profiles to be correct, the corresponding $\Delta=500$ values
were computed (Columns (8) to (10)).
The radial range of these fits is defined between the separations 
$\theta_{\mathrm{min}}$ and $\theta_{\mathrm{max}}$ in Columns (11) and (12).
$(S)$: Results based on Suprime-Cam. $(W)$: Results based on WFI.
}             
\label{table:res}      
\centering                          
\setlength\tabcolsep{3.3pt}
\begin{tabular}{c c c c c c c c c c c c c }        
\hline\hline                 
Cluster &  $\beta_{cut}$ & $p_{\mathrm{cut}}$ & $\langle \beta_g \rangle$ &$\langle p_g \rangle$ & $R_{200}$  & $c_{200}$ & $M_{200}$ & $R_{500}$  & $c_{500}$ & $M_{500}$ & $\Theta_{min}$ & $\Theta_{max}$ \\    
& & & &  & [Mpc] & & $[10^{14}\mathrm{M}_{\odot}]$ & [Mpc] & &  $[10^{14}\mathrm{M}_{\odot}]$ & $[``]$ & $[``]$ \\
\hline  
A2744 	&	0.47	&	0.5	&	0.55	&	0.91		&	$2.32^{+0.12}_{-0.12}$	&	$3.30^{+1.10}_{-1.10}$	&	$17.47^{+3.02}_{-2.71}$	&	$1.41^{+0.08}_{-0.08}$	&	$2.12^{+0.71}_{-0.71}$	&	$11.55^{+2.00}_{-1.79}$	&	120	&	1100	\\		
RXCJ$0019.0-2026 $	&	0.36	&	0.4	&	0.54	&	0.92		&	$2.05^{+0.12}_{-0.13}$	&	$3.45^{+1.15}_{-1.15}$	&	$11.51^{+2.28}_{-2.17}$	&	$1.24^{+0.08}_{-0.08}$	&	$2.22^{+0.74}_{-0.74}$	&	$	7.70^{+1.53}_{-1.45}$	&	30	&	1000	\\	
A2813 	&	0.33	&	0.3	&	0.53	&	0.9	&		$2.10^{+0.13}_{-0.13}$	&	$3.45^{+1.15}_{-1.15}$	&	$12.41^{+2.62}_{-2.30}$	&	$1.27^{+0.08}_{-0.08}$	&	$2.22^{+0.74}_{-0.74}$	&	$	8.30^{+1.75}_{-1.54}$	&	40	&	900	\\	
A209 	&	0.54	&	0.8	&	0.7	&	0.96	&		$2.18^{+0.09}_{-0.09}$	&	$3.55^{+1.18}_{-1.18}$	&	$13.49^{+1.82}_{-1.67}$	&	$1.35^{+0.06}_{-0.06}$	&	$2.29^{+0.76}_{-0.76}$	&	$	9.08^{+1.22}_{-1.12}$	&	90	&	1000	\\	
XLSSC $006 $	&	0.3	&	0.6	&	0.45	&	0.85	&		$1.71^{+0.13}_{-0.15}$	&	$3.35^{+1.12}_{-1.12}$	&	$	7.49^{+2.00}_{-1.94}$	&	$1.02^{+0.08}_{-0.10}$	&	$2.15^{+0.72}_{-0.72}$	&	$	4.97^{+1.33}_{-1.29}$	&	30	&	850	\\
RXCJ$0232.2-4420 $	&	0.4	&	0.5	&	0.52	&	0.93	&		$1.87^{+0.19}_{-0.21}$	&	$3.60^{+1.20}_{-1.20}$	&	$	7.59^{+2.88}_{-2.51}$	&	$1.09^{+0.12}_{-0.14}$	&	$2.33^{+0.78}_{-0.78}$	&	$	5.13^{+1.94}_{-1.69}$	&	50	&	800	\\
RXCJ$0245.4-5302 $	&	0.36	&	0.8	&	0.51	&	0.93	&		$1.64^{+0.18}_{-0.20}$	&	$3.70^{+1.23}_{-1.23}$	&	$	5.11^{+2.13}_{-1.82}$	&	$0.95^{+0.12}_{-0.13}$	&	$2.40^{+0.80}_{-0.80}$	&	$	3.47^{+1.45}_{-1.24}$	&	15	&	1000	\\
A383 	&	0.48	&	0.8	&	0.65	&	0.94	&		$1.92^{+0.11}_{-0.13}$	&	$3.70^{+1.23}_{-1.23}$	&	$	8.61^{+1.67}_{-1.73}$	&	$1.17^{+0.07}_{-0.08}$	&	$2.40^{+0.80}_{-0.80}$	&	$	5.86^{+1.13}_{-1.17}$	&	40	&	900	\\
RXCJ$0437.1+0043 $	&	0.25	&	0.5	&	0.51	&	0.91	&		$2.12^{+0.16}_{-0.18}$	&	$3.45^{+1.15}_{-1.15}$	&	$12.12^{+3.22}_{-3.04}$	&	$1.26^{+0.10}_{-0.12}$	&	$2.22^{+0.74}_{-0.74}$	&	$	8.10^{+2.15}_{-2.03}$	&	10	&	900	\\	
MS$0451.6-0305$ 	&	0.36	&	0.6	&	0.4	&	0.91	&		$1.79^{+0.17}_{-0.19}$	&	$3.15^{+1.05}_{-1.05}$	&	$	9.31^{+3.25}_{-2.91}$	&	$1.04^{+0.11}_{-0.12}$	&	$2.01^{+0.67}_{-0.67}$	&	$	6.08^{+2.12}_{-1.90}$	&	30	&	900	\\
A$520 $	&	0.43	&	0.7	&	0.66	&	0.94	&		$1.86^{+0.10}_{-0.11}$	&	$3.70^{+1.23}_{-1.23}$	&	$	8.02^{+1.45}_{-1.41}$	&	$1.14^{+0.06}_{-0.07}$	&	$2.40^{+0.80}_{-0.80}$	&	$	5.45^{+0.98}_{-0.96}$	&	5	&	900	\\
RXCJ$0516.6-5430 $	&	0.28	&	0.6	&	0.5	&	0.91	&		$2.13^{+0.15}_{-0.17}$	&	$3.40^{+1.13}_{-1.13}$	&	$12.64^{+3.10}_{-2.99}$	&	$1.27^{+0.10}_{-0.11}$	&	$2.19^{+0.73}_{-0.73}$	&	$	8.42^{+2.06}_{-1.99}$	&	60	&	1100	\\	
RXCJ$0528.9-3927 $	&	0.39	&	0.5	&	0.54	&	0.92	&		$1.74^{+0.15}_{-0.17}$	&	$3.65^{+1.22}_{-1.22}$	&	$	6.47^{+2.01}_{-1.86}$	&	$1.03^{+0.10}_{-0.11}$	&	$2.36^{+0.79}_{-0.79}$	&	$	4.38^{+1.36}_{-1.26}$	&	20	&	1000	\\
RXCJ$0532.9-3701 $	&	0.34	&	0.6	&	0.56	&	0.94	&		$1.98^{+0.13}_{-0.13}$	&	$3.50^{+1.17}_{-1.17}$	&	$10.08^{+2.28}_{-1.98}$	&	$1.19^{+0.08}_{-0.08}$	&	$2.26^{+0.75}_{-0.75}$	&	$	6.76^{+1.53}_{-1.33}$	&	45	&	1000	\\	
A3404 	&	0.54	&	0.6	&	0.69	&	0.97	&		$2.23^{+0.17}_{-0.17}$	&	$3.65^{+1.22}_{-1.22}$	&	$12.40^{+3.33}_{-2.82}$	&	$1.33^{+0.11}_{-0.11}$	&	$2.36^{+0.79}_{-0.79}$	&	$	8.40^{+2.26}_{-1.91}$	&	60	&	1000	\\	
Bullet 	&	0.31	&	0.5	&	0.5	&	0.9	&		$1.97^{+0.18}_{-0.18}$	&	$3.50^{+1.17}_{-1.17}$	&	$	9.34^{+3.11}_{-2.54}$	&	$1.15^{+0.12}_{-0.12}$	&	$2.26^{+0.75}_{-0.75}$	&	$	6.27^{+2.09}_{-1.71}$	&	10	&	1100	\\
A$907^{(W)}$ 	&	0.63	&	0.7	&	0.73	&	0.98	&		$1.62^{+0.17}_{-0.18}$	&	$4.00^{+1.33}_{-1.33}$	&	$	4.26^{+1.68}_{-1.40}$	&	$0.94^{+0.11}_{-0.12}$	&	$2.61^{+0.87}_{-0.87}$	&	$	2.95^{+1.16}_{-0.97}$	&	30	&	1000	\\
A$907^{(S)}$ 	&	0.63	&	0.8	&	0.75	&	0.97	&		$1.66^{+0.14}_{-0.14}$	&	$3.95^{+1.32}_{-1.32}$	&	$	4.91^{+1.48}_{-1.23}$	&	$0.99^{+0.09}_{-0.09}$	&	$2.57^{+0.86}_{-0.86}$	&	$	3.39^{+1.02}_{-0.85}$	&	60	&	900	\\
RXCJ$1023.6+0411$ 	&	0.4	&	0.6	&	0.56	&	0.9	&		$2.06^{+0.09}_{-0.10}$	&	$3.45^{+1.15}_{-1.15}$	&	$12.38^{+1.78}_{-1.79}$	&	$1.27^{+0.06}_{-0.06}$	&	$2.22^{+0.74}_{-0.74}$	&	$	8.28^{+1.19}_{-1.20}$	&	20	&	900	\\	
MS$1054.4-0321 $	&	0.15	&	0.3	&	0.21	&	0.78	&		$2.48^{+0.52}_{-0.66}$	&	$2.55^{+0.85}_{-0.85}$	&	$23.03^{+23.63}_{-16.31}$	&	$1.23^{+0.33}_{-0.41}$	&	$1.60^{+0.53}_{-0.53}$	&	$14.20^{+14.57}_{-10.06}$	&	100	&	660	\\		
MACSJ$1115.8+0129^{(W)}$ 	&	0.37	&	0.6	&	0.47	&	0.89	&		$1.73^{+0.20}_{-0.22}$	&	$3.55^{+1.18}_{-1.18}$	&	$	6.24^{+2.78}_{-2.32}$	&	$0.99^{+0.13}_{-0.14}$	&	$2.29^{+0.76}_{-0.76}$	&	$	4.20^{+1.87}_{-1.56}$	&	40	&	1000	\\
MACSJ$1115.8+0129^{(S)}$	&	0.33	&	0.5	&	0.48	&	0.9	&		$1.85^{+0.18}_{-0.19}$	&	$3.45^{+1.15}_{-1.15}$	&	$	8.12^{+2.92}_{-2.47}$	&	$1.08^{+0.12}_{-0.12}$	&	$2.22^{+0.74}_{-0.74}$	&	$	5.43^{+1.95}_{-1.65}$	&	30	&	900	\\
A$1300$ 	&	0.34	&	0.5	&	0.5	&	0.9	&		$1.90^{+0.16}_{-0.18}$	&	$3.50^{+1.17}_{-1.17}$	&	$	8.68^{+2.62}_{-2.42}$	&	$1.12^{+0.10}_{-0.12}$	&	$2.26^{+0.75}_{-0.75}$	&	$	5.82^{+1.76}_{-1.63}$	&	90	&	1000	\\
RXC$J1135.6-2019$ 	&	0.37	&	0.5	&	0.53	&	0.93	&		$1.73^{+0.14}_{-0.16}$	&	$3.60^{+1.20}_{-1.20}$	&	$	6.62^{+1.91}_{-1.80}$	&	$1.03^{+0.09}_{-0.10}$	&	$2.33^{+0.78}_{-0.78}$	&	$	4.47^{+1.29}_{-1.22}$	&	5	&	1000	\\
RXCJ$1206.2-0848 $	&	0.38	&	0.5	&	0.45	&	0.96	&	$2.09^{+0.19}_{-0.22}$	&	$3.20^{+1.07}_{-1.07}$	&	$13.21^{+4.37}_{-4.08}$	&	$1.22^{+0.12}_{-0.14}$	&	$2.05^{+0.68}_{-0.68}$	&	$	8.67^{+2.87}_{-2.68}$	&	100	&	800	\\	
MACSJ$1311.0-0311$ 	&	0.23	&	0.2	&	0.37	&	0.82	&	$1.88^{+0.15}_{-0.17}$	&	$3.20^{+1.07}_{-1.07}$	&	$10.61^{+3.01}_{-2.83}$	&	$1.11^{+0.10}_{-0.11}$	&	$2.05^{+0.68}_{-0.68}$	&	$	6.96^{+1.97}_{-1.86}$	&	10	&	900	\\	
A1689 	&	0.46	&	0.5	&	0.65	&	0.91	&	$2.85^{+0.07}_{-0.07}$	&	$3.35^{+1.12}_{-1.12}$	&	$30.98^{+2.40}_{-2.28}$	&	$1.79^{+0.04}_{-0.04}$	&	$2.15^{+0.72}_{-0.72}$	&	$20.56^{+1.59}_{-1.51}$	&	50	&	1200	\\		
RXJ$1347-1145^{(W)}$	&	0.2	&	0.4	&	0.35	&	0.87	&	$2.18^{+0.19}_{-0.20}$	&	$3.15^{+1.05}_{-1.05}$	&	$15.35^{+4.83}_{-4.18}$	&	$1.27^{+0.12}_{-0.13}$	&	$2.01^{+0.67}_{-0.67}$	&	$10.03^{+3.16}_{-2.73}$	&	40	&	1000	\\		
RXJ$1347-1145^{(S)}$ 	&	0.17	&	0.4	&	0.38	&	0.87	&	$2.38^{+0.15}_{-0.17}$	&	$3.05^{+1.02}_{-1.02}$	&	$21.60^{+4.66}_{-4.57}$	&	$1.42^{+0.10}_{-0.11}$	&	$1.94^{+0.65}_{-0.65}$	&	$14.00^{+3.02}_{-2.96}$	&	30	&	1000	\\		
MACSJ$1359.2-1929 $	&	0.24	&	0.3	&	0.37	&	0.87	&	$1.70^{+0.22}_{-0.26}$	&	$3.40^{+1.13}_{-1.13}$	&	$	6.28^{+3.24}_{-2.76}$	&	$0.95^{+0.14}_{-0.17}$	&	$2.19^{+0.73}_{-0.73}$	&	$	4.19^{+2.16}_{-1.84}$	&	20	&	900	\\
A$1835 $	&	0.44	&	0.6	&	0.6	&	0.96	&	$2.51^{+0.14}_{-0.15}$	&	$3.35^{+1.12}_{-1.12}$	&	$20.70^{+3.89}_{-3.69}$	&	$1.52^{+0.09}_{-0.10}$	&	$2.15^{+0.72}_{-0.72}$	&	$13.74^{+2.58}_{-2.45}$	&	60	&	900	\\		
RXJ$1504$ 	&	0.49	&	0.7	&	0.64	&	0.94	&	$1.87^{+0.14}_{-0.15}$	&	$3.70^{+1.23}_{-1.23}$	&	$	7.72^{+2.03}_{-1.84}$	&	$1.12^{+0.09}_{-0.10}$	&	$2.40^{+0.80}_{-0.80}$	&	$	5.25^{+1.38}_{-1.25}$	&	5	&	1000	\\
A$2163 $	&	0.5	&	0.8	&	0.62	&	0.95	&	$2.53^{+0.18}_{-0.18}$	&	$3.45^{+1.15}_{-1.15}$	&	$19.11^{+4.74}_{-4.06}$	&	$1.51^{+0.12}_{-0.12}$	&	$2.22^{+0.74}_{-0.74}$	&	$12.78^{+3.17}_{-2.72}$	&	90	&	900	\\		
A$2204$ 	&	0.54	&	0.8	&	0.7	&	0.97	&	$2.05^{+0.15}_{-0.16}$	&	$3.75^{+1.25}_{-1.25}$	&	$	9.58^{+2.45}_{-2.22}$	&	$1.32^{+0.10}_{-0.10}$	&	$2.43^{+0.81}_{-0.81}$	&	$	6.53^{+1.67}_{-1.51}$	&	5	&	1000	\\
RXCJ$2014.8-2430 $	&	0.69	&	0.8	&	0.71	&	0.98	&	$2.03^{+0.25}_{-0.29}$	&	$3.80^{+1.27}_{-1.27}$	&	$	7.86^{+3.80}_{-3.25}$	&	$1.16^{+0.16}_{-0.19}$	&	$2.47^{+0.82}_{-0.82}$	&	$	5.37^{+2.60}_{-2.22}$	&	60	&	1000	\\
RXCJ$2151.0-0736 $	&	0.31	&	0.7	&	0.51	&	0.94	&	$1.47^{+0.21}_{-0.25}$	&	$3.85^{+1.28}_{-1.28}$	&	$	3.24^{+1.90}_{-1.57}$	&	$0.82^{+0.14}_{-0.16}$	&	$2.50^{+0.83}_{-0.83}$	&	$	2.22^{+1.31}_{-1.08}$	&	30	&	1000	\\
A$2390 $	&	0.46	&	0.8	&	0.65	&	0.93	&	$2.20^{+0.10}_{-0.10}$	&	$3.50^{+1.17}_{-1.17}$	&	$14.01^{+2.10}_{-1.91}$	&	$1.35^{+0.06}_{-0.06}$	&	$2.26^{+0.75}_{-0.75}$	&	$	9.40^{+1.41}_{-1.28}$	&	50	&	1000	\\	
MACSJ$2214.9-1359$	&	0.27	&	0.3	&	0.39	&	0.83	&	$2.03^{+0.19}_{-0.20}$	&	$3.10^{+1.03}_{-1.03}$	&	$12.90^{+4.42}_{-3.77}$	&	$1.17^{+0.12}_{-0.13}$	&	$1.98^{+0.66}_{-0.66}$	&	$	8.39^{+2.88}_{-2.45}$	&	20	&	660	\\	
MACSJ$2243.3-0935 $	&	0.2	&	0.5	&	0.42	&	0.86	&	$2.18^{+0.17}_{-0.18}$	&	$3.15^{+1.05}_{-1.05}$	&	$15.74^{+4.34}_{-3.86}$	&	$1.29^{+0.11}_{-0.12}$	&	$2.01^{+0.67}_{-0.67}$	&	$10.29^{+2.84}_{-2.52}$	&	120	&	850	\\		
RXCJ$2248.7-4431 $	&	0.49	&	0.7	&	0.57	&	0.9	&	$2.19^{+0.14}_{-0.15}$	&	$3.30^{+1.10}_{-1.10}$	&	$14.89^{+3.26}_{-3.03}$	&	$1.32^{+0.09}_{-0.10}$	&	$2.12^{+0.71}_{-0.71}$	&	$	9.84^{+2.16}_{-2.01}$	&	45	&	890	\\	
A$2537 $	&	0.37	&	0.6	&	0.52	&	0.87	&	$2.25^{+0.12}_{-0.13}$	&	$3.35^{+1.12}_{-1.12}$	&	$15.77^{+2.82}_{-2.71}$	&	$1.37^{+0.08}_{-0.08}$	&	$2.15^{+0.72}_{-0.72}$	&	$10.46^{+1.87}_{-1.80}$	&	70	&	1000	\\		
RXCJ$2337.6+0016 $	&	0.19	&	0.7	&	0.55	&	0.88	&	$1.99^{+0.11}_{-0.13}$	&	$3.50^{+1.17}_{-1.17}$	&	$10.56^{+1.96}_{-2.04}$	&	$1.21^{+0.07}_{-0.08}$	&	$2.26^{+0.75}_{-0.75}$	&	$	7.08^{+1.32}_{-1.37}$	&	25	&	900	\\	

\hline  
\end{tabular}
\end{table*}

\subsubsection{Comparison to previous publications}
 Our cluster sample has a significant overlap with three other weak lensing cluster studies, the Canadian Cluster Comparison Project (CCCP,\cite{2015MNRAS.449..685H}), Weighing the Giants (WtG, \cite{2012arXiv1208.0605A}) and the Local Cluster Substructure Survey (LoCuSS, \cite{2016MNRAS.461.3794O}) cluster samples. While the WtG and LoCuSS samples partially use the same raw data as our work, the CCCP sample is based on imaging data from a different telescope and instrument.
 All publications assume the same reference cosmology with $\Omega_{m}\!=\!0.3$, $\Omega_{\Lambda}\!=\!0.7$ and $h\equiv H_{0}/(100 \mathrm{km s}^{-1} \mathrm{Mpc}^{-1})=0.7$. This allows us to directly compare the mass estimates at identical overdensities $\Delta$.
 \citet{2015MNRAS.449..685H} provide masses at $\Delta\!=\!500$, \citet{2016MNRAS.461.3794O} 
 additionally offer the estimates $\Delta\!=\!200$ to compare with our measurement.
 
For the ten clusters we have in common with the LoCuSS sample, we find a median ratio ($M_{\mathrm{LoCuSS}}/M_{\mathrm{APEX-SZ}}$) of $0.99$ for $\Delta\!=\!200$ and $1.08$ for $\Delta\!=\!500$. We see that one cluster, A907 strikes out from the distribution with a mass ratio $4.2$ at $\Delta\!=\!200$.
Omitting this cluster we obtain a average mass ratio of $0.96$ ($1.05$) for $\Delta\!=\!200$ ($\Delta\!=\!500$) and a standard deviation of $0.3$ for both overdensities. Under the assumption that the error on the mean is $0.3/\sqrt{9}$ and considering the additional 8\% systematic uncertainty, our mass estimates are in agreement with those given by \citet{2016MNRAS.461.3794O}.
 
 In case of the CCCP cluster sample, we find eleven clusters that are in common with the APEX-SZ sample.
\citet{2015MNRAS.449..685H} provide two different mass estimates, one is based on aperture mass, the other one is based on fitting an NFW profile to the shear.
 We find a median mass ratio ($M_{\mathrm{CCCP}}/M_{\mathrm{APEX-SZ}}$) of $0.86$ and $0.84$ for the aperture-based and NFW-based masses, respectively.
Furthermore, we find an average ratio of $0.95$ and $1.09$, respectively. The standard deviation is found to be $0.3$ for the aperture mass technique based masses and $0.45$ for the NFW profile fit. 
Using the same assumptions as in the previous paragraph, we find 0.09 and 0.14 as error on the mean for the aperture and the NFW based mass ratios. We therefore do not find any significant bias between our work and that of \citet{2015MNRAS.449..685H}.

We find ten clusters in common with WtG cluster sample. The masses given in WtG are given within a fixed radius of $1.5$ Mpc and were obtained using a fixed concentration of $4.0$ at $r_{200}$. We approximate the mass within $1.5$ Mpc using our NFW fit results at $r_{500}$. The median mass ratio for WtG is $1.12$ and the average is $1.25$.
The scatter of the mass ratio is 0.34, which yield under aforementioned assumptions an uncertainty of 0.13. 
The median mass ratio to WtG is therefore within one standard deviation, the mean mass ratio two standard deviations higher than our estimates. We note that we find concentrations $<4$ at $r_{200}$ for all clusters that overlap with WtG, which might contribute to the mass offset seen here.

To summarise, we see reasonable agreement between our mass estimates and those from literature, with the largest offset seen in the comparison with WtG. 

\subsubsection{Concentration parameter} \label{sec:c200}
\begin{figure*}
 \centering
\includegraphics[width=0.9\linewidth]{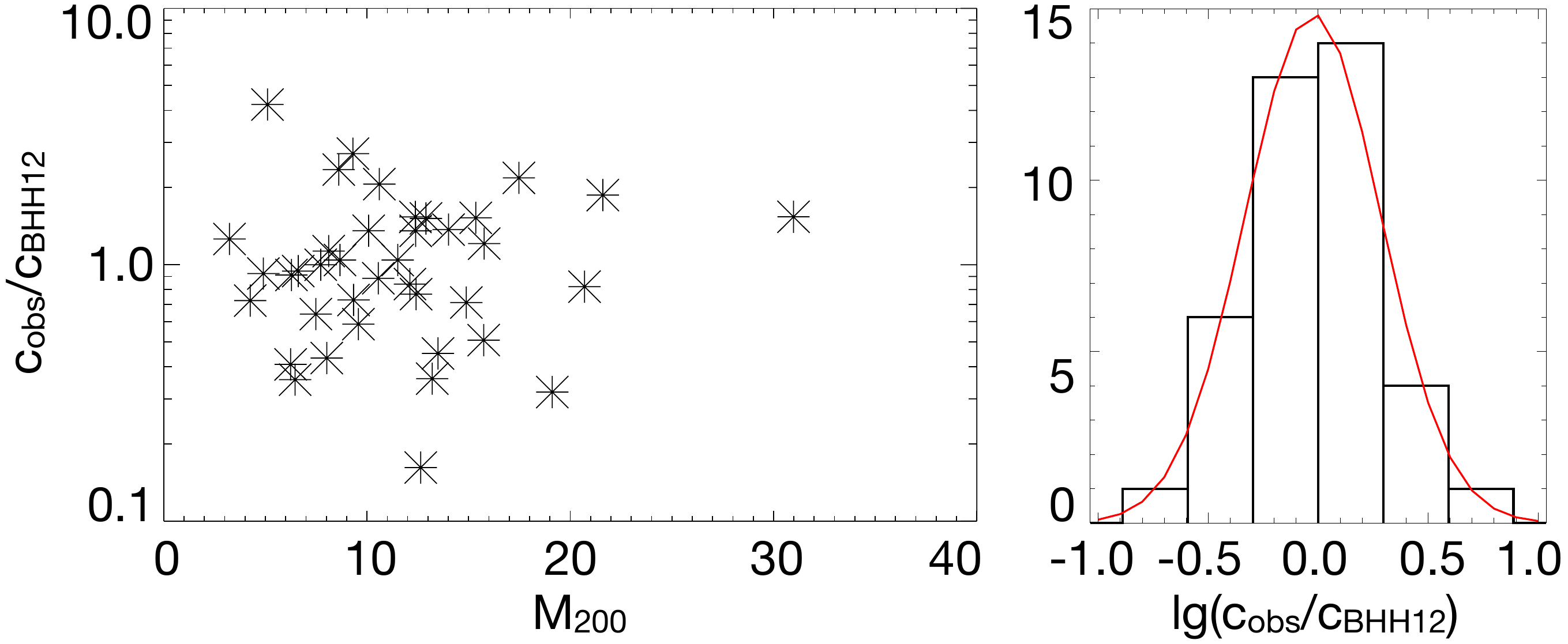}
\caption{\textit{Left panel}: Ratio of observed over predicted concentration versus measured $M_{200}$. \textit{Right panel}: Histogram of ratios of observed over predicted concentration in log space. The red line shows the fitted Gaussian function. The fit yields a mean of $-0.02\pm0.05$, which is corresponds to 0.95 in linear space.}\label{fig:masscon2}
\end{figure*}
In contrast with some weak lensing studies with large numbers of clusters \citep{2012arXiv1208.0605A, 2012MNRAS.427.1298H}, we provide weak lensing measurements with and without leaving the concentration free to vary. Since weak lensing generally offers only weak constrains on the concentration, we use the full sample of $39$ clusters to study the recovery of the average concentration.
This test does not aim to perform a detailed study of the mass-concentration relation but it allows to test for consistency between measured and predicted concentrations. Inconsistency between predicted and observed concentration would impact our lensing masses. This analysis is additionally motivated by the low concentrations found in the study of $8$ galaxy clusters by \citet{2012A&A...546A..79I}. Our analysis shares a significant amount of code and methods with \citet{2012A&A...546A..79I}. 
Furthermore, the measurement of low concentrations could be caused by remaining cluster contamination diluting the shear signal towards the cluster centre. The absence of such a underestimation therefore would further support the quality of our analysis even if the measurement accuracy of the concentration for individual clusters is poor.

As in \citet{2012A&A...546A..79I}, we compare our results with that of the $c-M$ scaling relation by \citet{2013ApJ...766...32B}, as one of only few simulations covering a sufficiently large volume to probe the high mass range covered by our sample.
The results in \citet{2013ApJ...766...32B} are based on three simulations with different box sizes and mass resolutions. The $\Lambda$CDM parameters used in that simulations are close to the results 
of WMAP-7 \citep{2011ApJS..192...18K} and the large box size allows better constraints for the high mass part of the scaling relation compared to previous studies. This is of special importance for our study since our clusters occupy the highest mass range studied in most simulations. 
\citet{2013ApJ...766...32B} found that the concentration parameter can be expressed as a function of the peak height parameter $\nu$ 
as follows:\footnote{We use the fitting formula for $\nu(M,z)$ from Table 2 of 
\citet{2013ApJ...766...32B}}
\begin{equation} \label{eq:cnurelation}
c_{200,BHH12}(\nu)\!=\!5.9\nu^{-0.41}D(z)^{0.54},
\end{equation}
for all systems and
\begin{equation} \label{eq:cnurelation2}
c_{200,BHH12}(\nu)\!=\!6.6\nu^{-0.35}D(z)^{0.53},
\end{equation}
for relaxed systems.
Here $D(z)$ is the linear growth factor for a flat $\Lambda$CDM universe. The distribution of the concentrations in the simulations can be described by a Gaussian with variance of $\sigma_{c} = 0.33c$. 
Using the connection between $\nu$ and the cluster mass as described in the above-mentioned paper, we can derive the expected concentration parameters for each of the clusters. 

The ratio between our $c_{200}$ measurements before applying a prior on concentration over the concentrations  obtained with the mass-concentration relation versus $M_{200}$ is shown in the top panel of Fig.~\ref{fig:masscon2}. The bottom panel of Fig.~\ref{fig:masscon2} shows the the histogram of the ratios between observed and predicted concentration in log space. The distribution is well described by a Gaussian function with a mean of $-0.02\pm0.05$. This corresponds to 0.95 in linear space and uncertainty of ~12\%.
We again stress that Fig.~\ref{fig:masscon2} is of indicative purpose only. A proper study of the mass-concentration relation would require significantly more effort such as the inclusion of the scatter in the applied $c-M$ relation, proper modeling of the errors in concentration, the correlation with mass, and the modeling of the selection function.

The mean ratio is consistent with one, further it is  close to the underestimation of $7$\% predicted by \citet{2012MNRAS.421.1073B} for weak lensing studies.
Given this results we do not expect that our weak lensing results are significantly affected by our choice of using a $c-M$ scaling for our our default mass estimates.

\subsection{Discussion of the example clusters}\label{indiclusters}
A detailed discussion of each cluster is beyond the scope of this paper and a small number of selected cluster will be further discussed in light of multi-frequency studies.
We therefore only discuss here the three clusters used as examples in this publication to some greater detail.
\subsubsection{A907}
\begin{figure*}
   \centering
    \includegraphics[width=0.99\linewidth]{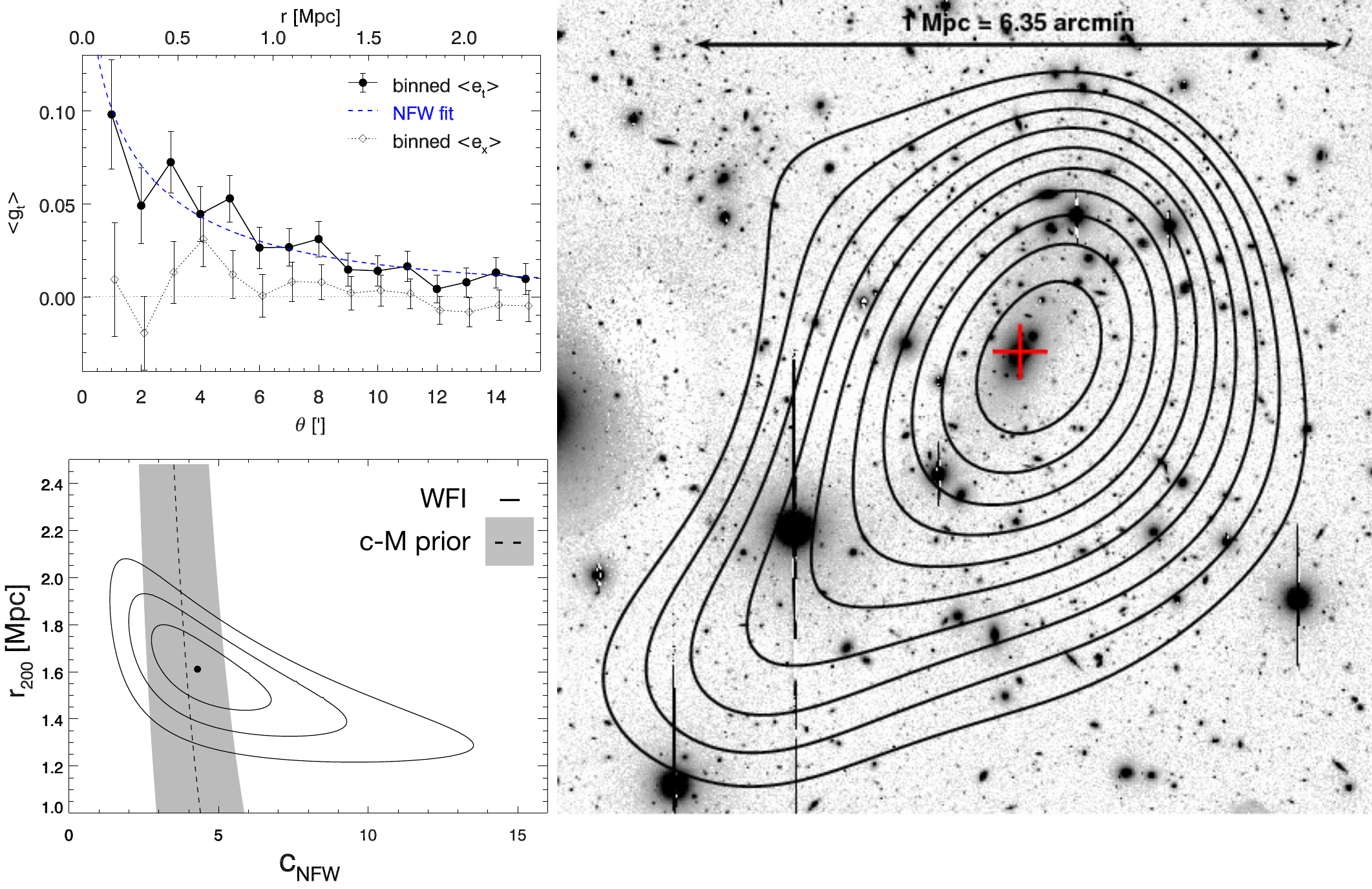}
      \caption{Lensing results for A907.
\textit{Top left panel:} Profiles of the binned
tangential ($\langle\varepsilon_{\mathrm{t}}\rangle$, filled circles) and binned cross ($\langle\varepsilon_{\mathrm{x}}\rangle$, open diamonds) ellipticities. Error bars reflect the bin dispersion.
\textit{Lower left panel:} $\Delta \chi^2(r_{200},c_{\mathrm{NFW}})$ with respect to its minimum, (filled circle); contours indicating $1\sigma$, $2\sigma$, $3\sigma$ confidence levels without c-M prior. Dashed line and gray shaded area shows the adopted c-M relation and one $\sigma$ uncertainty region}.\textit{Right panel:} $R$-band image of the central region overlaid with $\kappa$ contours using $\kappa=0.05$ in steps of $\Delta\kappa=0.01$.
The black contours show the detection 
optimizing background selection case (Table~\ref{table:res}). The red cross marks the BCG which we defined as the cluster centre.
         \label{FigA907}
 \end{figure*}

The excellent Suprime-Cam data for this cluster and its low cluster redshift result in a high number density of background galaxies of $n=20.5\,\text{arcmin}^{-2}$. 
Our results as listed in Table~\ref{table:res} and illustrated in Fig.~\ref{FigA907} are consistent with X-ray measurements by \citet{2008A&A...482..451Z}
and \citet{2010ApJ...722...55N} which derive masses between $M_{500}=3.2\times10^{14}$ and $M_{500}=4.7\times10^{14}\mathrm{M}_\odot$. 
~\citet{2010A&A...524A..68E} estimated $R_{200}$ using X-rays to be $R_{200}=2.18\pm0.17$ Mpc. This measurement disagrees with our estimate of $R_{200}=1.61^{+0.12}_{-0.11}$ Mpc.
The iso-density contours of the $\kappa$ reconstruction as seen in Fig.~\ref{FigA907} are elliptical and show a slight elongation to the South-East. The semi-major axis of the contours and that of the BCG are aligned into the same direction. 
  
\subsubsection{RXC0532}
\begin{figure*}
   \centering
   \includegraphics[width=\linewidth]{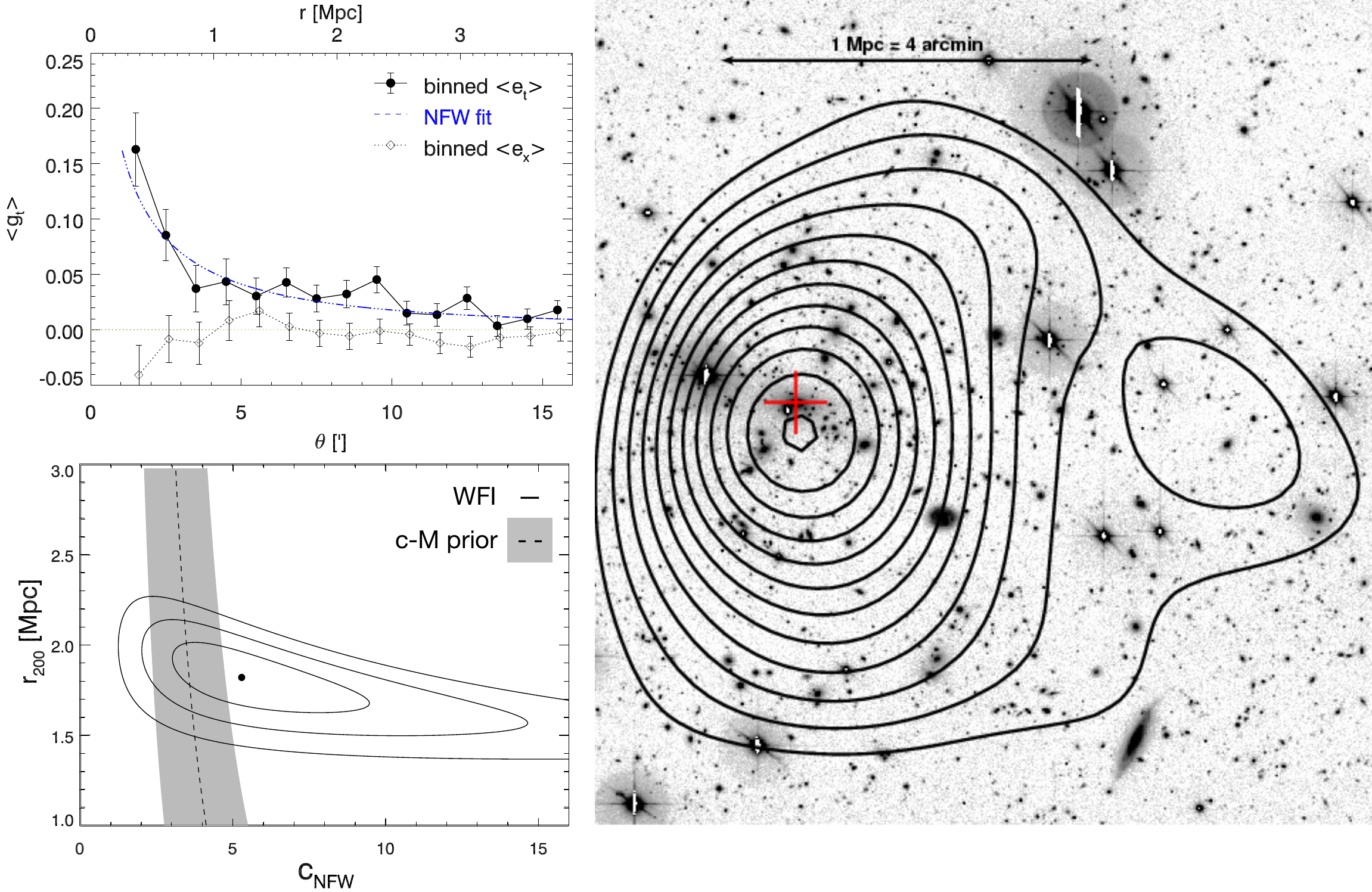}
      \caption{Lensing results for RXC0532.
\textit{Top left panel:} Profiles of the binned
tangential ($\langle\varepsilon_{\mathrm{t}}\rangle$, filled circles) and binned cross ($\langle\varepsilon_{\mathrm{x}}\rangle$, open diamonds) ellipticities. Error bars reflect the bin dispersion.
\textit{Lower left panel:} $\Delta \chi^2(r_{200},c_{\mathrm{NFW}})$ with respect to its minimum (filled circle), contours indicating $1\sigma$, $2\sigma$, $3\sigma$ confidence levels without c-M prior. Dashed line and gray shaded area shows the adopted c-M relation and one $\sigma$ uncertainty region.
\textit{Right panel:} $R$-band image of the central region overlaid with $\kappa$ contours using $\kappa=0.05$ in steps of $\Delta\kappa=0.01$.
The black contours show the detection optimizing 
background selection case Table~\ref{table:res}. The red cross marks the BCG which we defined as the cluster centre.}
         \label{FigRXC0532}
 \end{figure*}

The results of RXC 0532 as listed in table \ref{table:res} and shown in Fig.\ref{FigRXC0532} show the smallest relative errors on mass out of the three example clusters presented in this paper. Our estimated $R_{200}$ of $1.82^{+0.12}_{-0.12}$ Mpc fits well with the two results of ~\citet{2010A&A...524A..68E} of 
$1.78\pm0.18$ and $1.84\pm0.23$ Mpc based on X-ray observations. Using our best fit NFW parameters we get a mass of $M_{500}=7.4^{+1.6}_{-1.4}\times10^{14} \mathrm{M}_{\odot}$. This is slightly discrepant with the estimate
by the Planck Collaboration \citep{2011A&A...536A..11P} which, using a $M_{500}-Y_{\mathrm{X},500}$ scaling relation, derive a mass of 
$M_{500}=5.35\pm0.17\times 10^{14} \mathrm{M}_{\odot}$.

\subsubsection{RXC1347}
\begin{figure*}
   \centering
   \includegraphics[width=\linewidth]{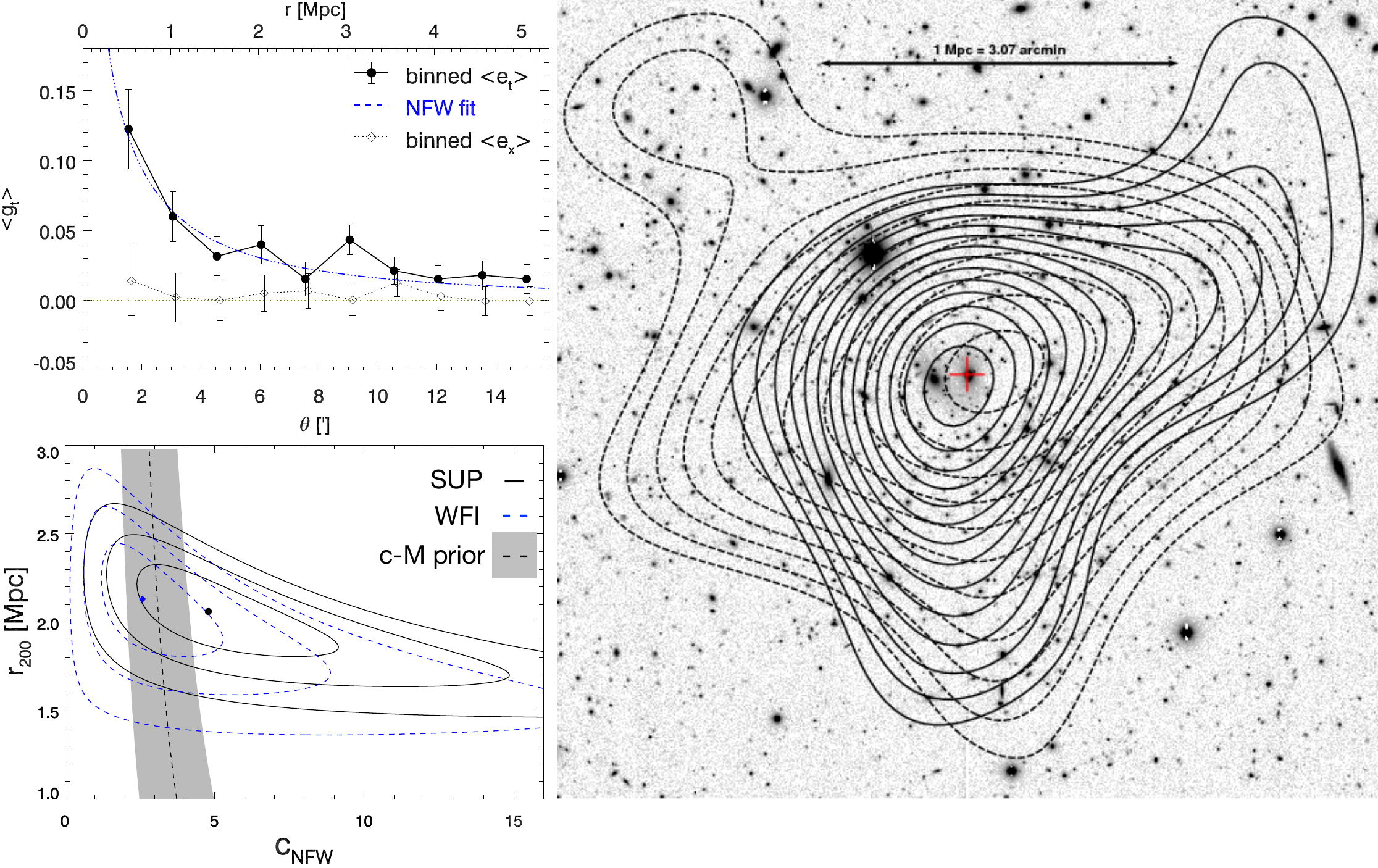}
      \caption{Lensing results for RXC1347.
\textit{Top left panel:} Profile of the binned
tangential ($\langle\varepsilon_\mathrm{t}\rangle$, filled circles) and binned cross ($\langle\varepsilon_\mathrm{x}\rangle$, open diamonds) ellipticities. Error bars reflect the bin dispersion. Based on the SuprimeCam 
data. 
\textit{Lower left panel:} $\Delta \chi^2(r_{200},c_{\mathrm{NFW}})$ with respect to its minimum. Black lines correspond to SuprimeCam measurements, blue dashed lines to WFI based measurements.
Dashed line and gray shaded area shows the adopted c-M relation and one $\sigma$ uncertainty region.
\textit{Right panel:} $R$-band image of the central region overlaid with $\kappa$ contours using $\kappa=0.05$ in steps of $\Delta\kappa=0.01$.
Solid lines represent the results
obtained with SuprimeCam the dashed lines the results obtained with WFI. The red cross marks the BCG which we defined as the cluster centre.}
         \label{FigRXC1347}
 \end{figure*}

RXCJ1347, as the most X-ray luminous galaxy cluster known, was the object of several detailed studies. Our results shown in Table~\ref{table:res} and 
Fig.~\ref{FigRXC1347} show excellent agreement with the result of \citet{2011ApJ...729..127U} which obtain of 
$M_{200}=14.3\pm3.5\times 10^{14} \mathrm{M}_{\odot}$ for a non-parametric de-projection analysis and $M_{200}=19.1^{+3.6}_{-3.3}\times 10^{14} \mathrm{M}_{\odot}$
for a NFW fit on the deeper Suprime-Cam data. Also the weak lensing mass from \citet{2010MNRAS.403.1787L} of $M_{\mathrm{vir}}=18.9^{+5.9}_{-5.5}\times10^{14} \mathrm{M}_{\odot}$ 
using the Megaprime instrument on CFHT agrees well with our measurement of ${M}_{\mathrm{vir}}=18.3^{+4.9}_{-4.0}\times10^{14} \mathrm{M}_{\odot}$ extrapolated the virial radius.

Our result also agrees with the X-ray mass estimates of ~\citet{2007A&A...472..383G} which
derive $M_{200}=11.0\pm2.8\times 10^{14} \mathrm{M}_{\odot}$ for the best \color{blue} \citet{1978A&A....70..677C} $\beta$ profile \color{black}
and $M_{200}=17.9\pm2.1\times 10^{14} \mathrm{M}_{\odot}$ for the best NFW profile fit.
The WFI data are shallower than the Subaru ones but show excellent agreement with them. We  highlight that these results have been obtained with independent instruments and different filter sets.
It is remarkable that we have a similar or higher number density of background sources for the lensing analysis as \cite{2011ApJ...729..127U} despite using shallower data. This can partially be explained by the
different data reduction pipelines yielding different seeing in the co-added image, but the lower mean lensing depth of $\langle\beta\rangle=0.35$ compared to $\langle\beta\rangle=0.49$ 
indicates that we use a background sample with lower mean redshift in our sample than \citet{2011ApJ...729..127U}.

\section{Conclusion and future perspectives}\label{sec:conclu}
In this paper we presented the weak lensing analysis of $39$ galaxy clusters, observed as part of the APEX-SZ cluster sample.
We presented and discussed a background selection method which calculates the angular diameter ratio $\beta$ for each galaxy based on
observations in three broad-band filters.
We investigated the remaining contamination by cluster galaxies using the binned shear profiles of two different subsamples of background galaxies, a 
method insensitive to holes in the data set. Additionally, we investigated the radial number density profiles of the background galaxy sample.
Both methods did not show significant hints of contamination by cluster members.
Based on the COSMOS photo-$z$ catalogue we showed that cosmic variance can result in a scatter of the mean lensing depth $\langle\beta\rangle$ of $1-3$\% for clusters in the
redshift range of our sample and typical image depth of our data. Due to correlated structures this estimate likely gives a lower limit of the true scatter in $\langle\beta\rangle$.
We further showed that our method of colour-based distance estimates can reduce this scatter by up to $40$\%.
Using the full cluster sample, we showed that our overall mass estimates are robust against moderate changes in the selection parameter $\beta_{\mathrm{cut}}$ and galaxy signal-to-noise threshold.
Further, it could be shown that the derived concentrations are consistent with those predicted by the mass concentration relation presented in \cite{2013ApJ...766...32B}, with no significant hint of an underestimation of the derived concentrations.
The comparison with literature shows overall agreement with the masses derived for the CCCP or LoCuSS samples, but show lower masses compared to the results in WtG.

As examples for the whole cluster sample we discussed the results of three clusters in a redshift range from $z=0.15$ to $z=0.45$ in greater detail. The weak lensing analyses 
yielded results which are consistent with previous X-ray and lensing mass estimates. 

In case of A907, the weak lensing data disfavor the recent X-ray estimate of ~\citet{2010A&A...524A..68E}, favoring the results of ~\citet{2010ApJ...722...55N} 
and ~\citet{2008A&A...482..451Z}.
We also found agreement of the mass estimates based on Suprime-Cam and WFI for the massive cluster RXCJ1347 using very similar cuts. 

The weak-lensing masses presented in this work are used in the mass calibration of the integrated Comptonization using the APEX-SZ clusters presented in \cite{2018Nagarajan}. Together with our SZ analysis (\citealt{2016MNRAS.460.3432B,2018Nagarajan}) and our upcoming X-ray analysis (F. Pacaud et al., in prep.) this work builds one major part of our multi-frequency study of APEX-SZ galaxy clusters.

\section*{Acknowledgments}

The authors thank Mischa Schirmer for providing us with the coadded images of the MS0451.6$-$0305 field.
We acknowledge the support of the Max Planck Gesellschaft Faculty Fellowship program and the High Energy Group at MPE. Further, we acknowledge the support of the DFG Cluster of Excellence ``Origin and Structure of the Universe", the Transregio program TR33 ``The Dark Universe", the Ludwig-Maximilians-Universit\"at, the University of Bonn and the SFB 956 ``Conditions and Impact of Star Formation".
This publication makes use of data products from the Wide-field Infrared Survey Explorer, which is a joint project of the University of California, Los Angeles, and the Jet Propulsion Laboratory/California Institute of Technology, funded by the National Aeronautics and Space Administration.
FP acknowledges support by the German Aerospace
Agency (DLR) with funds from the Ministry of Economy
and Technology (BMWi) through grants 50 OR 1514 and 50
OR 1608.




\bibliographystyle{mnras}
\bibliography{APEXWLpap.bib} 



\appendix
\section{Measurements without prior on concentration parameter}
In addition to the weak lensing measurements using the default setting shown in Table~\ref{table:res}, we present in Table~\ref{table:resAP} the results without applying a prior on the concentration parameter. Table~\ref{table:resAP2} shows the results using the S/N optimized background selection without prior on the concentration.

\begin{table*}
\caption{Weak lensing mass estimates for the conservative background selection, for the overdensities $\Delta=200$ and $\Delta=500$, without prior on concentration. For the clusters
A907, MACSJ1115 and RXCJ1347, the WFI based masses are listed in the first line, the SUP based masses are listed in the following line without reciting the cluster name.
}             
\label{table:resAP}      
\centering                          
\setlength\tabcolsep{3.3pt}
\begin{tabular}{c c c c c c c }        
\hline\hline                 
Cluster & $R_{200}$  & $c_{200}$ & $M_{200}$ & $R_{500}$  & $c_{500}$ & $M_{500}$  \\    
&  [Mpc] & & $[10^{14}\mathrm{M}_{\odot}]$ & [Mpc] & &  $[10^{14}\mathrm{M}_{\odot}]$ \\
\hline  
A2744	&	$	2.09	_{-	0.13	}^{+	0.13	}$	&	$	7.2	_{-	2.95	}^{+	6.95	}$	&	$	14.16	_{-	2.48	}^{+	2.81	}$	&	$	1.41	_{-	0.07	}^{+	0.07	}$	&	$	4.86	_{-	2.08	}^{+	4.92	}$	&	$	10.87	_{-	1.59	}^{+	1.72	}$\\
RXCJ0019.0-2026	&	$	1.94	_{-	0.15	}^{+	0.15	}$	&	$	3.6	_{-	1.1	}^{+	1.5	}$	&	$	11.05	_{-	2.37	}^{+	2.77	}$	&	$	1.25	_{-	0.09	}^{+	0.08	}$	&	$	2.33	_{-	0.76	}^{+	1.05	}$	&	$	7.47	_{-	1.43	}^{+	1.51	}$\\
A2813	&	$	2.07	_{-	0.18	}^{+	0.19	}$	&	$	2.65	_{-	0.95	}^{+	1.25	}$	&	$	13.61	_{-	3.25	}^{+	4.1	}$	&	$	1.3	_{-	0.09	}^{+	0.09	}$	&	$	1.67	_{-	0.65	}^{+	0.87	}$	&	$	8.49	_{-	1.67	}^{+	1.83	}$\\
A209	&	$	2.39	_{-	0.16	}^{+	0.19	}$	&	$	1.6	_{-	0.5	}^{+	0.65	}$	&	$	19.07	_{-	3.58	}^{+	4.92	}$	&	$	1.43	_{-	0.07	}^{+	0.07	}$	&	$	0.96	_{-	0.33	}^{+	0.44	}$	&	$	10.12	_{-	1.44	}^{+	1.61	}$\\
XLSSC 006	&	$	1.68	_{-	0.23	}^{+	0.25	}$	&	$	2.15	_{-	1.65	}^{+	2.8	}$	&	$	8.51	_{-	3.04	}^{+	4.39	}$	&	$	1.04	_{-	0.14	}^{+	0.09	}$	&	$	1.33	_{-	1.07	}^{+	1.95	}$	&	$	4.98	_{-	1.71	}^{+	1.47	}$\\
RXCJ0232.2-4420	&	$	1.48	_{-	0.2	}^{+	0.2	}$	&	$	15.95	_{-	7.2	}^{+	22.35	}$	&	$	4.91	_{-	1.73	}^{+	2.27	}$	&	$	1.03	_{-	0.12	}^{+	0.12	}$	&	$	11.05	_{-	5.1	}^{+	15.97	}$	&	$	4.08	_{-	1.31	}^{+	1.67	}$\\
RXCJ0245.4-5302	&	$	1.47	_{-	0.22	}^{+	0.22	}$	&	$	10.35	_{-	5.2	}^{+	5.65	}$	&	$	4.93	_{-	1.9	}^{+	2.56	}$	&	$	1.01	_{-	0.14	}^{+	0.13	}$	&	$	7.08	_{-	3.67	}^{+	4.01	}$	&	$	3.95	_{-	1.43	}^{+	1.78	}$\\
A383	&	$	1.6	_{-	0.13	}^{+	0.13	}$	&	$	8.7	_{-	2.5	}^{+	3.8	}$	&	$	5.62	_{-	1.26	}^{+	1.49	}$	&	$	1.09	_{-	0.08	}^{+	0.08	}$	&	$	5.92	_{-	1.76	}^{+	2.69	}$	&	$	4.42	_{-	0.91	}^{+	1.05	}$\\
RXCJ0437.1+0043	&	$	2.02	_{-	0.23	}^{+	0.24	}$	&	$	2.9	_{-	1.3	}^{+	2.35	}$	&	$	12.54	_{-	3.81	}^{+	5.02	}$	&	$	1.28	_{-	0.12	}^{+	0.12	}$	&	$	1.84	_{-	0.89	}^{+	1.64	}$	&	$	8.01	_{-	2.04	}^{+	2.39	}$\\
MS0451.6-0305	&	$	1.52	_{-	0.17	}^{+	0.17	}$	&	$	8.55	_{-	3.55	}^{+	7.45	}$	&	$	7.27	_{-	2.18	}^{+	2.72	}$	&	$	1.03	_{-	0.11	}^{+	0.1	}$	&	$	5.81	_{-	2.5	}^{+	5.28	}$	&	$	5.7	_{-	1.59	}^{+	1.9	}$\\
A520	&	$	2.09	_{-	0.18	}^{+	0.21	}$	&	$	1.6	_{-	0.6	}^{+	0.75	}$	&	$	12.7	_{-	3.01	}^{+	4.22	}$	&	$	1.25	_{-	0.09	}^{+	0.09	}$	&	$	0.96	_{-	0.39	}^{+	0.51	}$	&	$	6.74	_{-	1.32	}^{+	1.49	}$\\
RXCJ0516.6-5430	&	$	2.44	_{-	0.27	}^{+	0.45	}$	&	$	0.55	_{-	0.5	}^{+	0.7	}$	&	$	22.37	_{-	6.63	}^{+	14.8	}$	&	$	1.26	_{-	0.22	}^{+	0.15	}$	&	$	0.28	_{-	0.26	}^{+	0.44	}$	&	$	7.74	_{-	3.37	}^{+	3.07	}$\\
RXCJ0528.9-3927	&	$	1.87	_{-	0.25	}^{+	0.25	}$	&	$	1.3	_{-	0.8	}^{+	1.35	}$	&	$	9.94	_{-	3.48	}^{+	4.54	}$	&	$	1.09	_{-	0.14	}^{+	0.12	}$	&	$	0.76	_{-	0.5	}^{+	0.91	}$	&	$	4.89	_{-	1.7	}^{+	1.72	}$\\
RXCJ0532.9-3701	&	$	1.82	_{-	0.15	}^{+	0.14	}$	&	$	4.75	_{-	1.6	}^{+	2.35	}$	&	$	9.07	_{-	2.06	}^{+	2.26	}$	&	$	1.2	_{-	0.09	}^{+	0.08	}$	&	$	3.13	_{-	1.12	}^{+	1.66	}$	&	$	6.5	_{-	1.29	}^{+	1.34	}$\\
A3404	&	$	1.97	_{-	0.22	}^{+	0.25	}$	&	$	4.95	_{-	1.9	}^{+	2.85	}$	&	$	10.25	_{-	3.06	}^{+	4.42	}$	&	$	1.3	_{-	0.13	}^{+	0.13	}$	&	$	3.27	_{-	1.33	}^{+	2.01	}$	&	$	7.4	_{-	1.96	}^{+	2.53	}$\\
Bullet	&	$	1.92	_{-	0.23	}^{+	0.24	}$	&	$	2.55	_{-	1.15	}^{+	1.55	}$	&	$	10.9	_{-	3.47	}^{+	4.62	}$	&	$	1.2	_{-	0.13	}^{+	0.12	}$	&	$	1.6	_{-	0.78	}^{+	1.08	}$	&	$	6.72	_{-	1.93	}^{+	2.31	}$\\
A907	&	$	1.54	_{-	0.23	}^{+	0.24	}$	&	$	2.9	_{-	1.4	}^{+	2.55	}$	&	$	4.82	_{-	1.85	}^{+	2.62	}$	&	$	0.98	_{-	0.13	}^{+	0.12	}$	&	$	1.84	_{-	0.95	}^{+	1.78	}$	&	$	3.08	_{-	1.05	}^{+	1.28	}$\\
	&	$	1.56	_{-	0.19	}^{+	0.2	}$	&	$	3.65	_{-	1.5	}^{+	2.6	}$	&	$	5.01	_{-	1.62	}^{+	2.19	}$	&	$	1.01	_{-	0.11	}^{+	0.1	}$	&	$	2.36	_{-	1.04	}^{+	1.83	}$	&	$	3.4	_{-	0.96	}^{+	1.13	}$\\
RXCJ1023.6+0411	&	$	1.88	_{-	0.11	}^{+	0.1	}$	&	$	5.3	_{-	1	}^{+	1.3	}$	&	$	10.17	_{-	1.68	}^{+	1.71	}$	&	$	1.25	_{-	0.07	}^{+	0.06	}$	&	$	3.52	_{-	0.7	}^{+	0.92	}$	&	$	7.44	_{-	1.1	}^{+	1.11	}$\\
MS1054.4-0321	&	$	1.99	_{-	0.44	}^{+	0.33	}$	&	$	16	_{-	12.5	}^{+	43	}$	&	$	22.79	_{-	12.02	}^{+	13.32	}$	&	$	1.38	_{-	0.31	}^{+	0.23	}$	&	$	11.09	_{-	8.83	}^{+	0	}$	&	$	18.97	_{-	10.07	}^{+	11.09	}$\\
MACSJ1115.8+0129	&	$	1.65	_{-	0.25	}^{+	0.24	}$	&	$	1.45	_{-	1.05	}^{+	2.1	}$	&	$	7.4	_{-	2.88	}^{+	3.72	}$	&	$	0.97	_{-	0.23	}^{+	0.15	}$	&	$	0.86	_{-	0.66	}^{+	1.44	}$	&	$	3.79	_{-	2.09	}^{+	2.03	}$\\
	&	$	1.67	_{-	0.21	}^{+	0.2	}$	&	$	3.9	_{-	1.8	}^{+	2.85	}$	&	$	7.67	_{-	2.55	}^{+	3.1	}$	&	$	1.09	_{-	0.12	}^{+	0.11	}$	&	$	2.54	_{-	1.25	}^{+	2	}$	&	$	5.28	_{-	1.6	}^{+	1.79	}$\\
A1300	&	$	1.74	_{-	0.2	}^{+	0.2	}$	&	$	3.65	_{-	2	}^{+	5.75	}$	&	$	8.2	_{-	2.52	}^{+	3.17	}$	&	$	1.13	_{-	0.12	}^{+	0.11	}$	&	$	2.36	_{-	1.37	}^{+	4.05	}$	&	$	5.56	_{-	1.54	}^{+	1.72	}$\\
RXCJ1135.6-2019	&	$	1.62	_{-	0.19	}^{+	0.18	}$	&	$	3.4	_{-	1.6	}^{+	2.1	}$	&	$	6.61	_{-	2.07	}^{+	2.46	}$	&	$	1.04	_{-	0.11	}^{+	0.1	}$	&	$	2.19	_{-	1.1	}^{+	1.47	}$	&	$	4.41	_{-	1.21	}^{+	1.32	}$\\
RXCJ1206.2-0848	&	$	2.04	_{-	0.25	}^{+	0.26	}$	&	$	1.15	_{-	0.7	}^{+	1.15	}$	&	$	15.46	_{-	5.01	}^{+	6.69	}$	&	$	1.17	_{-	0.2	}^{+	0.15	}$	&	$	0.66	_{-	0.43	}^{+	0.77	}$	&	$	7.26	_{-	3.14	}^{+	3.23	}$\\
MACSJ1311.0-0311	&	$	1.63	_{-	0.17	}^{+	0.17	}$	&	$	6.6	_{-	2.35	}^{+	3.9	}$	&	$	8.39	_{-	2.36	}^{+	2.91	}$	&	$	1.1	_{-	0.1	}^{+	0.1	}$	&	$	4.43	_{-	1.65	}^{+	2.76	}$	&	$	6.36	_{-	1.59	}^{+	1.85	}$\\
A1689	&	$	2.62	_{-	0.09	}^{+	0.1	}$	&	$	5.15	_{-	0.8	}^{+	0.9	}$	&	$	24.51	_{-	2.44	}^{+	2.92	}$	&	$	1.74	_{-	0.05	}^{+	0.05	}$	&	$	3.41	_{-	0.56	}^{+	0.63	}$	&	$	17.84	_{-	1.46	}^{+	1.7	}$\\
RXJ1347-1145	&	$	1.91	_{-	0.29	}^{+	0.26	}$	&	$	4.8	_{-	2.7	}^{+	9.55	}$	&	$	12.83	_{-	5	}^{+	5.98	}$	&	$	1.26	_{-	0.15	}^{+	0.13	}$	&	$	3.17	_{-	1.88	}^{+	6.75	}$	&	$	9.22	_{-	2.88	}^{+	3.06	}$\\
	&	$	2.1	_{-	0.18	}^{+	0.19	}$	&	$	5.7	_{-	2.3	}^{+	3.75	}$	&	$	17.05	_{-	4.02	}^{+	5.06	}$	&	$	1.4	_{-	0.1	}^{+	0.1	}$	&	$	3.8	_{-	1.61	}^{+	2.65	}$	&	$	12.63	_{-	2.61	}^{+	2.95	}$\\
MACSJ1359.2-1929	&	$	1.51	_{-	0.28	}^{+	0.25	}$	&	$	3.1	_{-	2.15	}^{+	3.8	}$	&	$	6.31	_{-	2.9	}^{+	3.68	}$	&	$	0.96	_{-	0.18	}^{+	0.15	}$	&	$	1.98	_{-	1.45	}^{+	2.67	}$	&	$	4.11	_{-	1.9	}^{+	2.15	}$\\
A1835	&	$	2.49	_{-	0.19	}^{+	0.22	}$	&	$	2.75	_{-	0.75	}^{+	0.9	}$	&	$	22.69	_{-	4.81	}^{+	6.56	}$	&	$	1.57	_{-	0.1	}^{+	0.11	}$	&	$	1.74	_{-	0.51	}^{+	0.63	}$	&	$	14.3	_{-	2.61	}^{+	3.14	}$\\
RXJ1504	&	$	1.75	_{-	0.2	}^{+	0.2	}$	&	$	3.7	_{-	1.5	}^{+	2.6	}$	&	$	7.56	_{-	2.31	}^{+	2.9	}$	&	$	1.13	_{-	0.11	}^{+	0.1	}$	&	$	2.4	_{-	1.04	}^{+	1.83	}$	&	$	5.14	_{-	1.34	}^{+	1.48	}$\\
A2163	&	$	3.07	_{-	0.45	}^{+	0.32	}$	&	$	1.1	_{-	0.6	}^{+	0.85	}$	&	$	40.28	_{-	15.24	}^{+	13.95	}$	&	$	1.75	_{-	0.18	}^{+	0.17	}$	&	$	0.63	_{-	0.37	}^{+	0.56	}$	&	$	18.59	_{-	5.19	}^{+	5.77	}$\\
A2204	&	$	2.23	_{-	0.28	}^{+	0.36	}$	&	$	2.2	_{-	0.95	}^{+	1.1	}$	&	$	14.63	_{-	4.85	}^{+	8.29	}$	&	$	1.38	_{-	0.14	}^{+	0.16	}$	&	$	1.36	_{-	0.64	}^{+	0.76	}$	&	$	8.63	_{-	2.41	}^{+	3.26	}$\\
RXCJ2014.8-2430	&	$	1.76	_{-	0.38	}^{+	0.46	}$	&	$	3.05	_{-	2	}^{+	6.35	}$	&	$	7.18	_{-	3.72	}^{+	7.23	}$	&	$	1.12	_{-	0.2	}^{+	0.2	}$	&	$	1.95	_{-	1.35	}^{+	4.47	}$	&	$	4.65	_{-	2.03	}^{+	2.96	}$\\
RXCJ2151.0-0736	&	$	1.26	_{-	0.25	}^{+	0.22	}$	&	$	4.85	_{-	2.95	}^{+	8	}$	&	$	3.06	_{-	1.48	}^{+	1.9	}$	&	$	0.83	_{-	0.16	}^{+	0.13	}$	&	$	3.2	_{-	2.05	}^{+	5.65	}$	&	$	2.2	_{-	1.04	}^{+	1.21	}$\\
A2390	&	$	2.03	_{-	0.12	}^{+	0.12	}$	&	$	4.8	_{-	1.15	}^{+	1.45	}$	&	$	11.96	_{-	2	}^{+	2.25	}$	&	$	1.34	_{-	0.07	}^{+	0.07	}$	&	$	3.17	_{-	0.81	}^{+	1.02	}$	&	$	8.59	_{-	1.27	}^{+	1.37	}$\\
MACSJ2214.9-1359	&	$	1.76	_{-	0.25	}^{+	0.24	}$	&	$	4.7	_{-	2.1	}^{+	4.6	}$	&	$	10.67	_{-	3.93	}^{+	4.99	}$	&	$	1.16	_{-	0.14	}^{+	0.13	}$	&	$	3.1	_{-	1.46	}^{+	3.24	}$	&	$	7.63	_{-	2.46	}^{+	2.75	}$\\
MACSJ2243.3-0935	&	$	2.1	_{-	0.21	}^{+	0.21	}$	&	$	1.6	_{-	0.85	}^{+	1.6	}$	&	$	16.97	_{-	4.6	}^{+	5.62	}$	&	$	1.25	_{-	0.16	}^{+	0.12	}$	&	$	0.96	_{-	0.55	}^{+	1.09	}$	&	$	9.01	_{-	3	}^{+	2.9	}$\\
RXCJ2248.7-4431	&	$	2.16	_{-	0.18	}^{+	0.18	}$	&	$	2.35	_{-	0.8	}^{+	1.05	}$	&	$	16.46	_{-	3.78	}^{+	4.47	}$	&	$	1.34	_{-	0.11	}^{+	0.1	}$	&	$	1.46	_{-	0.54	}^{+	0.73	}$	&	$	9.91	_{-	2.15	}^{+	2.29	}$\\
A2537	&	$	2.1	_{-	0.15	}^{+	0.16	}$	&	$	4.05	_{-	1.4	}^{+	2.1	}$	&	$	14.28	_{-	2.85	}^{+	3.52	}$	&	$	1.37	_{-	0.08	}^{+	0.08	}$	&	$	2.64	_{-	0.97	}^{+	1.48	}$	&	$	9.91	_{-	1.65	}^{+	1.84	}$\\
RXCJ2337.6+0016	&	$	1.94	_{-	0.16	}^{+	0.17	}$	&	$	3.05	_{-	1.05	}^{+	1.3	}$	&	$	10.97	_{-	2.5	}^{+	3.14	}$	&	$	1.24	_{-	0.09	}^{+	0.09	}$	&	$	1.95	_{-	0.72	}^{+	0.91	}$	&	$	7.11	_{-	1.41	}^{+	1.61	}$\\

\hline  
\end{tabular}
\end{table*}

\begin{table*}
\caption{Weak lensing mass estimates for the S/N optimized background selection, for the overdensities $\Delta=200$ and $\Delta=500$, without prior on concentration. For the clusters
A907, MACSJ1115 and RXCJ1347, the WFI based masses are listed in the first line, the SUP based masses are listed in the following line without reciting the cluster name.
}             
\label{table:resAP2}      
\centering                          
\setlength\tabcolsep{3.3pt}
\begin{tabular}{c c c c c c c }        
\hline\hline                 
Cluster & $R_{200}$  & $c_{200}$ & $M_{200}$ & $R_{500}$  & $c_{500}$ & $M_{500}$  \\    
&  [Mpc] & & $[10^{14}\mathrm{M}_{\odot}]$ & [Mpc] & &  $[10^{14}\mathrm{M}_{\odot}]$ \\
\hline  
A2744	&	$	2.04	_{-	0.13	}^{+	0.12	}$	&	$	6.65	_{-	2.65	}^{+	6.6	}$	&	$	13.17	_{-	2.36	}^{+	2.46	}$	&	$	1.37	_{-	0.07	}^{+	0.06	}$	&	$	4.47	_{-	1.86	}^{+	4.67	}$	&	$	10	_{-	1.42	}^{+	1.47	}$\\
RXCJ0019.0-2026	&	$	2.01	_{-	0.13	}^{+	0.13	}$	&	$	3.75	_{-	1.05	}^{+	1.1	}$	&	$	12.29	_{-	2.23	}^{+	2.54	}$	&	$	1.3	_{-	0.08	}^{+	0.07	}$	&	$	2.43	_{-	0.73	}^{+	0.77	}$	&	$	8.38	_{-	1.4	}^{+	1.33	}$\\
A2813	&	$	1.96	_{-	0.17	}^{+	0.18	}$	&	$	2.15	_{-	0.85	}^{+	1.1	}$	&	$	11.56	_{-	2.75	}^{+	3.49	}$	&	$	1.21	_{-	0.09	}^{+	0.09	}$	&	$	1.33	_{-	0.57	}^{+	0.76	}$	&	$	6.77	_{-	1.44	}^{+	1.55	}$\\
A209	&	$	2.35	_{-	0.15	}^{+	0.17	}$	&	$	1.70	_{-	0.50	}^{+	0.65	}$	&	$	18.13	_{-	3.25	}^{+	4.23	}$	&	$	1.41	_{-	0.07	}^{+	0.07	}$	&	$	1.02	_{-	0.33	}^{+	0.44	}$	&	$	9.83	_{-	1.3	}^{+	1.49	}$\\
XLSSC 006	&	$	1.46	_{-	0.16	}^{+	0.16	}$	&	$	4.90	_{-	2.20	}^{+	3.90	}$	&	$	5.58	_{-	1.64	}^{+	2.04	}$	&	$	0.97	_{-	0.09	}^{+	0.09	}$	&	$	3.24	_{-	1.54	}^{+	2.75	}$	&	$	4.03	_{-	1.07	}^{+	1.17	}$\\
RXCJ0232.2-4420	&	$	1.52	_{-	0.2	}^{+	0.22	}$	&	$	9.1	_{-	4.75	}^{+	6.9	}$	&	$	5.32	_{-	1.83	}^{+	2.66	}$	&	$	1.04	_{-	0.12	}^{+	0.12	}$	&	$	6.2	_{-	3.35	}^{+	4.89	}$	&	$	4.2	_{-	1.33	}^{+	1.67	}$\\
RXCJ0245.4-5302	&	$	1.48	_{-	0.22	}^{+	0.2	}$	&	$	9.35	_{-	4.5	}^{+	13.95	}$	&	$	5.03	_{-	1.93	}^{+	2.33	}$	&	$	1.01	_{-	0.14	}^{+	0.12	}$	&	$	6.38	_{-	3.17	}^{+	9.91	}$	&	$	3.99	_{-	1.4	}^{+	1.58	}$\\
A383	&	$	1.6	_{-	0.13	}^{+	0.13	}$	&	$	8.4	_{-	2.35	}^{+	3.65	}$	&	$	5.62	_{-	1.26	}^{+	1.49	}$	&	$	1.09	_{-	0.08	}^{+	0.08	}$	&	$	5.71	_{-	1.66	}^{+	2.58	}$	&	$	4.4	_{-	0.91	}^{+	0.98	}$\\
RXCJ0437.1+0043	&	$	1.86	_{-	0.2	}^{+	0.22	}$	&	$	3.7	_{-	1.7	}^{+	2.75	}$	&	$	9.79	_{-	2.83	}^{+	3.9	}$	&	$	1.21	_{-	0.11	}^{+	0.11	}$	&	$	2.4	_{-	1.17	}^{+	1.93	}$	&	$	6.65	_{-	1.64	}^{+	1.89	}$\\
MS0451.6-0305	&	$	1.58	_{-	0.16	}^{+	0.16	}$	&	$	8.3	_{-	3.1	}^{+	6.1	}$	&	$	8.16	_{-	2.24	}^{+	2.74	}$	&	$	1.07	_{-	0.1	}^{+	0.09	}$	&	$	5.63	_{-	2.19	}^{+	4.32	}$	&	$	6.38	_{-	1.62	}^{+	1.84	}$\\
A520	&	$	2.09	_{-	0.17	}^{+	0.2	}$	&	$	1.7	_{-	0.6	}^{+	0.75	}$	&	$	12.7	_{-	2.85	}^{+	4	}$	&	$	1.26	_{-	0.08	}^{+	0.08	}$	&	$	1.02	_{-	0.4	}^{+	0.51	}$	&	$	6.88	_{-	1.18	}^{+	1.44	}$\\
RXCJ0516.6-5430	&	$	2.34	_{-	0.25	}^{+	0.55	}$	&	$	0.5	_{-	0.45	}^{+	0.65	}$	&	$	19.73	_{-	5.67	}^{+	17.44	}$	&	$	1.19	_{-	0.22	}^{+	0.14	}$	&	$	0.26	_{-	0.23	}^{+	0.4	}$	&	$	6.56	_{-	3.02	}^{+	2.67	}$\\
RXCJ0528.9-3927	&	$	1.9	_{-	0.2	}^{+	0.25	}$	&	$	0.9	_{-	0.85	}^{+	0.9	}$	&	$	10.43	_{-	2.96	}^{+	4.68	}$	&	$	1.05	_{-	0.16	}^{+	0.13	}$	&	$	0.5	_{-	0.48	}^{+	0.59	}$	&	$	4.44	_{-	1.77	}^{+	1.8	}$\\
RXCJ0532.9-3701	&	$	1.89	_{-	0.13	}^{+	0.13	}$	&	$	4	_{-	1.25	}^{+	1.75	}$	&	$	10.16	_{-	1.96	}^{+	2.24	}$	&	$	1.23	_{-	0.08	}^{+	0.07	}$	&	$	2.61	_{-	0.87	}^{+	1.23	}$	&	$	7.03	_{-	1.22	}^{+	1.3	}$\\
A3404	&	$	1.94	_{-	0.2	}^{+	0.21	}$	&	$	5.85	_{-	2	}^{+	3.15	}$	&	$	9.79	_{-	2.73	}^{+	3.53	}$	&	$	1.3	_{-	0.12	}^{+	0.11	}$	&	$	3.91	_{-	1.4	}^{+	2.22	}$	&	$	7.28	_{-	1.8	}^{+	2.1	}$\\
Bullet	&	$	1.84	_{-	0.19	}^{+	0.22	}$	&	$	3.25	_{-	1.3	}^{+	1.55	}$	&	$	9.59	_{-	2.67	}^{+	3.87	}$	&	$	1.18	_{-	0.11	}^{+	0.11	}$	&	$	2.08	_{-	0.89	}^{+	1.08	}$	&	$	6.32	_{-	1.62	}^{+	2	}$\\
A907	&	$	1.6	_{-	0.21	}^{+	0.21	}$	&	$	3.1	_{-	1.35	}^{+	2.25	}$	&	$	5.41	_{-	1.86	}^{+	2.42	}$	&	$	1.02	_{-	0.11	}^{+	0.11	}$	&	$	1.98	_{-	0.92	}^{+	1.58	}$	&	$	3.52	_{-	1.05	}^{+	1.2	}$\\
	&	$	1.53	_{-	0.2	}^{+	0.23	}$	&	$	2.45	_{-	1.2	}^{+	2	}$	&	$	4.73	_{-	1.62	}^{+	2.47	}$	&	$	0.96	_{-	0.1	}^{+	0.1	}$	&	$	1.53	_{-	0.81	}^{+	1.39	}$	&	$	2.88	_{-	0.83	}^{+	1.03	}$\\
RXCJ1023.6+0411	&	$	1.83	_{-	0.1	}^{+	0.09	}$	&	$	5.65	_{-	1.05	}^{+	1.35	}$	&	$	9.38	_{-	1.46	}^{+	1.45	}$	&	$	1.22	_{-	0.06	}^{+	0.06	}$	&	$	3.77	_{-	0.74	}^{+	0.95	}$	&	$	6.94	_{-	0.97	}^{+	0.99	}$\\
MS1054.4-0321	&	$	2.07	_{-	0.36	}^{+	0.39	}$	&	$	6.65	_{-	5.3	}^{+	9.35	}$	&	$	25.65	_{-	11.19	}^{+	17.4	}$	&	$	1.39	_{-	0.22	}^{+	0.19	}$	&	$	4.47	_{-	3.68	}^{+	6.62	}$	&	$	19.47	_{-	7.74	}^{+	8.82	}$\\
MACSJ1115.8+0129	&	$	1.57	_{-	0.23	}^{+	0.21	}$	&	$	1.5	_{-	1.05	}^{+	2.1	}$	&	$	6.37	_{-	2.41	}^{+	2.92	}$	&	$	0.93	_{-	0.22	}^{+	0.14	}$	&	$	0.89	_{-	0.66	}^{+	1.44	}$	&	$	3.31	_{-	1.84	}^{+	1.7	}$\\
	&	$	1.61	_{-	0.19	}^{+	0.17	}$	&	$	3.7	_{-	1.6	}^{+	2.7	}$	&	$	6.87	_{-	2.16	}^{+	2.42	}$	&	$	1.04	_{-	0.11	}^{+	0.1	}$	&	$	2.4	_{-	1.11	}^{+	1.9	}$	&	$	4.67	_{-	1.35	}^{+	1.47	}$\\
A1300	&	$	1.82	_{-	0.19	}^{+	0.19	}$	&	$	2.5	_{-	1.35	}^{+	2.9	}$	&	$	9.38	_{-	2.64	}^{+	3.26	}$	&	$	1.14	_{-	0.12	}^{+	0.1	}$	&	$	1.57	_{-	0.91	}^{+	2.02	}$	&	$	5.75	_{-	1.57	}^{+	1.63	}$\\
RXCJ1135.6-2019	&	$	1.53	_{-	0.17	}^{+	0.16	}$	&	$	4.15	_{-	1.7	}^{+	2.45	}$	&	$	5.57	_{-	1.66	}^{+	1.94	}$	&	$	1	_{-	0.1	}^{+	0.09	}$	&	$	2.71	_{-	1.18	}^{+	1.72	}$	&	$	3.89	_{-	1.04	}^{+	1.13	}$\\
RXCJ1206.2-0848	&	$	2.07	_{-	0.19	}^{+	0.2	}$	&	$	1.9	_{-	0.85	}^{+	1.5	}$	&	$	16.15	_{-	4.05	}^{+	5.15	}$	&	$	1.26	_{-	0.13	}^{+	0.12	}$	&	$	1.16	_{-	0.56	}^{+	1.03	}$	&	$	9.09	_{-	2.49	}^{+	2.74	}$\\
MACSJ1311.0-0311	&	$	1.61	_{-	0.17	}^{+	0.16	}$	&	$	4	_{-	1.6	}^{+	2.4	}$	&	$	8.08	_{-	2.3	}^{+	2.66	}$	&	$	1.05	_{-	0.1	}^{+	0.09	}$	&	$	2.61	_{-	1.11	}^{+	1.69	}$	&	$	5.59	_{-	1.44	}^{+	1.64	}$\\
A1689	&	$	2.58	_{-	0.08	}^{+	0.08	}$	&	$	5.45	_{-	0.75	}^{+	0.95	}$	&	$	23.41	_{-	2.11	}^{+	2.25	}$	&	$	1.72	_{-	0.05	}^{+	0.05	}$	&	$	3.62	_{-	0.53	}^{+	0.67	}$	&	$	17.21	_{-	1.36	}^{+	1.43	}$\\
RXJ1347-1145	&	$	2.06	_{-	0.22	}^{+	0.21	}$	&	$	2.7	_{-	1.3	}^{+	2.55	}$	&	$	16.09	_{-	4.63	}^{+	5.44	}$	&	$	1.3	_{-	0.12	}^{+	0.11	}$	&	$	1.7	_{-	0.88	}^{+	1.78	}$	&	$	10.09	_{-	2.47	}^{+	2.64	}$\\
	&	$	2.08	_{-	0.17	}^{+	0.16	}$	&	$	4.85	_{-	1.6	}^{+	2.2	}$	&	$	16.57	_{-	3.74	}^{+	4.13	}$	&	$	1.37	_{-	0.1	}^{+	0.09	}$	&	$	3.2	_{-	1.12	}^{+	1.55	}$	&	$	11.93	_{-	2.31	}^{+	2.43	}$\\
MACSJ1359.2-1929	&	$	1.38	_{-	0.28	}^{+	0.24	}$	&	$	2.7	_{-	2.25	}^{+	4.95	}$	&	$	4.82	_{-	2.38	}^{+	2.97	}$	&	$	0.87	_{-	0.24	}^{+	0.14	}$	&	$	1.7	_{-	1.48	}^{+	3.47	}$	&	$	3.02	_{-	1.89	}^{+	1.69	}$\\
A1835	&	$	2.42	_{-	0.17	}^{+	0.2	}$	&	$	3	_{-	0.8	}^{+	0.95	}$	&	$	20.83	_{-	4.09	}^{+	5.6	}$	&	$	1.54	_{-	0.1	}^{+	0.1	}$	&	$	1.91	_{-	0.55	}^{+	0.66	}$	&	$	13.44	_{-	2.33	}^{+	2.77	}$\\
RXJ1504	&	$	1.68	_{-	0.17	}^{+	0.18	}$	&	$	4.05	_{-	1.55	}^{+	2.2	}$	&	$	6.69	_{-	1.83	}^{+	2.39	}$	&	$	1.1	_{-	0.09	}^{+	0.09	}$	&	$	2.64	_{-	1.08	}^{+	1.55	}$	&	$	4.64	_{-	1.07	}^{+	1.29	}$\\
A2163	&	$	2.87	_{-	0.36	}^{+	0.52	}$	&	$	1.4	_{-	0.75	}^{+	0.9	}$	&	$	32.91	_{-	10.9	}^{+	21.33	}$	&	$	1.69	_{-	0.15	}^{+	0.18	}$	&	$	0.82	_{-	0.48	}^{+	0.61	}$	&	$	16.65	_{-	4.07	}^{+	5.91	}$\\
A2204	&	$	2.19	_{-	0.26	}^{+	0.19	}$	&	$	2.5	_{-	0.6	}^{+	1.2	}$	&	$	13.86	_{-	4.37	}^{+	3.93	}$	&	$	1.37	_{-	0.14	}^{+	0.1	}$	&	$	1.57	_{-	0.41	}^{+	0.83	}$	&	$	8.5	_{-	2.3	}^{+	1.94	}$\\
RXCJ2014.8-2430	&	$	1.59	_{-	0.31	}^{+	0.33	}$	&	$	3.1	_{-	1.9	}^{+	4.05	}$	&	$	5.29	_{-	2.53	}^{+	4.03	}$	&	$	1.02	_{-	0.18	}^{+	0.17	}$	&	$	1.98	_{-	1.29	}^{+	2.84	}$	&	$	3.44	_{-	1.52	}^{+	1.99	}$\\
RXCJ2151.0-0736	&	$	1.21	_{-	0.23	}^{+	0.2	}$	&	$	6.25	_{-	3.55	}^{+	9.75	}$	&	$	2.71	_{-	1.27	}^{+	1.58	}$	&	$	0.81	_{-	0.14	}^{+	0.12	}$	&	$	4.19	_{-	2.49	}^{+	6.9	}$	&	$	2.04	_{-	0.89	}^{+	1.04	}$\\
A2390	&	$	2.06	_{-	0.11	}^{+	0.11	}$	&	$	4.75	_{-	1.05	}^{+	1.35	}$	&	$	12.5	_{-	1.9	}^{+	2.11	}$	&	$	1.36	_{-	0.06	}^{+	0.06	}$	&	$	3.13	_{-	0.74	}^{+	0.95	}$	&	$	8.96	_{-	1.2	}^{+	1.31	}$\\
MACSJ2214.9-1359	&	$	1.89	_{-	0.22	}^{+	0.22	}$	&	$	4.35	_{-	1.70	}^{+	3.10	}$	&	$	13.21	_{-	4.10	}^{+	5.17	}$	&	$	1.24	_{-	0.12	}^{+	0.11	}$	&	$	2.85	_{-	1.18	}^{+	2.18	}$	&	$	9.31	_{-	2.44	}^{+	2.81	}$\\
MACSJ2243.3-0935	&	$	2.13	_{-	0.20	}^{+	0.20	}$	&	$	1.30	_{-	0.65	}^{+	1.20	}$	&	$	17.71	_{-	4.53	}^{+	5.47	}$	&	$	1.24	_{-	0.15	}^{+	0.12	}$	&	$	0.76	_{-	0.41	}^{+	0.81	}$	&	$	8.72	_{-	2.77	}^{+	2.79	}$\\
RXCJ2248.7-4431	&	$	2.29	_{-	0.17	}^{+	0.16	}$	&	$	2.05	_{-	0.6	}^{+	0.75	}$	&	$	19.61	_{-	4.05	}^{+	4.41	}$	&	$	1.41	_{-	0.1	}^{+	0.08	}$	&	$	1.26	_{-	0.4	}^{+	0.51	}$	&	$	11.32	_{-	2.15	}^{+	2.14	}$\\
A2537	&	$	2.16	_{-	0.13	}^{+	0.13	}$	&	$	4.3	_{-	1.25	}^{+	1.85	}$	&	$	15.54	_{-	2.64	}^{+	2.98	}$	&	$	1.42	_{-	0.07	}^{+	0.07	}$	&	$	2.82	_{-	0.87	}^{+	1.3	}$	&	$	10.92	_{-	1.53	}^{+	1.71	}$\\
RXCJ2337.6+0016	&	$	1.94	_{-	0.16	}^{+	0.17	}$	&	$	3.05	_{-	1.05	}^{+	1.30	}$	&	$	10.97	_{-	2.50	}^{+	3.14	}$	&	$	1.24	_{-	0.09	}^{+	0.09	}$	&	$	1.95	_{-	0.72	}^{+	0.91	}$	&	$	7.11	_{-	1.4	}^{+	1.61	}$\\
\hline  
\end{tabular}
\end{table*}

\section{Color composite images with weak lensing convergence contours}
\begin{figure*}
{\includegraphics[width=1.00\textwidth]{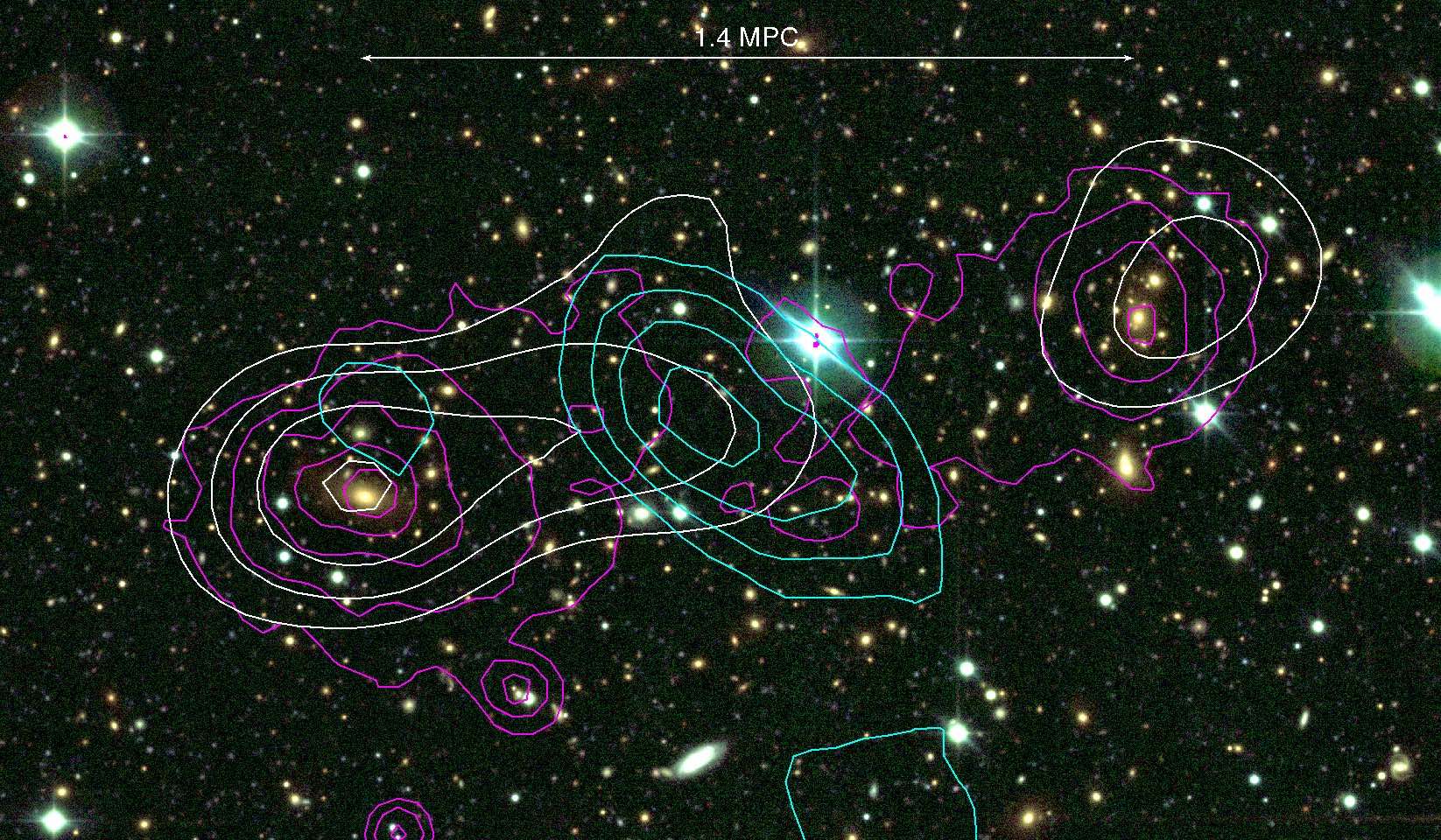}}
\caption{RXCJ1135.6-2019: Color composite image with lensing convergence contours in white, XMM surface brightness contours in magenta and SZ contours in cyan (top panel), XMM surface brightness map with lensing convergence contours. Note: SZ contours only indicative. This cluster is a SZ non-detection when centered on the BCG.}\label{im:optcontAP}
\end{figure*}

We provide RGB color images with weak lensing contours
for the central regions of all clusters presented in this work as online only supplementary data.
As example for the available material we show in Fig.~\ref{im:optcontAP}
the RGB color image, with contours for the cluster RXCJ1135.6-2019. One of the two outliers discussed in \cite{2018Nagarajan}.

%


\bsp	
\label{lastpage}
\end{document}